\documentclass{aa} 
\usepackage{graphicx} 
\usepackage{txfonts} 
\usepackage{hyperref} 
\usepackage{natbib}
\bibpunct{(}{)}{;}{a}{}{,} 
\usepackage{float}		
\usepackage{amssymb}
\usepackage[table,dvipsnames]{xcolor}
\usepackage{multirow} 
\usepackage{placeins}  

\usepackage{subcaption}
\usepackage{comment}

\newcommand{\ro}[1]{{#1}}  

\newcommand{\rt}[1]{{#1}}  

\def\nrao{\object{NRAO\,530}\xspace}
\def\nraonum{\object{1730$-$130}\xspace}
\def\sgra{\object{Sgr\,A$^{*}$}\xspace}

\newcommand{\kr}{$k_{\rm r} $}

\graphicspath{{./}{figs/}}

\begin{document}
	
\title{Kilogauss magnetic field and jet dynamics in the quasar \nrao}
\titlerunning{Kilogauss B-field in quasar \nrao}
\authorrunning{M. Lisakov}

\author{Mikhail Lisakov\inst{1, 2, 3}\fnmsep\thanks{mikhail@lisakov.com}    
\and    Svetlana Jorstad\inst{4,5}
\and    Maciek Wielgus\inst{6,2}
\and    Evgeniya V. Kravchenko\inst{3,7}
\and    Aleksei S. Nikonov\inst{2}
\and    Ilje Cho\inst{8,9,6}
\and    Sara Issaoun\inst{10,11}
\and    Juan-Carlos Algaba\inst{12}
\and    Thomas P. Krichbaum\inst{2}
\and    Uwe Bach\inst{2}
\and    Eduardo Ros\inst{2}
\and    Helge Rottmann\inst{2}
\and    Salvador S\'anchez\inst{13}
\and    Jan Wagner\inst{2}
\and    Anton Zensus\inst{2}
}
\institute{
    Instituto de F\'{i}sica, Pontificia Universidad Cat\'{o}lica de Valpara\'{i}so, Casilla 4059, Valpara\'{i}so, Chile
    \and
    Max-Planck-Institut f\"ur Radioastronomie, Auf dem H\"ugel 69, D-53121 Bonn, Germany 
    \and
    Astro Space Center, Lebedev Physical Institute, Profsouznaya 84/32, Moscow 117997, Russia
    \and
    Institute for Astrophysical Research, Boston University, 725 Commonwealth Ave., Boston, MA 02215, USA
    \and
    Saint Petersburg State University, Universitetskaya nab. 7-9, St. Petersburg, 199034, Russia
    \and
    Instituto de Astrof\'{\i}sica de Andaluc\'{\i}a-C\'{\i}SIC, Glorieta de la Astronom\'{\i}a s/n, E-18008 Granada, Spain
    \and
    Moscow Institute of Physics and Technology, Institutsky per. 9, Moscow region, Dolgoprudny, 141700, Russia
    \and
    Korea Astronomy and Space Science Institute, Daedeok-daero 776, Yuseong-gu, Daejeon 34055, Republic of Korea
    \and
    Department of Astronomy, Yonsei University, Yonsei-ro 50, Seodaemun-gu, Seoul 03722, Republic of Korea
    \and
    Center for Astrophysics | Harvard \& Smithsonian, 60 Garden Street, Cambridge, MA 02138, USA
    \and
    NASA Hubble Fellowship Program, Einstein Fellow
    \and
    Department of Physics, Faculty of Science, University of Malaya, 50603 Kuala Lumpur, Malaysia
    \and   
    IRAM (Instituto de Radioastronomía Milimétrica) Avda. Divina Pastora 7, Local 20 18012  Granada, Spain 
}

\date{Received  , 2024; accepted , 2024}


\abstract
{The advancement of the Event Horizon Telescope has enabled the study of relativistic jets in active galactic nuclei down to sub-parsec linear scales even at high redshift. Quasi-simultaneous multifrequency observations provide insights into the physical conditions in compact regions and allow testing accretion theories. 
}{
Initially we aimed at measuring the magnetic field strength close to the central supermassive black hole in \nrao (\nraonum) by studying frequency-dependent opacity of the jet matter, Faraday rotation and the spectral index in the mm-radio bands. 
}{
\nrao was observed quasi-simultaneously at 15, 22, 43, 86, and 227\,GHz at four different very long baseline interferometer (VLBI) networks . By the means of imaging and model-fitting, we aligned the images, taken at different frequencies. We explored opacity along the jet and distribution of the linearly polarized emission in it.
}{
Our findings reveal that the jet of \nrao at 86 and 227\,GHz is transparent down to its origin, with 70\,mJy emission detected at 227\,GHz potentially originating from the accretion disk. The magnetic field strength near the black hole, estimated at $5r_\mathrm{g}$, is $3\times10^3-3\times10^4$\,G (depending on the central black hole mass). These values represent some of the highest magnetic field strengths reported for active galaxies. We also report the first ever VLBI measurement of the Faraday rotation at 43--227\,GHz, which reveals rotation measure values as high as -48000 rad/m2 consistent with higher particle density and stronger magnetic fields at the jet's outset. The complex shape of the jet in \nrao is in line with the expected behavior of a precessing jet, with a period estimated to be around $6\pm4$~years.
}{}
\maketitle

\section{Introduction}
\label{sec:intro} 
The quasar \nrao (1730$-$130, J1733$-$1304) is a radio-loud compact source that belongs to the sub-class of blazars, which exhibit highly variable non-thermal emission. It possesses highly relativistic jet with an apparent speed exceeding 30~c \citep{Weaver22}, bright $\gamma$-ray emission \citep{4FGL}, and prominent variability of the optical linear polarization\footnote{\url{http://www.bu.edu/blazars/VLBA\_GLAST/1730.html}}. On parsec scales the jet is directed to the north \citep[e.g.,][]{2019ApJ...874...43L}, while the kilo-parsec scale jet is perpendicular to this direction, extending from east to west over $3''$\citep{Kharb2010}. Although quasars are generally associated with the Fanaroff-Riley II (FRII) type of extragalactic radio sources, the \nrao kilo-parsec scale jet has a hybrid morphology, with a diffuse eastern portion displaying FRI-type features, but compact emission on the western side agrees better with the FRII properties. This complex structure of the jet can be explored by multi-frequency radio observations. Such a probe of the properties of the radio jet was presented in \cite{Rusen2010} using quasi-simultaneous observations obtained in 2007 at 86, 43, 22 GHz, and 15 GHz. That author identified the most southern component of the jet as the apparent ``core'' based on its flat spectral index, while components farther down the jet have optically thin spectral indices that evolve with distance from the core in a quasi-sinusoidal pattern with increasing characteristic scale. The data also indicate a frequency dependent core shift not only along the jet but also transverse to the jet. 
\cite{2012AJ....144..105H,Hovatta2014} have analyzed the spectral characteristics and Faraday Rotation Measure (RM) in the parsec-scale jet of \nrao at four lower frequencies between 8.1 and 15.4\,GHz within the MOJAVE (Monitoring of Jets in Active Galactic Nuclei with VLBA Experiments) sample. They have also found a flat spectral index of the core ($\alpha=-0.10$, $S_\nu\propto\nu^\alpha$) and a steep spectral index of the extended jet ($\alpha=-1.14$), as well as relatively low values of RM both \ro{around the} core \ro{region} (371~rad~m$^{-2}$) and in the jet (0.9~rad~m$^{-2}$). 
\cite{AN2013} have investigated periodicity in the radio light curves of \nrao at 14.5, 8, and 4.5 GHz using the UMRAO (University of Michigan Radio Astronomy Observatory) data \citep{Aller1985} from 1967 to 2012. They have detected two strong and persistent periods of $\sim$10~yr and $\sim$6~yr and several weaker shorter timescale periodicity of $\sim$3.5, $\sim$3, and $\sim$2 yr. The authors found that the characteristic frequencies of these periodicity have a harmonic relationship. \cite{AN2013} suggested that the multiplicity and apparent harmonic relation of the periodicity can be interpreted by the global p-mode oscillation of the accretion disk, implying a disk-jet connection. However, it can be explained also by magneto-hydrodynamic (MHD) instabilities \citep{istomin1996stability, Narayan_2009, mizuno2012three} or Lense–Thirring precession \citep{thirring1918effect} observed by \cite{2023ApJ...951..106B, 2023A&A...672L...5V, cui_etal2023}.

Owing to its brightness, compact nature, and position on the sky, \nrao is often used as a calibrator for Very Long Baseline Interferometric (VLBI) observations of the radio source Sagittarius A$^*$ (\sgra) located in the center of our Galaxy \citep[e.g.,][]{Lu2011}. This was also the case for the Event Horizon Telescope observations of \sgra in 2017 April \citep{SgraP1}. \cite{Jorstad2023} presented the total and polarized intensity images of \nrao obtained during the EHT campaign at 227\,GHz. The source has a multi-component structure extended to the north-west, with linear polarization detected in the core and jet. For the purpose of this research, we have collected VLBI data of the quasar at 86, 43, 22, and 15~GHz contemporaneous to the EHT observation. With an EHT resolution of $\sim$20 micro-arcsecond ($\mu$as) and the extension of the jet at 15 GHz out to 25~mas from the core, the combined data set covers a wide range of spacial scales, from sub-parsec ($\sim$0.15 pc) to $\sim$200 pc projected distances, based on the quasar redshift of z=0.902 \citep{Junkkarinen1984}. This corresponds to the linear scale of 7.9 pc~mas$^{-1}$ with the cosmological parameters $H_\circ=67.7$~km~s$^{-1}$Mpc$^{-1}$, $\Omega_M=0.307$, and $\Omega_\Lambda=0.693$ \citep{CosmoCalc2006, Planck2016}. In accord with \cite{Jorstad2023}, we adopt the following parameters of the source: Doppler factor at mas-scales $\delta=9$, viewing angle $\alpha=3^\circ$, and bulk Lorentz factor $\Gamma=8$ \citep{Weaver22}.

The multi-frequency VLBI data allow us to study the structure, spectral characteristics, and polarization properties of the jet of \nrao on multiple scales, from the vicinity of black hole (BH) to hundreds of parsecs down the jet. The outline of the paper is as follows: data sources, their issues, and data reduction methods are discussed in Sect.~\ref{sec:data};  the magnetic field is estimated in Sect.~\ref{sec:bfield}; spacial structure of \nrao at multiple frequencies is covered in Sect.~\ref{sec:fitting_pr_instab}; brightness temperature is discussed in Sect.~\ref{sec:tb}; source structure in linear polarization and Faraday rotation measure are presented in Sect.~\ref{sec:polar}; and, finally, a coherent model of the source \ro{is discussed in Sect.~\ref{sec:discussion}.}

\section{Data and Methods}
\label{sec:data}
\subsection{227 GHz}
\nrao was observed by the Event Horizon Telescope (EHT) on 2017 April 5-7, as a calibrator of Sagittarius A$^*$ \citep{SgraP1,SgraP2}. The source was observed with the full EHT array composed of 8 stations at 6 geographical sites: the Atacama Large Millimeter/submillimeter Array \citep[ALMA, operating as a phased array;][]{Goddi2019} and the Atacama Pathfinder Experiment (APEX) telescope in Chile; the Large Millimeter Telescope Alfonso Serrano (LMT) in Mexico; the IRAM (Instituto de Radioastronom\'ia Milim\'etrica) 30\,m telescope (PV) in Spain; the Submillimeter Telescope (SMT) in Arizona; the James Clerk Maxwell Telescope (JCMT) and the Submillimeter Array (SMA) in Hawai'i; and the South Pole Telescope (SPT) in Antarctica. Two 2\,GHz-wide frequency bands, centered at 227.1\,GHz, and 229.1\,GHz were recorded with full polarization, with an exception of a single-polarization JCMT. Subsequent data reduction, calibration and validation were described in \citet{Blackburn_2019,Janssen2019, EHT_M87_P3}, with minor updates of the data reduction pipeline described in \citet{SgraP2}, and the detailed polarimetric leakage calibration follows \citet{PaperVII} and \citet{Issaoun2022}. The absolute calibration of the electric vector position angle (EVPA) follows the highly accurate polarimetric calibration of ALMA \citep{Goddi2019,PaperVII}. Detailed analysis of the \nrao EHT data set and resulting total intensity and linear polarization images was performed by \citet{Jorstad2023}.
In this study we use the fiducial total and polarized intensity images presented in \citet{Jorstad2023} which are obtained by combining data over all days and both frequencies, since no unambiguous temporal of spectral variability has been detected during the 3 consecutive days of the EHT observations.

     
\begin{table*}
    \caption{EHT and complementary VLBI observations of \nrao in Spring 2017.}
    \label{tab:observ}
    \centering
    \begin{tabular}{lccccc} 
    \hline
    $\nu_{\rm obs}$ & Program & Obs. date & Antennas & Beam & Reference\\
    (GHz)           &         &           &          & (mas)\\
    \hline
    15 & MOJAVE & 2017-01-03 & VLBA & 0.50x1.26, -0.9 & \citet{2018ApJS..234...12L}\\
       &        & 2017-05-25 &      & 0.48x1.16, -2.7 & \\
    \rowcolor{gray!10}
    22 & EAVN & 2017-04-03 &  KYS,\,KUS,\,KTN,\,MIZ,\,IRK,\,OGA,\,ISG,\,TIA & 2.07$\times$0.99, 19.5$^\circ$ & \citet{Cho2022}\\
    43 & BEAM-ME & 2017-03-19 & VLBA & 0.17x0.48, -9.3 & \citet{Jorstad2017}\\
       &         & 2017-04-16 &      & 0.17x0.48, -12.2 & \\
    \rowcolor{gray!10}
      86 & GMVA & 2017-04-03 & VLBA, ALMA, GB, YS, PV, EB & 0.112$\times$0.085, 30.6$^\circ$ & \citet{Issaoun_2019}\\
    227 & EHT & 2017-04-06 & ALMA,\,APEX,\,SPT,\,PV,\,LMT,\,SMT,\,JCMT,\,SMA & 0.026$\times$0.016, 71.1$^\circ$ & \citet{Jorstad2023} \\
     \hline
    \end{tabular}
\end{table*}

\subsection{86 GHz}
    
The source was observed with the Global Millimeter VLBI Array (GMVA), consisting of the eight Very Long Baseline Array (VLBA) stations equipped with 86\,GHz receivers, the Green Bank Telescope (GB), the Yebes 40-m telescope (YS), the IRAM 30-m telescope (PV), the Effelsberg 100-m telescope (EB), and the ALMA phased array. The observations were conducted on 2017 April 3 over 7.5 hr, as a part of a Sgr A$^*$ observing campaign (GMVA project MB007), for which \nrao served as a calibrator \citep{Issaoun_2019}. The data were recorded at four intermediate frequency bands (IFs) centered at 86.06, 86.12, 86.18, and 86.24\,GHz, with a total bandwidth of 256\,MHz per polarization. Corresponding total intensity images of \nrao, obtained through closure-only imaging method \citep{Chael_2018} were published by \citet{Issaoun_2019}. For the consistency with other images obtained at low frequencies (see below) we have performed both total and polarized intensity imaging of the 86\,GHz data using the traditional VLBI \textsc{CLEAN} method using \textsc{Difmap}\citep{Difmap} software package, with the approach for high frequency VLBI data described in \citet{Casadio2019}. Due to uncertain amplitude calibration of some GMVA stations in this observation, the total VLBI-flux at 86\,GHz was scaled to 2.74~Jy  measured by ALMA on 2017 April 3 \citep{Goddi2019}. The data were corrected for the polarimetric leakage D-terms  following the prescription defined in \citet{Leppanen1995} using \textsc{LPCAL} task in Astronomical Imaging Processing System (AIPS) supplied by NRAO (National Radio Astronomy Observatory). We have performed several analyses to search for D-terms: 1) using \nrao for determining D-terms for each IF; 2) using \nrao with data combined over all IFs; and 3) averaging D-term solutions derived for \nrao and OJ~287 observed in the same run \citep{Zhao2022} with data combined over all IFs. We have found that the lowest level of the noise (rms) for Stokes~Q and U maps is achieved in the case of data corrected with D-terms obtained for the \nrao solutions with data combined over IFs. The EVPA calibration was obtained by comparison of the polarization position angle integrated over the \nrao image with that measured by ALMA \citep{Goddi2019} which provides an accuracy of EVPA measurements of $\pm$6$^\circ$.


\subsection{43 GHz}
The dataset at 43\,GHz ($\lambda=7$~mm) is obtained as part of the VLBA-BU-BLAZAR program\footnote{\url{https://www.bu.edu/blazars/VLBAproject.html}}, which includes monthly observations of a sample of 38 radio and $\gamma$-ray bright AGN over 24~hrs at each epoch. The data and calibration details description are given in \citet{Jorstad2017}. \nrao was observed in 2017 March 19 and April 16 during 50 and 45\,min integration time per source (9 and 10 scans of ~5-min duration), respectively, at a recording rate of 2048\,Mb per second (Mbps). Observations were performed in full polarization mode in 256\,MHz bandwidth, which is split into four IFs centered at 43.0075, 43.0875, 43.1515, and 43.2155\,GHz. A final set of D-terms used at each epoch was obtained by averaging solutions over 15 sources in the sample for which there is the best agreement between the D-terms. The EVPA calibration was obtained by different methods: comparison between the Very Large Array(VLA) and VLBA integrated EVPAs at quasi-simultaneous epochs, the D-term method \citep{Gomez2002}, and using EVPA-stable features (for a period of several months) in the images of the jets of 3C 279, OJ 287, 3C 446, and 3C 454.3, which allow us to reach an accuracy of $\pm$5$^\circ$.

\subsection{22 GHz}
The 22\,GHz ($\lambda=1.3$~cm) data is obtained from the East Asian VLBI Network (EAVN) which consists of the KaVA 
(KVN\footnote{Korean VLBI Network: 21\,m telescopes of Yonsei (KYS), Ulsan (KUS), and Tamna (KTN) in Korea} 
and VERA\footnote{VLBI Exploration of Radio Astrometry: 20\,m telescopes of Mizusawa (MIZ), Iriki (IRK), Ogasawara (OGA), and Ishigakijima (ISG) in Japan} Array; e.g., \citealt{Niinuma_2015, Hada_2017, Park_2019}) 
and additional East-Asian telescopes (e.g., Tianma-65m, Nanshan-26m, and Hitachi-32m telescopes; \citealt{Cui_2021}). 
In this study, the EAVN observation on April~3 has been analyzed which is a part of the KaVA/EAVN Large Program towards \sgra \citep{Cho2022}. The \nrao has been observed as one of the calibrators with on-source time $\sim$30\,minutes. The data are recorded with 256\,MHz (32\,MHz$\,\times\,$8\,channels) total bandwidth, single RCP polarization. 
Total 10 telescopes have participated (7 telescopes of KaVA, Tianma-65m, Nanshan-26m, and Hitachi-32m), but the Nanshan-26m and Hitachi-32m were flagged out due to a severe problem with \ro{their antenna gains}. 

\subsection{15\,GHz}
\nrao is included \ro{in the} regular monitoring of the 2\,cm VLBA survey \citep{1998AJ....115.1295K}, which nowadays is followed by the MOJAVE\footnote{\url{https://www.cv.nrao.edu/MOJAVE/allsources.html}} program \citep{2021ApJ...923...30L}. The closest observations to the 2017 EHT session were performed on January~3 and May~25. The data were taken in dual polarization mode at a central frequency of 15.369\,GHz. The data were recorded at 2048 Mbps with 2-bit sampling in a bandwidth of 256\,MHz per polarization. The electric vector position angle (EVPA) and flux density accuracy are estimated to be $\thicksim 5$\degr and $\thicksim 5$\%, respectively \citep{2018ApJS..234...12L}.

\subsection{Modelling and cross-frequency alignment of images}
\label{sec:alignment}

The first essential step that allows us to extract physical information from the images and UV-data is fitting the structure of the jet with a small number of 2D Gaussians, hereafter named components. This fitting was performed in the UV-domain using the \textsc{modelfit} procedure in \textsc{Difmap}. \ro{Parameters of all model components are listed in Table~\ref{tab:models}.} Optically thin components at different frequencies were used to determine the shift between images at these frequencies, a crucial step for all subsequent analysis. This step is required since the information about the absolute coordinates of the source is lost during the data reduction and imaging processes.

\ro{To obtain the most accurate image alignment}, we performed it in two steps. First, for each pair of adjacent frequencies, we have fitted the models using only a common range of UV-distances in order to equalize resolution. This was possible for all frequency pairs except for 86-227\,GHz, which have only 5\% of their visibilities in the common UV-range. This approach allowed us to get similar structure of the jet at both frequencies in the pair and provided a robust estimate of the alignment shift between the maps. For most of the frequency pairs there were several pairs of cross-identified components. The final shift was calculated as an average of all pairs of cross-identified components weighted with the ratio of their flux density and size\rt{, as follows:}

\rt{
\begin{equation}
    \mathbf{MS} = \frac{\sum_{i=1}^{n} (\mathbf{R}_i^{\nu_1} - \mathbf{R}_i^{\nu_2}) w_{\nu_1} w_{\nu_2} }{\sum_{i=1}^{n} w_{\nu_1} w_{\nu_2}},
\end{equation}
}
\rt{where $\mathbf{R}_i^{\nu_\mathrm{x}}$ is component's position at frequency $\nu_\mathrm{x}$. The weights are calculated as:
}
\rt{
\begin{equation}
    w_{\nu_\mathrm{x}} = \frac{F_i^{\nu_\mathrm{x}} / S_i^{\nu_\mathrm{x}} }{\sum_{i=1}^{n} F_i^{\nu_\mathrm{x}} / S_i^{\nu_\mathrm{x}}}
\end{equation}
}
\rt{where $F_i^{\nu_\mathrm{x}}$ and $S_i^{\nu_\mathrm{x}}$ are component's flux density and size at frequency $\nu_\mathrm{x}$. Normalizing weights to the average at a given frequency accounts for different angular resolution and total flux at different frequencies.  With this approach, the brightest and the most compact components contributed most to the derived shift.} \ro{Since we have used distinct, bright, and compact components for this analysis, the uncertainty of the derived shift is better than $10\,\mu$as in all cases.} Examples of the alignment are collected in Appendix~\ref{app:image_alignment}.

\begin{figure}
    \centering
    \includegraphics[width=\columnwidth]{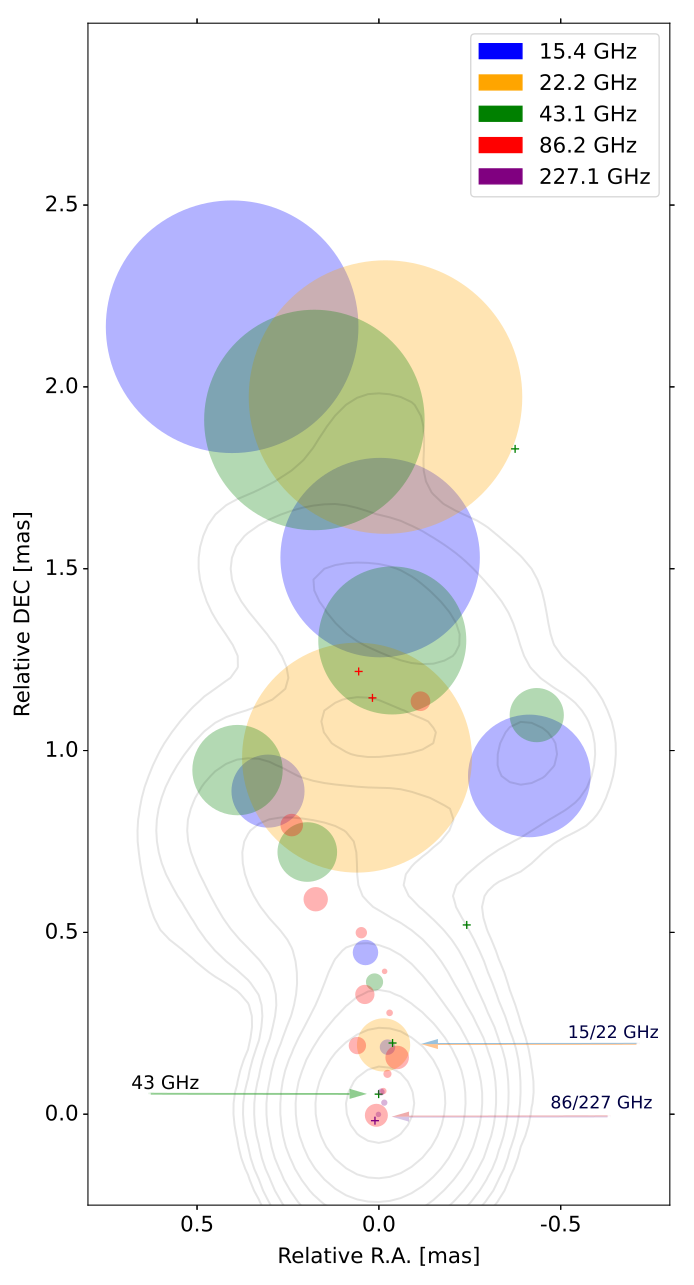}
    \caption{2D Gaussian best-fit models of \nrao at five frequencies, after alignment. Each resolved Gaussian component is shown as a circle with a diameter of the full width at half maximum (FWHM) of the component. Unresolved components are shown as crosses. Different colors correspond to different frequencies in the dataset as shown in the legend. Grey contours represent the source structure at 43~GHz \ro{convolved with a circular beam}. Horizontal arrows indicate position of the apparent jet beginning at 227 and 86~GHz, 43~GHz, 15 and 22~GHz (bottom to top).}
    \label{fig:all_models}
\end{figure}

Second, at each frequency, we fitted the best model to the data using the full available range of UV-distances. At all frequencies, these \textsc{modelfit} models are of comparable quality to the \textsc{CLEAN} models. The shifts derived in the first step were used to align these models and reveal the multifrequency structure of the jet, as presented in Fig.~\ref{fig:all_models}.

Since 15 and 43\,GHz observations were not coordinated with the EHT campaign in April 2017, we had to account for possible changes in the source structure between the EHT epoch (around April 5th, 2017) and corresponding observations. From the kinematics of the source \citep[e.g, ][]{Weaver22} the rate of structural changes is expected to be approximately $1\,\mu$as per day. The source structure has not changed significantly between these observations, that is the number of components was the same for both observations, their position, size, and flux changed moderately. Hence we have interpolated the source models between two adjacent observations at 15\,GHz and 43\,GHz separately, see Table~\ref{tab:observ} for the exact dates.
Possible uncertainties of the model parameters are typically small, as we show in Appendix~\ref{app:data_interpolation}. The uncertainty of the component position introduced by interpolation is not exceeding typical uncertainty of the model component position determination. 
The best-fit model components at each frequency are collected in Appendix~\ref{app:all_models}. 

After applying image shifts, we plotted all model components from all five frequencies together in Fig.~\ref{fig:all_models}. It is immediately visible that the jet of \nrao shows overall consistent and complex structure across frequencies. Most importantly, shifted models at all frequencies allow to study spacial scales from sub-parsecs to tens of parsec.

\section{Magnetic field}
\label{sec:bfield}

\ro{Multifrequency analysis of total intensity data allows us to estimate the magnetic field in the jet based on using frequency dependent shift of the apparent jet beginning (VLBI core) due to synchrotron opacity \citep{lobanov_ultracompact_1998}.}


After aligning models at different frequencies, we immediately measured the distance between apparent VLBI cores at different frequencies, which we refer to later on as the core-shift. It was calculated separately in R.A. and DEC. 
Then we convert the core-shifts into the map shifts using $\mathbf{MS} = \mathbf{\Delta r} - \mathbf{R}_{\nu_1} + \mathbf{R}_{\nu_2}$ \citep{lisakov_connection_2017}, where $\mathbf{MS}$ is the shift between the map phase centers, $\mathbf{\Delta r}$ is the measured core shift, and $\mathbf{R}$ are position of the core at each frequency. The values of the map shift are used to align single-frequency images for subsequent spectral index or Faraday Rotation measure analysis. All measured values for each pair of frequencies are collected in Table~\ref{tab:coreshift}.

\begin{table}
\caption{Map shifts and core shifts}
\label{tab:coreshift}
\centering
\begin{tabular}{c|c|c|c|c|c} 
\multirow{2}{*}{Freq pair} & \multicolumn{2}{c|}{map shift}&
  \multicolumn{3}{c}{core shift} \\
\cline{2-6}
&
  \multicolumn{1}{c|}{$\Delta$RA} &
  \multicolumn{1}{c|}{$\Delta$DEC} &
  \multicolumn{1}{c|}{$\Delta$RA} &
  \multicolumn{1}{c|}{$\Delta$DEC} &
  \multicolumn{1}{c}{$\Delta$r} \\
\hline
15-22\tablefootmark{a}&  14 & 14  & 0  & 0 & 0  \\
15-43  &  2 & -167 & 24 & -129 & 131  \\
43-86  & 12 & -95  & 6  & -58 & 48  \\ 
86-227 & -6 & 17   & -6 & 3 & 6  \\ 
\end{tabular}
\tablefoot{
The values were derived using optically thin model components as a reference and are provided in $\mu$as.
\tablefoottext{a}{for frequencies 15-22\,GHz there were insufficient number of cross-identified components, hence the core shift was assumed to be zero.}
}
\end{table}

\subsection{Detection of the core shift and de-projected distance scale}
A frequency-dependent shift of the apparent core of a jet naturally arises in case of a frequency-dependent opacity acting in the jet. For a simple case of a conical jet with synchrotron opacity and equipartition between magnetic field and particle energy density \citep{1981ApJ...243..700K,lobanov_ultracompact_1998}, position of the VLBI core is inversely proportional to frequency $r=a \nu^{-1/k_{\rm r}} +b$, where typically $k_{\rm r}\approx1$, i.e. at higher frequencies the jet's apparent beginning is closer to the black hole. This dependency for \nrao is shown in Fig.~\ref{fig:coreshift}. \ro{We note that the assumption of the conical jet shape likely holds, see Appendix~\ref{app:jet_shape}}. Moreover, within the uncertainties of our measurements, even such a simple model \ro{described the data well}.

\begin{figure}
  \centering
  \includegraphics[width=0.95\columnwidth]{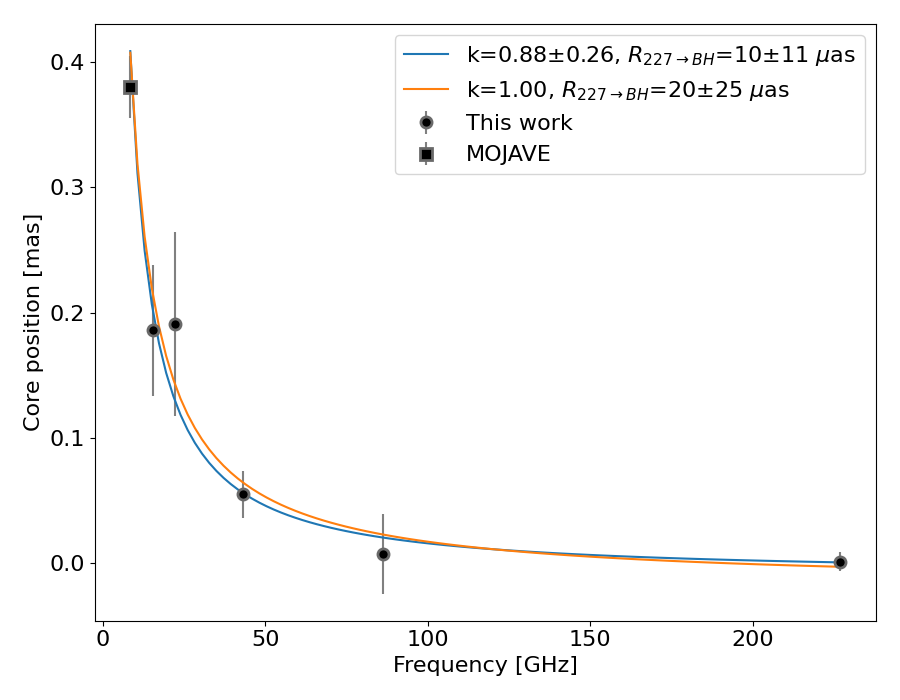}
  \caption{Apparent core position as a function of frequency. Circles represent data from this research. The measurement at 8~GHz (square) is taken from \cite{pushkarev_mojave_2012} and was not used in the fit. The \ro{blue} curve indicates the best fit of the form $r=a \nu^{-1/k_{\rm r}} +b$ using 227\,GHz as the reference frequency. The \ro{orange} curve represents the fit adopting the power law index $k_{\rm r}=1$. Error bars include component position uncertainty as well as interpolation uncertainty at 15 and 43\,GHz. $R_{227\rightarrow BH}$ is the angular distance between the 227~GHz core and the central black hole. }
  \label{fig:coreshift}
\end{figure}

For the best fit, the coefficient $k_{\rm r}=0.88 \pm 0.26$. Within the errors, it is indistinguishable from $k_{\rm r}=1$ expected in the case of a conical jet. The value of \kr~is consistent with the results of \citet{Lu2011} who obtained \kr$=0.78\pm0.29$ between 15 and 86\,GHz, using the latter as a reference frequency.
Within an assumption that for infinite frequency the matter of the jet is totally transparent, we are able to locate the black hole and establish the distance scale in the jet. 
\ro{
For $k_{\rm r}=1$, the black hole is located 20~$\mu$as from the 227~GHz core, while for $k_{\rm r}=0.88$ this angular distance is 10~$\mu$as. 
This value corresponds to 0.08~pc in the image plane, or 1.6~pc of distance along the jet. According to \cite{2014MNRAS.437.3396K} (eq.~4), we calculated the magnetic field strength at a distance of 1~pc from the jet apex $B_1 = 1.5$~G, given the opening angle of the jet $\phi=0.5^\circ$ \citep{2009A&A...507L..33P}. We adopted the range of electron energies to be $\gamma_\mathrm{min} = 10$ and $\gamma_\mathrm{max} = 10^4$. Since magnetic field in the jet is likely toroidal and scales as $B\propto B_1\, r^{-1}$, we can calculate it at any distance along the jet. The distances to VLBI cores and magnetic field values at all frequencies are listed in Table~\ref{tab:magnetic_field}.
The uncertainty of the magnetic field is mainly propagated from the uncertainty of $k_\mathrm{r}$ and is quite large. Improving the accuracy of component position measurements and extending frequency coverage in future experiments will help improving these uncertainties and search for non-power-law dependence of the core position on frequency. 
}


The distance between the apparent jet beginning at 86~GHz and 227~GHz falls below the level of uncertainty of the core position at these frequencies. This distance might be non-zero but too small to be firmly detected. On the other hand, at these high frequencies the apparent core might be not a surface of a unit optical depth, but rather a bright recollimation shock, which is discussed in Sect.~\ref{sec:spectral} and Sect.~\ref{sec:discussion}. In this case, no core shift due to synchrotron opacity is expected at all. 

\ro{The EHT array is currently undergoing developments to enable VLBI observations at 345\,GHz \citep{Doeleman2023,Raymond2024}.} It should be noted that within the model, if core-shift is non-zero between 227~GHz and 345~GHz, its amplitude would be $\Delta r_{227-345} \approx 4\,\mu$as and would likely not be measurable with the current EHT setup. However, spectral information in this frequency range would help to deduce the real nature of the bright origin of the jet of \nrao at 227~GHz and beyond.

\subsection{Spectral properties}
\label{sec:spectral}

\begin{figure}
    \centering
    \includegraphics[width=\columnwidth]{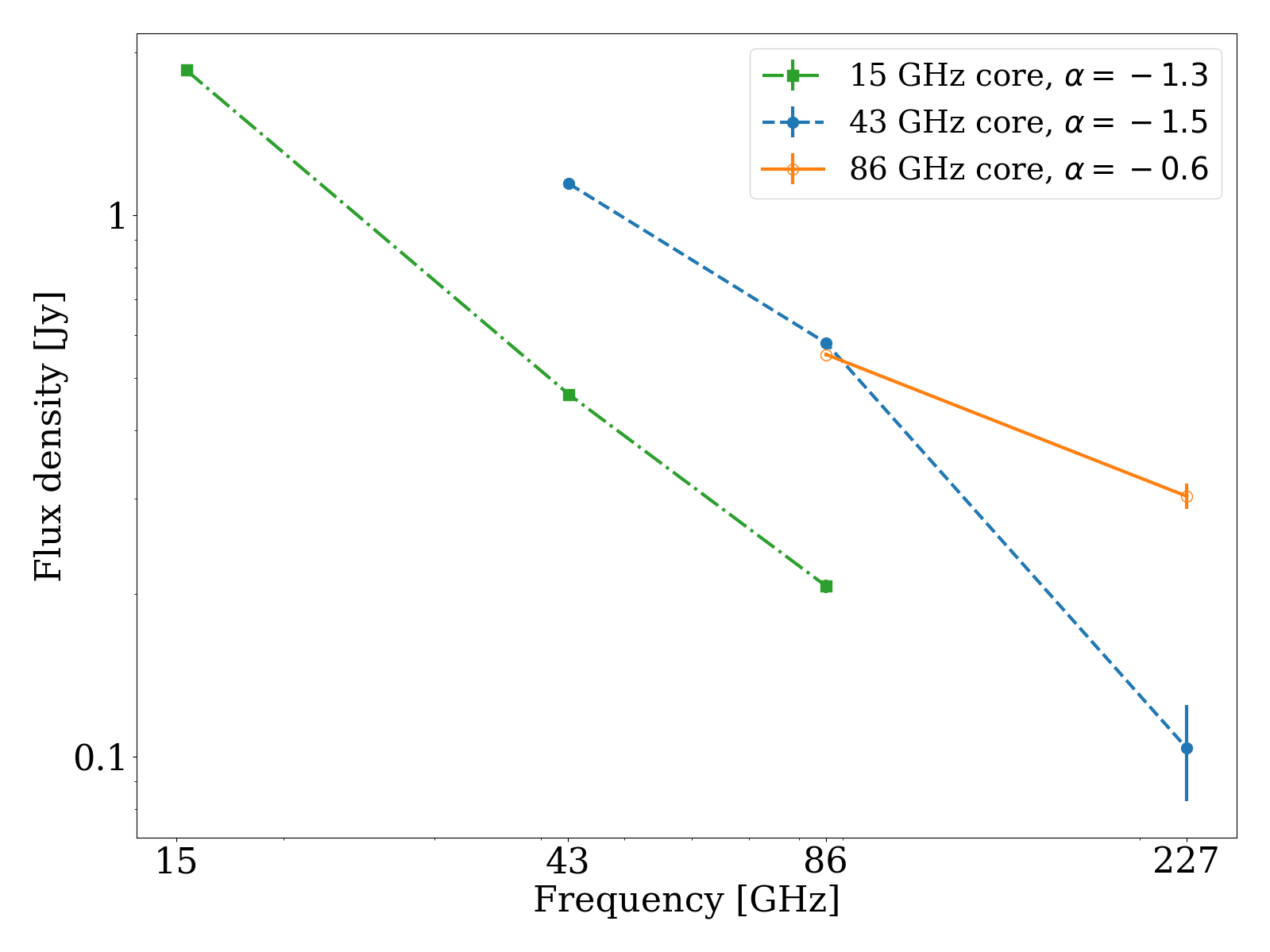}
    \caption{Spectra of the core regions at 15, 43\,GHz, and 86\,GHz.}
    \label{fig:spectra}
\end{figure}

With multifrequency models aligned, we have access to the spectral information of individual components. We have identified which components describe the same spatial regions of the jet at different frequencies. Special regions of interest are: apparent core at 86 and 227\,GHz (components G1, W0); apparent core at 43\,GHz (G3, W1, Q0); apparent core at 15\,GHz (W3, Q1, U0). All parameters of fitted components are presented in Appendix~\ref{app:all_models}. In all these regions there are components at higher frequencies, coinciding with the apparent core at the lower one.  The spectra for these regions are plotted in Fig.~\ref{fig:spectra}. We have checked for two sources of systematic effects that can affect derived spectral indices. Firstly, the sizes of model-fitted components are not strictly equal at different frequencies. However, the difference in sizes is usually within the uncertainty of the measurements. Hence we assume that at all frequencies corresponding components describe the same physical volume. Secondly, there might be a systematic uncertainty associated with flux scaling at 86\,GHz, since zero-baseline VLBI flux density was scaled to that measured at ALMA. However, if ALMA catches some flux which is resolved at VLBI scales, then the flux of each component in the 86\,GHz model should be lower. Hence, spectral index of the 227\,GHz core will remain that of the optically thin region regardless of the flux scaling uncertainties.

Since we do not detect a significant core-shift between 86 and 227\,GHz and the spectral index is negative ($S_\nu\propto\nu^\alpha$), we conclude that the jet matter is optically thin at these frequencies throughout the whole observed jet. In this case, the apparent core at 86\,GHz and 227\,GHz could be a standing shock feature, \ro{which} makes it stand out in its brightness in comparison to the rest of the jet. If the apparent core at 345~GHz is co-spatial with that at 227~GHz, its flux density is estimated on the level of 0.2~Jy.


Since we use fitted Gaussian components, we compare emission of the same volume in the jet at different frequencies. Hence, measured spectral index is a proper proxy to the underlying power-law spectrum of the electron energy distribution. 
In both 15 and 43\,GHz core regions measured optically thin spectral index is $\alpha\approx-1.4$ which yields a steep electron energy spectrum $N(E)=N_\mathrm{0} E^{-p}$ with $p = 1 - 2\alpha = 3.8$. The same argument yields $p=2.2$ in the apparent core at 227\,GHz, which requires rapid energy losses for the higher energy electrons in the region between 1 and 10~pc from the black hole.


\begin{table}[]
    \caption{Magnetic field estimates from the core-shift measurements}
    \label{tab:magnetic_field}
    \centering
    \begin{tabular}{c|c|c}
        Freq & Distance & B$_\mathrm{coreshift}$  \\
        (GHz) & (pc) & (mG)  \\
        \hline
        15.4  & 33.7 & $45^{+72}_{-45}$     \\
        \rowcolor{gray!10}
        22.2  & 22.2 & $69^{+112}_{-69}$    \\
        43.1  & 10.5 & $145^{+239}_{-145}$  \\
        \rowcolor{gray!10}
        86.2  & 4.8  & $319^{+525}_{-319}$  \\
        227.1 & 1.6  & $959^{+1580}_{-959}$ \\
    \end{tabular}
    \tablefoot{The values are provided at the position of the apparent jet beginning at different frequencies. Columns are: Freq -- frequency, Distance -- deprojected distance along the jet, B$_\mathrm{coreshift}$ -- magnetic field estimated using $B(r)=B_1\, r^{-1}$, where $r$ is the distance along the jet and $B_1 = 1.5$~G.}
    \label{tab:magnetic_field}
\end{table}

\section{Helical structure of the jet}
\label{sec:fitting_pr_instab}


\ro{The internal structure of the source, within a 1~mas radius, exhibits a curved jet with a downstream spiral structure. This pattern is likely due to either a precessing jet nozzle \citep[e.g.,][]{2004ApJ...616L..99C, 2022ApJ...933...71F}, or instabilities within the jet itself, such as the Kelvin-Helmholtz instability \citep[e.g.,][]{1978A&A....64...43F, 2004A&A...427..415P}.}
To investigate these structures further, we utilized a three-dimensional helix model, denoted as $\textrm{H}(A, \lambda, \phi)$, which was fitted to the measured positions \ro{of jet components}. The parametric equation describing the helix model is given as follows:

\begin{equation}
    \textrm{H}(A, \lambda, \phi) = 
    \begin{cases}
            x = A(d)\, \sin \left( 2 \pi d / \lambda + \phi \right), \\
            y = A(d)\, \cos \left( 2 \pi d / \lambda + \phi \right), \\
            z = d,
    \end{cases}
    \label{eq:helix_eq}
\end{equation}
where $A$ is an amplitude, $\lambda$ is a wavelength of the helix, $\phi$ is a phase, and $d$ is a distance from the origin along the jet direction. 


In order to account for distortion effects caused by the small viewing angle $\theta = 3^{\circ}$ \citep{Jorstad2017, Weaver22} of the jet, the helix model was projected onto the sky plane prior to fitting, incorporating an additional parameter, the jet position angle $\Psi$, to account for the average position angle of the jet. The complete model \ro{is} defined as:
\begin{equation}
    \textrm{M}(A, \lambda, \phi, \theta, \Psi) = R(\Psi) \times R(\theta) \times \textrm{H}(A, \lambda, \phi),
    \label{eq:helix_rotate} 
\end{equation}
where $R(\Psi)$ and $R(\theta)$ are rotational matrices. We employed bootstrapping to obtain probability distributions and confidence intervals for the parameters.

It is important to note that although the helix structure appears similar in both the instability and precession cases, these patterns arise from distinct phenomena. Consequently, different assumptions were made for each case, requiring modification of  Eq.~\ref{eq:helix_eq}. In the case of a precession-based pattern, we assume that all jet matter trajectories were ballistic. Therefore in snapshot images one sees a narrow curved jet, contained within a wider cone. The fitted precession-helix should increase its amplitude with distance with a linear law $A \propto d$ and have constant wavelength $\lambda$. 

\begin{figure}
    \centering
    \includegraphics[width=\columnwidth, trim=0cm 1cm 0cm 1cm, clip]{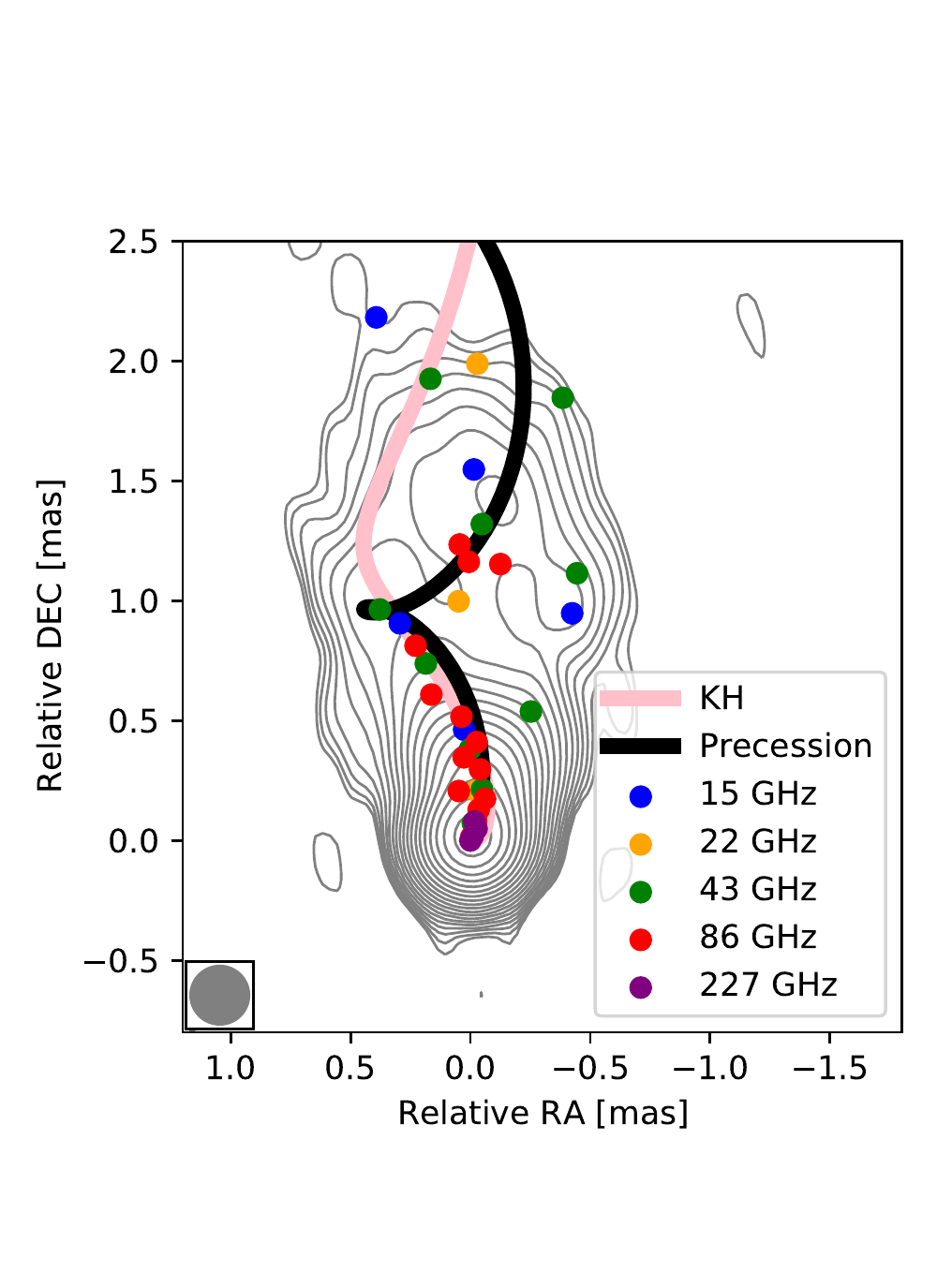}
    \caption{\nrao jet stack intensity \textsc{CLEAN} image at 43\,GHz, positions of the Gaussian modelfit components overlaid by the precession model (black curve) and the Kelvin-Helmholtz (KH) instability model (pink curve). All image model components are displayed here as \rt{same-size circles for clarity}. The colors correspond to frequencies 15, 22, 43, 86 and 227\,GHz as in Fig.~\ref{fig:all_models}. The stacked 43\,GHz intensity map is presented here with contours at increasing powers of 2 starting from 3\,mJy/beam. Map peak is 1.6\,Jy/beam. The \textsc{CLEAN} beam is shown in the lower left corner as a 0.25\,mas circle. The stacked map was made using component-based method alignment within 14 years (60 epochs) of observations by the BU Blazar program. }
    \label{fig:pres_instab}
\end{figure}

In the other case, the amplitude of the helix grows proportionally to the jet radius~$R_{\textrm{jet}}$. Thus, the amplitude and the wavelength of the helix should mirror the jet geometry profile, consequently, $A \propto d^{k}$, $\lambda \propto d^{k}$, where k is determined by the jet expansion profile $R_{\textrm{jet}} \propto d^{k}$ \citep{2000ApJ...533..176H, LOBANOV2003629}. Since we used combined data from all five frequencies together for fitting, we assigned a weight for each data point $W$ that depends on the flux \ro{density} $F$, size $S$ and observed frequency of the component, thus $W = F (\nu / \nu_0)^{-\alpha} S^{-1}$, where \ro{we chose spectral index $\alpha = -0.75$, which is common value in AGN jets. In this case, the spectral index value does not correspond to a steeper one estimated from the VLBI cores of the current observations where $\alpha \approx -1.5$ Fig.~\ref{fig:spectra}. However since steeper spectra do not change weights significantly, we decided to take a more conservative value, which is supported by MOJAVE observations \cite{Hovatta2014} at 8 and 15\,GHz.} The final fitting results are shown in Table~\ref{tab:helix_fits} and Fig.~\ref{fig:pres_instab}.

\begin{table}[]
    \caption{Parameters of helices in precession and KH instability cases}
    \label{tab:helix_fits}
    \centering
    \begin{tabular}{ l | l | l }
        Parameter & Precession & Instability \\
        \hline
        $A$ & 0.03 $\pm$ 0.02 & (0.09 $\pm$ 0.05)$R_{\textrm{jet}}$ \\
        \rowcolor{gray!10}
        $\lambda$ $(\textrm{pc})$ & 273 $\pm$ 99 &  (13 $\pm$ 10)$R_{\textrm{jet}}$ \\
        $\phi$ ($^{\circ}$) & 246 $\pm$ 35 & 320 $\pm$ 99\\
        \rowcolor{gray!10}
        k & 0.8 $\pm$ 0.1 & 0.5 $\pm$ 0.2 \\
        P.A.($^{\circ}$) & 11 $\pm$ 10 &  6 $\pm$ 10 \\
    \end{tabular}
    \tablefoot{Parameters $A, \lambda$ are the amplitude and the de-projected wavelength, $R_{\textrm{jet}}$ is a jet radius, $\phi$ is a phase and $k$ is a parameter in power law jet geometry function $R_{\textrm{jet}} \propto d^k$, where $d$ is a distance from the jet origin.}
\end{table}

\section{Brightness temperature}
\label{sec:tb}

\begin{figure}
    \centering
    \includegraphics[width=\columnwidth]{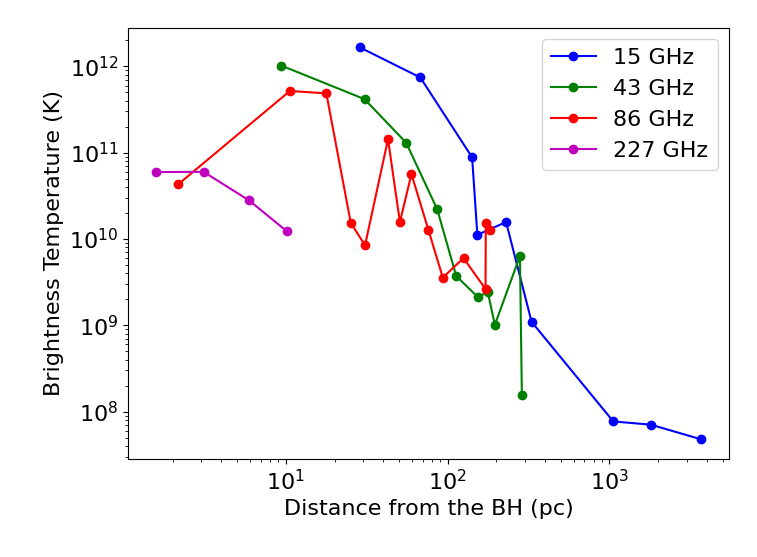}
    \caption{Observed brightness temperature of model components as a function of the position along the jet, de-projected. Colors represent different frequencies. 
    }
    \label{fig:tb_r}
\end{figure}

After decomposing the source structure into distinct Gaussian components we can directly calculate observed brightness temperature for each of them using Eq.~\ref{eq:tb}
\begin{equation}
    \label{eq:tb}
    T_\mathrm{b}^\mathrm{obs} = \delta T_\mathrm{b}^\mathrm{int}  = \delta \frac{2\ln{2}}{\pi k_\mathrm{B}} \frac{(1+z)S \lambda^2}{\theta^2},
\end{equation}
where $k_\mathrm{B}$ is the Boltzmann's constant, $S$ is the component flux \ro{density}, $\lambda$ is the wavelength, $\delta$ is the Doppler-factor, $z$ -- redshift, and $\theta$ is the angular size of the component, defined as the maximum of the FWHM of the fitted Gaussian and the resolution limit \citep{lobanov_resolution_2005}. Since we have aligned all single-frequency models of the source, we can study how brightness temperature changes along the jet, as is shown in Fig.~\ref{fig:tb_r}.

Eq.~\ref{eq:tb} implies that the brightness temperature at a given distance from the central engine depends on the spectral index as $T_\mathrm{b} \propto \nu^{-2+\alpha}$, if optically thin synchrotron emission is assumed. A two-order difference between the brightness temperatures values at 43 and 227\,GHz at 10~pc implies $\alpha\approx-1.2$. This is consistent with the optically thin part of the spectrum of the 43\,GHz core region, $\alpha= -1.5$ (blue line in Fig.~\ref{fig:spectra}).

15 and 43\,GHz analysis of the brightness-temperature gradients in a sample of 28 AGN show that they generally are well described by a single power law \citet[][Kravchenko et al. in prep.]{burd2022}. This is applicable in assumption of a straight constant-speed jet with a power-law distribution of the magnetic field and the particle density along the outflow. Regular deviation of the decline in the $T_\mathrm{b}$ from a pure power-law dependency seen in \nrao in Fig.~\ref{fig:tb_r} indicates departure from this scenario.
We suggest that change in local orientation of the jet, resulting in a variations of the viewing angle of $2^{\circ} < \theta < 4^{\circ}$ due to precession, lead to change of the Doppler factor and, hence, causes non-linear gradient of the brightness-temperature with the distance along the jet. 


\section{Linear polarization and rotation measure structure}
\label{sec:polar}

\begin{figure*}
    \centering
    \includegraphics[width=0.8\columnwidth]{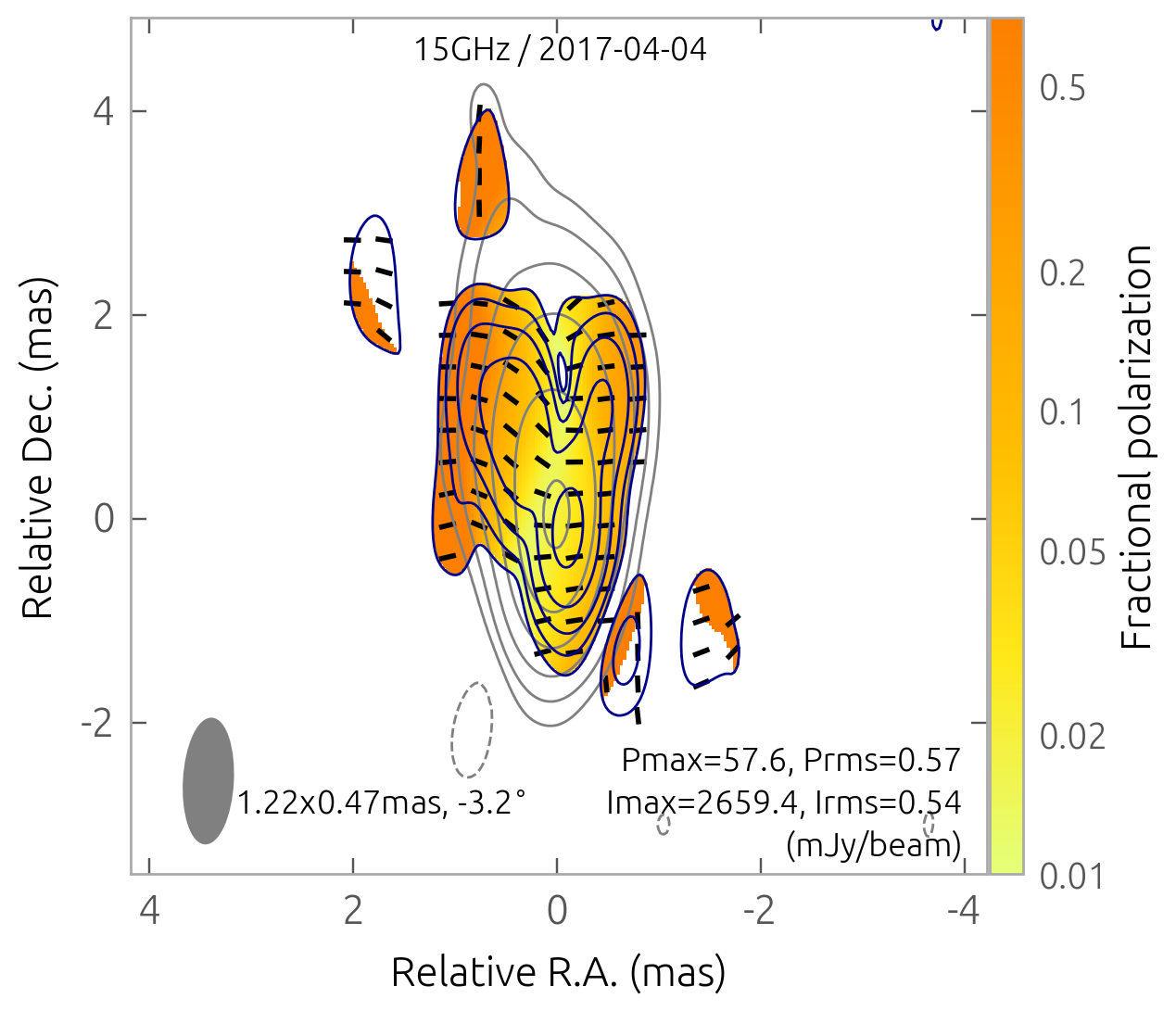}
    \includegraphics[width=0.8\columnwidth]{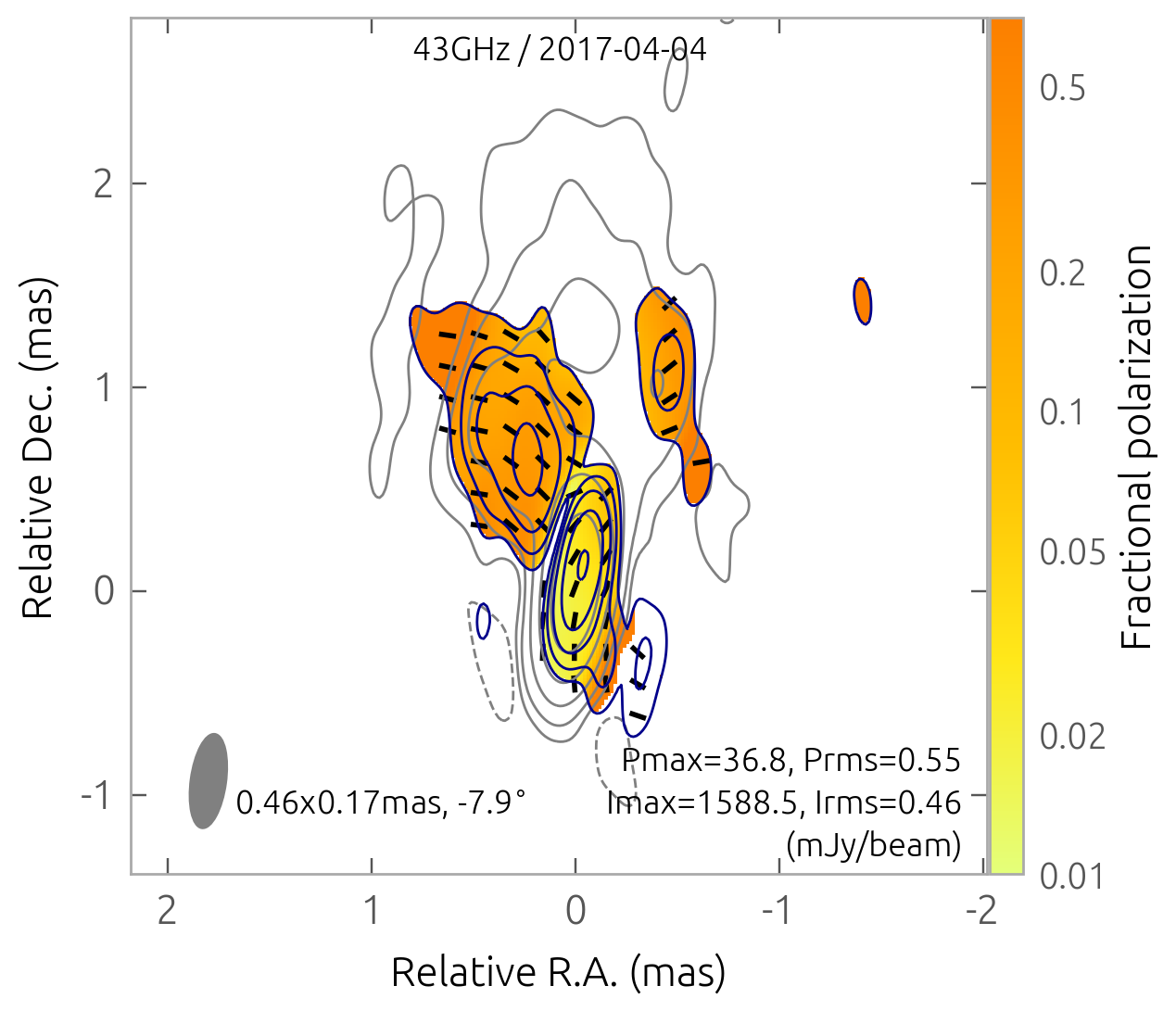}
    \includegraphics[width=0.8\columnwidth]{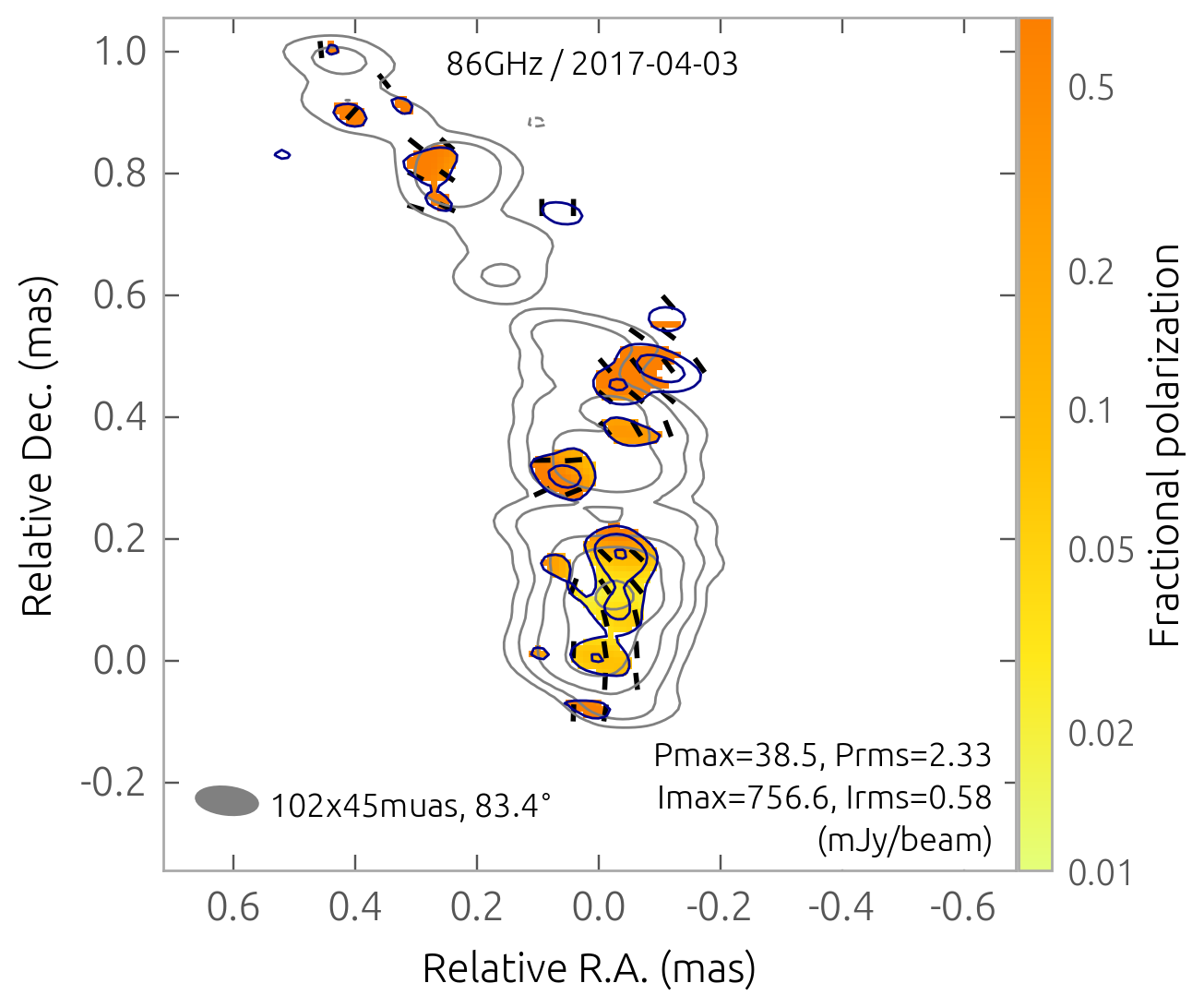}
    \includegraphics[width=0.8\columnwidth]{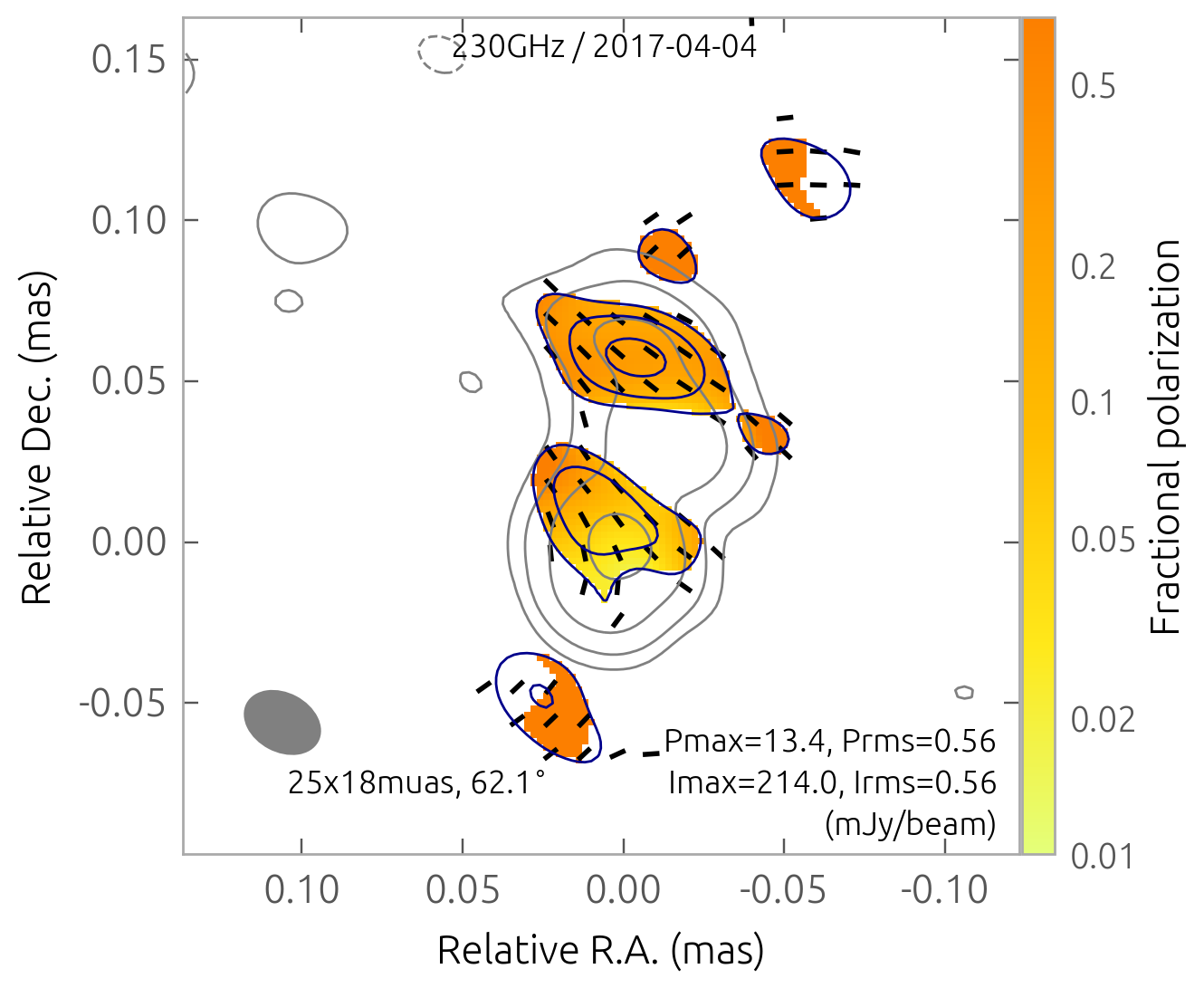}
    \caption{Total intensity and linear polarization 15, 43, 86, and 227\,GHz CLEAN images. Color denotes distribution of the fractional linear polarization overlaid with the sticks indicating the electric polarization vector directions, not corrected for Faraday rotation. Synthesized beam is shown by a shaded ellipse  in the lower left corner and its size is given. The Stokes $I$ contours are plotted at increasing powers of 4, starting from the corresponding 3~rms level. The $P$ contours are drawn at 4~rms level.
    Here we plot interpolated 15 and 43\,GHz images (See section~\ref{sec:alignment} for details). The original MOJAVE and BU-BLAZAR images are given in Appendix~\ref{app:data_interpolation}. }
    \label{fig:pol_images}
 \end{figure*}

Polarimetric images of \nrao at 15, 43 and 86\,GHz are shown in Fig.~\ref{fig:pol_images}, alongside with the 227\,GHz EHT image.
The structure is dominated by the core and the jet that is visible at low frequencies. 
At 15\,GHz, the jet polarization is characterized by a two-sided structure, which is also marginally detected at 43\,GHz. The fractional polarization in the jet center is of a few per cent and increases up to 50 per cent toward jet edges. The latter is partially caused by systematic errors of the \textsc{CLEAN} procedure that can be corrected for from the results of simulations \citep{2023MNRAS.520.6053P}.
Linearly polarized emission in the inner $\thicksim0.2$~mas is represented by an extended region at 86\,GHz, that is resolved into two separate components at 227\,GHz.
At all frequencies, fractional polarization is weak in the core region (about 2\%) and increases up to 30\% downstream and toward the jet edges.

Multi-wavelength observations can be combined to obtain a Faraday rotation measure (RM) image of \nrao.
The RM is determined by $\chi = \chi_0 + RM \times \lambda^2$, where $\lambda$ is the observed wavelength, $\chi_0$ is the intrinsic and $\chi$ is the observed electric vector position angle of the emitting region.
Broad frequency coverage enables us to construct RM maps on a range of different scales. Due to progressively larger difference in the wavelengths that leads to the significant smearing of polarized structure when restoring with the large beam size, in the RM analysis we consider separately two frequency ranges: 15~--~43\,GHz and 43~--~227\,GHz.
For the 43 -- 86 -- 227\,GHz range, the images were tapered and convolved with a common circular beam size of 0.1\,mas, which slightly over-resolves the 43\,GHz image while still preserving a fraction of the higher resolution achieved at 86\,GHz and 227\,GHz. 
Images at different frequencies were aligned using map shifts quoted in Table~\ref{tab:coreshift}.

In the RM analysis, we consider only pixels in the images for which linearly polarized intensity \ro{at all frequencies} was higher than 3~times~rms. 
The RM map is computed by performing a linear fit to the dependence of the EVPA($\lambda^2$) at each pixel, blanking pixels with a poor ﬁt based on a $\chi^2$ criterion. 
\ro{The $\chi^2$ of the fit is calculated using the formula:
\begin{equation}
    \chi^2 = \sum_{i=1}^{N}\frac{(O_i-E_i)^2}{\sigma_i^2},
\end{equation}
where $N$ is the number of data points, $O_i$ are the observed data, $E_i$ are the expected data based on the model, and $\sigma_i$ is the measurement error of the individual data point.}

Since 22\,GHz data is single polarization and cannot be used in this analysis, the $\chi - \lambda^2$ alignment between two frequencies of 15 and 43\,GHz is the subject to $\pm n\pi$-ambiguity. Considering the value of the 86\,GHz EVPA integrated over an image, we see a \ro{robust} linear dependence over 15--86\,GHz, which implies no $\pm n\pi$ EVPA rotation to be made. 

\begin{figure}
    \centering
    \includegraphics[width=0.8\columnwidth,angle=270]{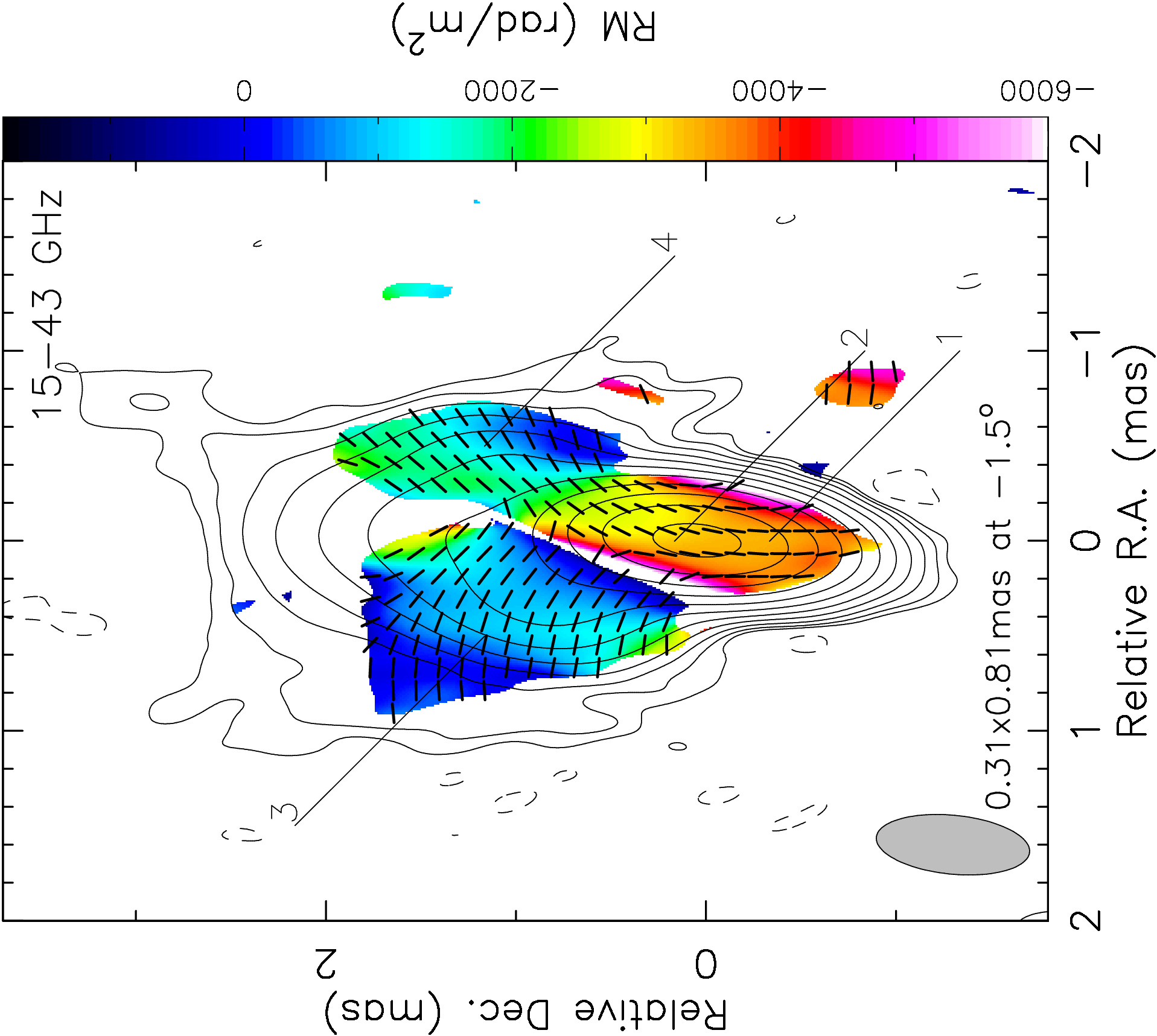}\\
    \includegraphics[width=0.45\columnwidth,angle=270]{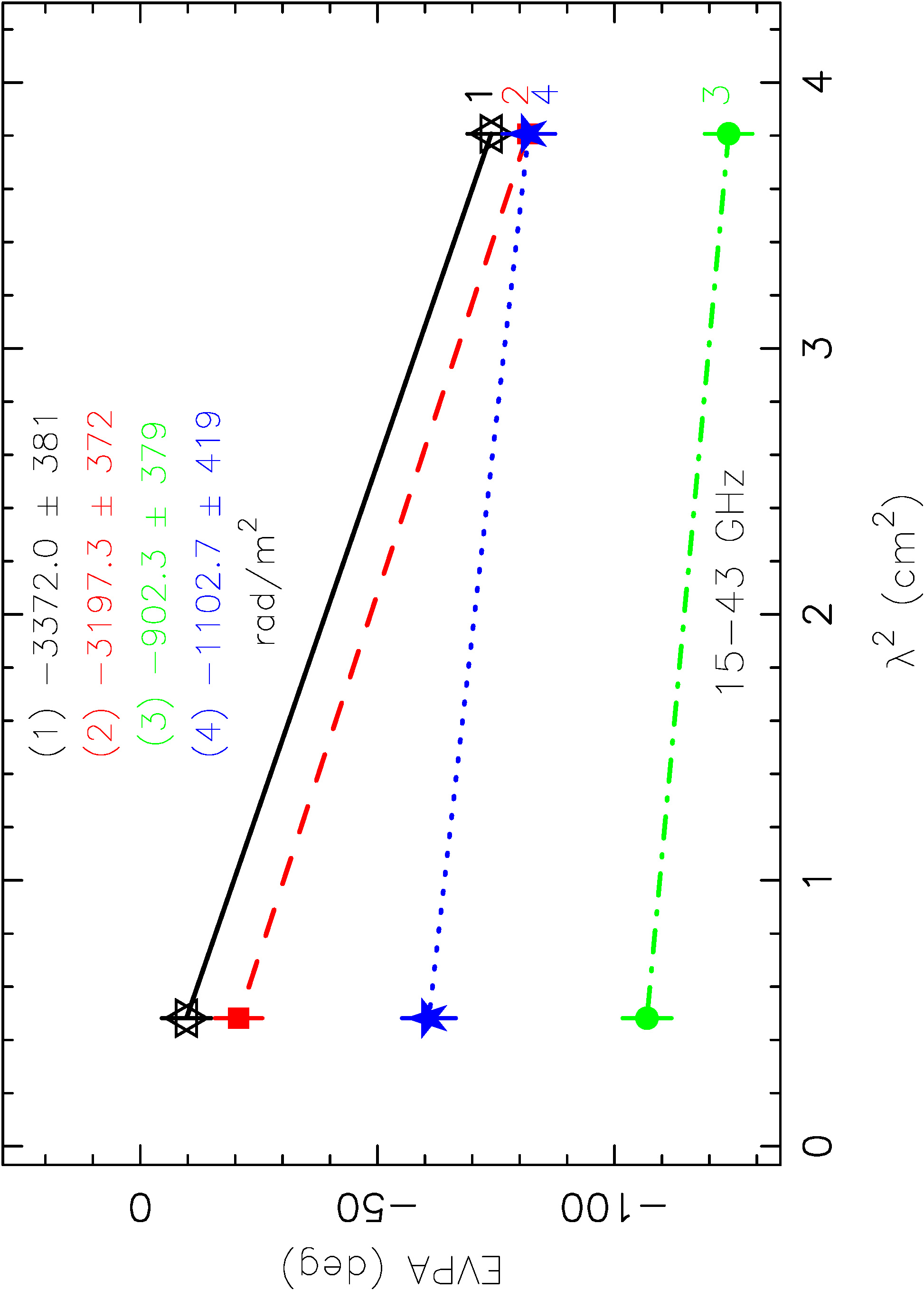}\\
    \caption{(top panel) 15 -- 43\,GHz rotation measure map with 43\,GHz total intensity contours at increasing powers of 2, lowest contour 1.40~mJy~beam$^{-1}$. Sticks denote linear polarization orientation at 43\,GHz corrected for Faraday rotation. Numbers indicate locations where the EVPA($\lambda^2$)-fits (bottom panel) are taken. Faraday rotation regression is shown by lines, and estimated values are given in rad~m$^{-1}$.}
    \label{fig:frm_1543}
\end{figure}

The resultant rotation measure map in the 15--43\,GHz range with $\lambda^2$-fits at four locations in the jet are given in Fig.~\ref{fig:frm_1543}. 
The RM values in the core (associated with the self-synchrotron absorbed region) amount to $\thicksim -3 000$~rad~m$^{-2}$, that decreases to few hundreds rad~m$^{-2}$ downstream, in both eastern and western parts of the jet.

\ro{The RM value in the jet is comparable with estimates by \citet{2012AJ....144..105H} defined in a range of 8-15\,GHz in amount of $371$\,rad\,m$^{-2}$. There is an indication on the RM map of \citet{2012AJ....144..105H} for RM of about $-700$\,rad\,m$^{-2}$ toward the core region. This value is consistent with our measurements, \ro{however} we probe this region in more detail.}
Our RM values are in good agreement with the multi wavelength polarimetric observations of \nrao by \citet{2010MNRAS.408..841C}, who estimated RM of $-1062\pm0.2$~rad~m$^{-2}$ between 15 and 43\,GHz.

\begin{figure}
    \centering
    \includegraphics[width=0.8\columnwidth,angle=270]{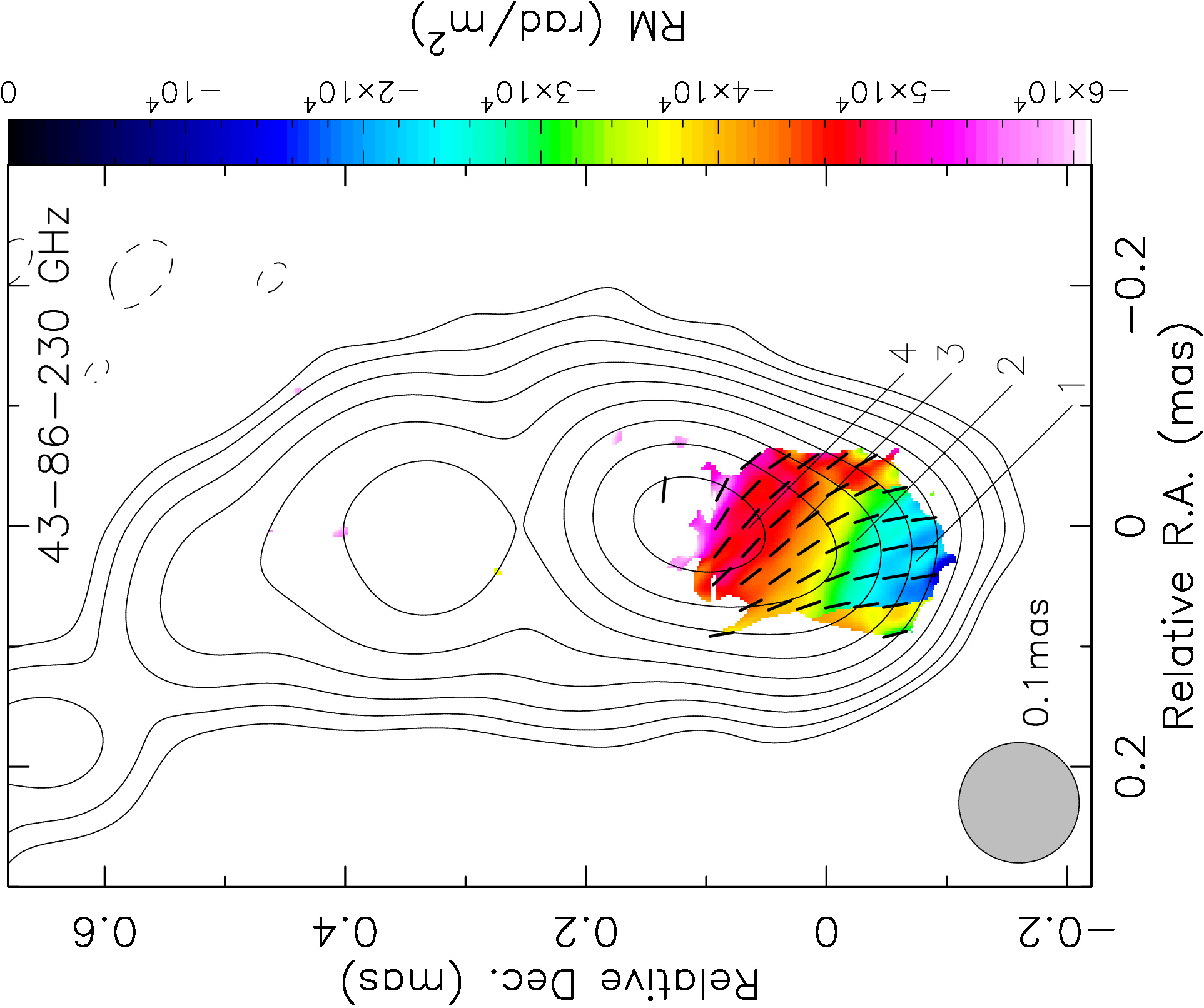}
    \includegraphics[width=0.45\columnwidth,angle=270]{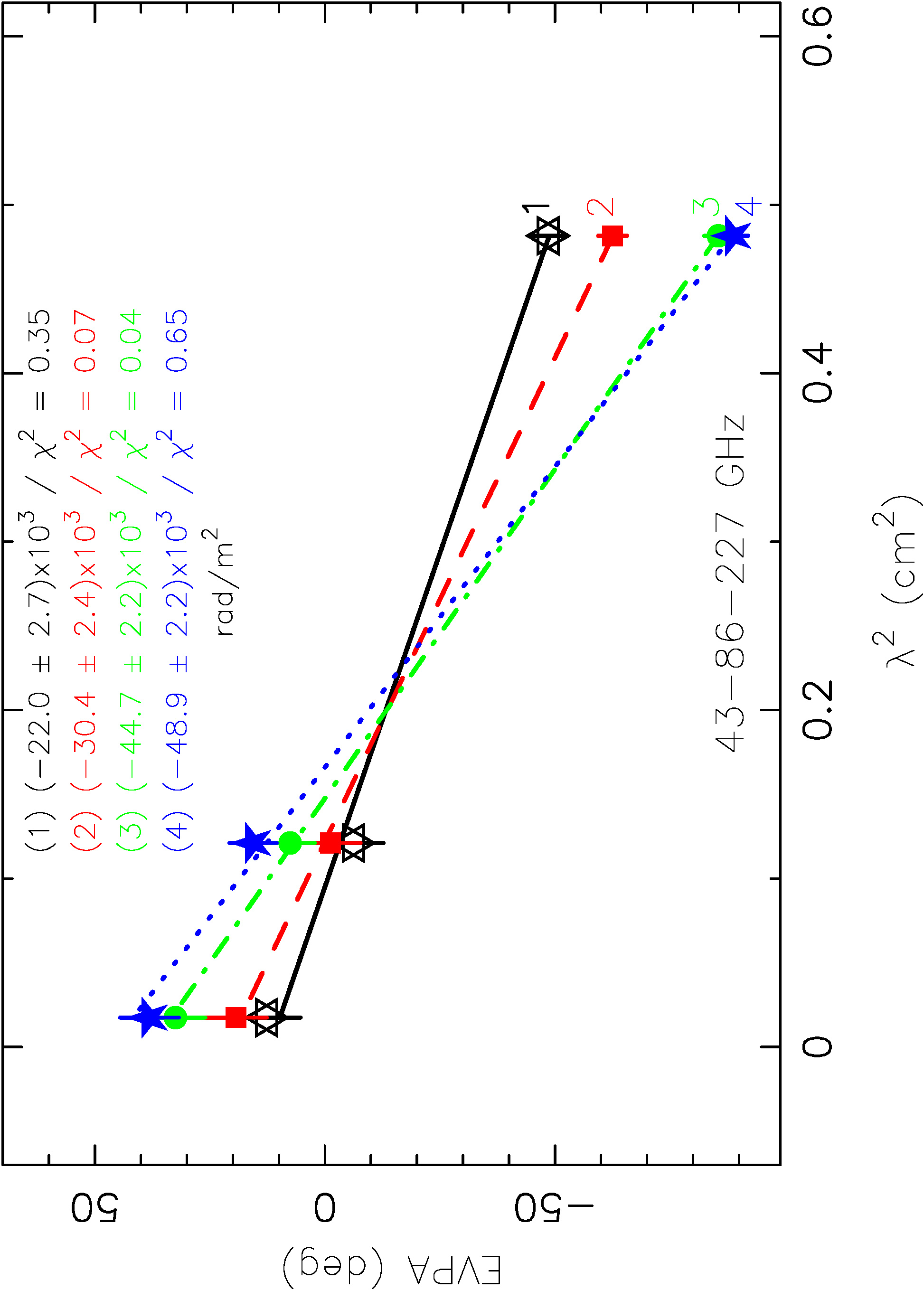}
    \caption{(top panel) 43 -- 227\,GHz rotation measure map with 86\,GHz total intensity contours at increasing powers of 2, lowest contour 3.34~mJy~beam$^{-1}$. Sticks denote linear polarization orientation at 86\,GHz corrected for Faraday rotation. Numbers indicate locations where the EVPA($\lambda^2$)-fits (bottom panel) are taken. Faraday rotation regression is shown by lines, and estimated values are given in rad~m$^{-1}$. }
    \label{fig:frm_43230}
 \end{figure}

The 43--227\,GHz RM map with $\lambda^2$-fits at four locations in the jet are shown in Fig.~\ref{fig:frm_43230}. 
We estimate RM value in the inner region of $\thicksim -4\times10^4$~rad~m$^{-2}$. 
\citet{2021ApJ...910L..14G} provide a RM estimates of $(-2.1\pm0.6)\times 10^4$\,rad\,m$^{-2}$ at 3\,mm, which is comparable to the inter-band RM between 1 and 3\,mm ($-3.3\times10^4$\,rad\,m$^{-2}$). Also, these and our RM estimates are in agreement with the RM values at 1.3\,mm reported by \citet{2018ApJ...868..101B}, in the amount of $(-3.1 \pm 1.3)\times10^4$~rad~m$^{-2}$.
  
\begin{figure}
    \centering
    \includegraphics[width=0.8\columnwidth,angle=270]{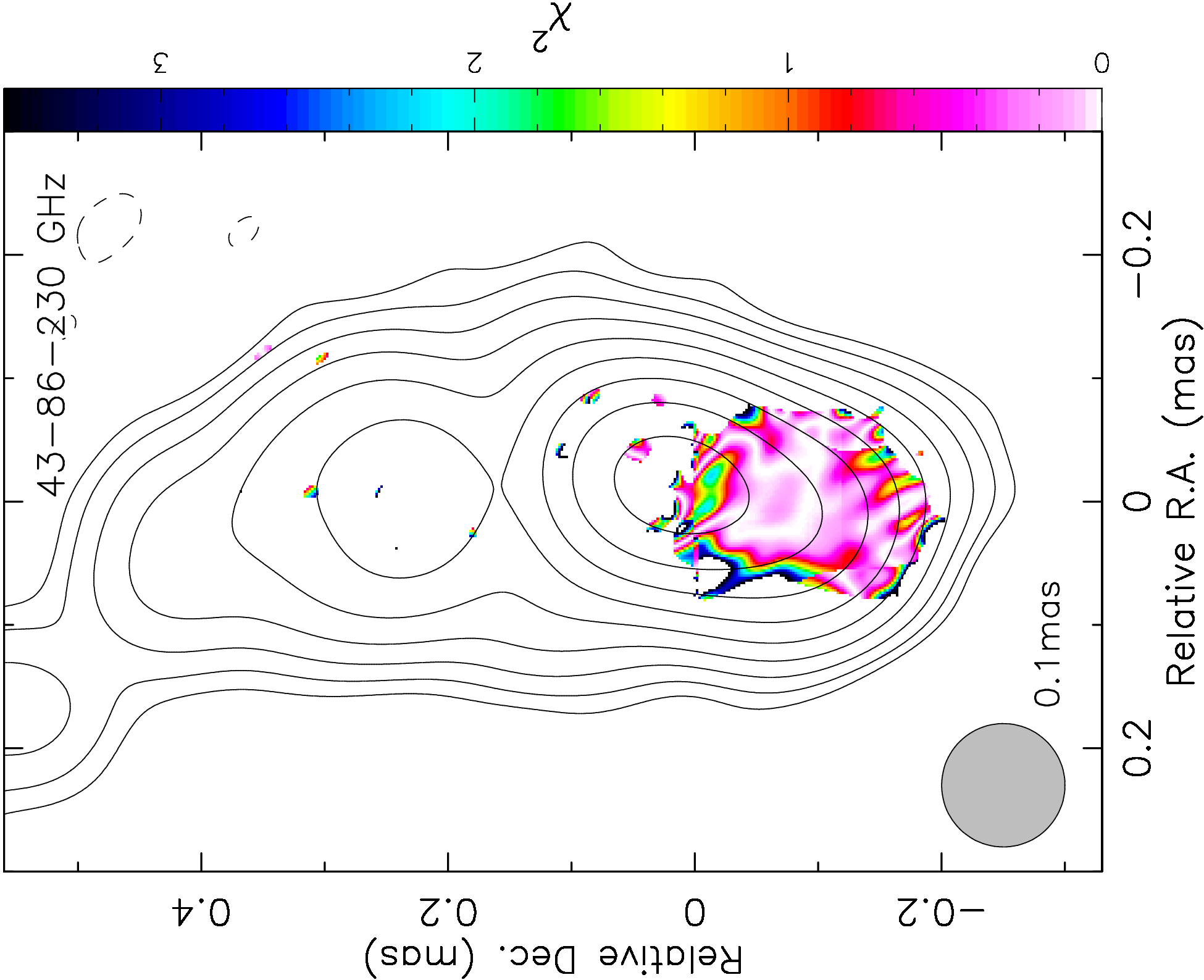}
    \caption{The map of $\chi^2$-values of the EVPA-$\lambda^2$ fit for a frequency range 43 -- 227\,GHz. Grey contours are the same as in Fig.~\ref{fig:frm_43230}.}
    \label{fig:chi_43230}
 \end{figure}

\ro{In Fig.~\ref{fig:chi_43230} we plot the map of $\chi^2$-values obtained from the EVPA-$\lambda^2$ fit. We fit a two-parameter model to three data points, therefore we have 1 degree of freedom. From the $\chi^2$-distribution, the corresponding 95\% confidence limit is $\chi^2 < 3.841$.}
\rt{Linear fits of $\chi$ vs. $\lambda^2$ are found in all pixels for which linearly polarized intensity in all bands was higher than 3 times its rms.}
The observed large value of RM is consistent with the rotating medium being external to the emitting region. 
To our knowledge, this is the first RM map of \nrao obtained at such high frequencies in VLBI observations.
  
\cite{2007AJ....134..799J} derived the following relation of RM as a function of frequency for a conical jet under equipartition $|RM| \propto \nu^{a}$, where the value of $a$ depends on the power-law change in the electron density $n_{\rm e}$ as a function of distance $r$ from the black hole, $n_{\rm e} \propto r^{-a}$. Typical values obtained in the literature vary in the range $a = 1 - 4$ \citep{2007AJ....134..799J, 2013MNRAS.429.3551A, 2017MNRAS.467...83K}, while the value $a = 2$ would imply that the Faraday rotation is occurring in a sheath around a conically expanding jet.
For \nrao, the linear fit between our 15--43\,GHz and 43--227\,GHz RM estimates in the core region results in $a \thicksim 1.7$, consistent with the model of a sheath surrounding a conically expanding flow.
In the source frame, $RM_{\rm int} = (1+z)^2 RM$. Given $z=0.902$, $RM_{\rm int}\thicksim10^5$\,rad\,m$^{-2}$.

Intrinsic orientation of polarization (corrected for the RM) is shown in Fig.~\ref{fig:frm_1543} and \ref{fig:frm_43230}.
The EVPAs at cm-wavelengths are pointing towards the external region of the jet, reminding fountain-like pattern. 
Zero polarization along the jet spine, visible in Fig.~\ref{fig:pol_images} and in the stack 15\,GHz images of \nrao \citep{2023MNRAS.520.6053P} is accompanied by a high variability of EVPAs observed on a time interval of two decades \citep{2023MNRAS.523.3615Z}.
\cite{Jorstad2017} estimated an intrinsic viewing angle $\delta_0>60$\degr, which suggests helical magnetic field pitch angle for this source $\gamma_0\sim40-70$\degr.

\begin{figure}
    \centering
    \includegraphics[width=0.8\columnwidth]{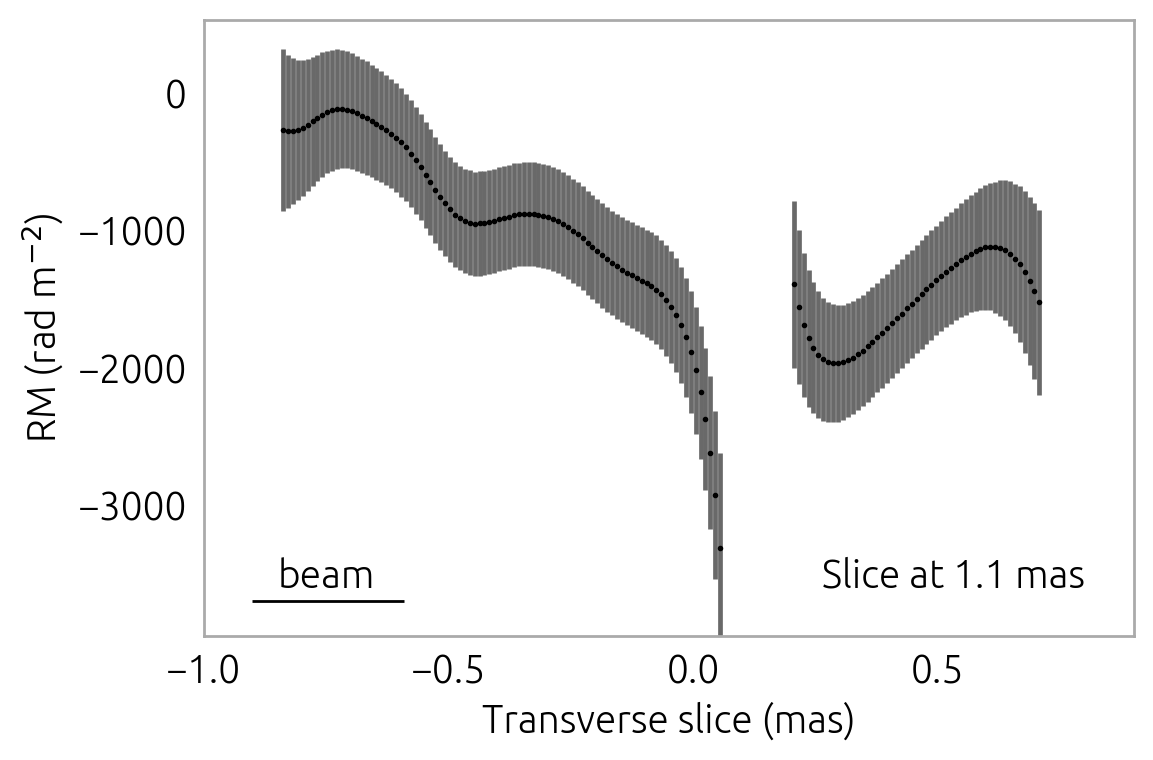}
    \caption{The transverse slice taken across the RM map (Fig.~\ref{fig:frm_1543}) at 1.1~mas. The black dots show the RM fit at every pixel across the slice with the grey area indicating the error in the fit for each point.}
    \label{fig:rm_trslice}
 \end{figure} 

As we point out in Sect.~\ref{sec:fitting_pr_instab}, the jet appears wide at 1~mas from the core most likely due to the intrinsic changes in the jet direction amplified by the projection effects, see Fig.~\ref{fig:pres_instab}. In this case, intrinsically narrow relativistic jet is precessing inside a wider cone cleared by the jet itself. 
\ro{The RM, taken transverse the outflow direction at 1.1~mas down from the core, changes its value} (Fig.~\ref{fig:rm_trslice}). Although transverse RM values do not cross zero, almost monotonic behaviour of RM transverse to the jet hints on the presence of the helical magnetic field in the plasma outside of the wide cone, similar to that discovered in 3C~273 by \cite{lisakov_oversized_2021}.
The scenario of a precessing jet threaded by the helical magnetic field can also explain the observed polarization structure with the EVPAs pointing toward jet edges, accompanied by a low polarization and high variability of EVPAs along the jet spine \citep{2024AJrus.100.12}. In this scenario, due to the limited resolution of the VLBI observations, the initially strong polarization along the jet spine is smeared in the projection to the sky due to the overlap of regions with the different EVPAs orientations.


\section{Discussion}
\label{sec:discussion}
   \subsection{Support for the presence of component C2}
\begin{figure*}
\sidecaption
  \includegraphics[width=12cm]{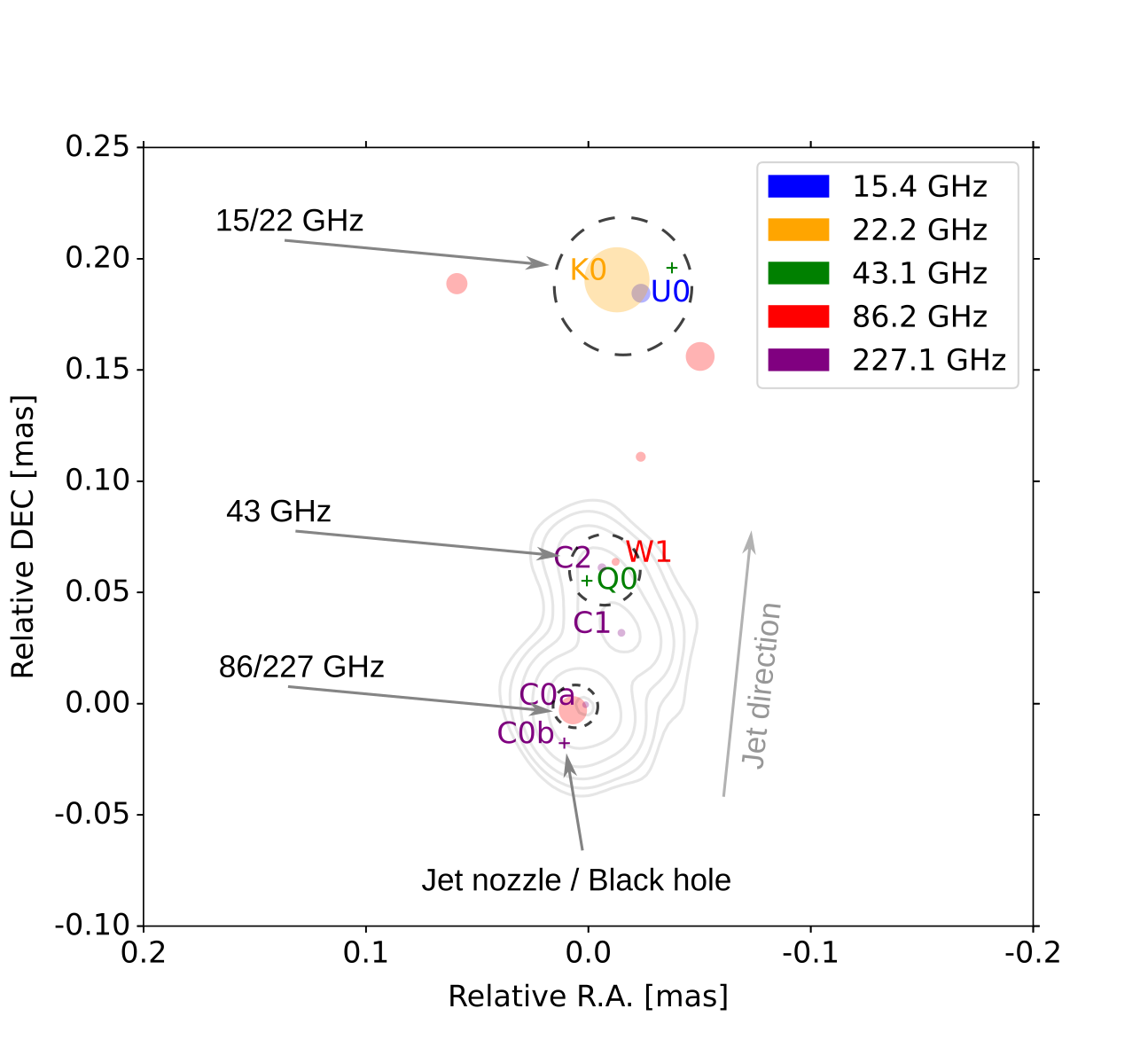}
     \caption{Central region of \nrao. Contours represent total intensity at 227\,GHz. Model components at 227\,GHz are named according to \cite{Jorstad2023}, at other frequencies -- according to Table~\ref{tab:models}. Size of Gaussian components is reduced by a factor of 5 with respect to Fig.~\ref{fig:all_models} for visual clarity. Dashed circles show the location of the apparent VLBI core at different frequencies according to the labels. Component C2 spatially coincides with the apparent core of the 43\,GHz model Q0 and an 86\,GHz component W1. A weak component C0b spatially coincides with the true jet origin calculated using core shift.}
    \label{fig:nrao530_c2}
\end{figure*}

   Data at 227\,GHz provide a hint of the dim C2 component, the most downstream one, see Fig.~5 of \cite{Jorstad2023}. However, it was detected with \textsc{Difmap}, \textsc{eht-imaging} and \textsc{DMC}, but not with \textsc{SMILI} or \textsc{Themis}. We have found a counterpart of this component at other frequencies, namely at 86\,GHz and 43\,GHz. Zoom-in map of the region is presented in Fig.~\ref{fig:nrao530_c2}. Here for 227\,GHz we adopted naming from \cite{Jorstad2023}.   
   It is apparent that component C2 at 227\,GHz spatially coincides with the core of the 43\,GHz jet and component W1 from the 86\,GHz jet. It reassures the detection of C2 \ro{that was} made with 227\,GHz data only.


\subsection{Jet structure and its evolution in two possible scenarios}
  
The jet structure analysis can reveal intrinsic physical parameters. In the first scenario of precession, one can evaluate the period and the opening angle $\alpha$ of the \ro{precession} cone. The opening angle was obtained directly from the fitted helix, thus $\alpha = 3^{\circ} \pm 1^{\circ}$. In the assumption of ballistic trajectories of plasma in a jet, we can estimate the distance of a one-period turn of a precessing jet $d_{\textrm{period}} \approx \lambda_{\textrm{proj}} = 1.6 \pm 0.5$\,mas (Table~\ref{tab:helix_fits}). Using weighted average Lorenz factor $\Gamma = 8.9 \pm 2.5$ \citep{Jorstad2017}, we found an apparent speed $\beta_{\textrm{app}} \approx 7$ that corresponds to angular speed of $\mu \approx 0.3$\,mas/yr. We estimated the period of precession $P = d_{\textrm{period}}/\beta_{\textrm{app}} = 6 \pm 4$\,yr, which does not contradict the case, where the precession is initiated by a double supermassive black hole system like in OJ~287 \citep{1988ApJ...325..628S, 2009ApJ...698..781V, 2018MNRAS.478.3199B, 2022ApJ...924..122G} \ro{or M~81 \citep{2011A&A...533A.111M, 2023A&A...672L...5V}}. The result corresponds to the period found in the radio light curve \citep{AN2013}. An additional support for the obtained period can be demonstrated by the 14-year stack image of the jet \ro{at 43~GHz} (Fig.~\ref{fig:pres_instab}). The image shows a symmetrical jet experiencing a dramatic expansion until 1\,mas, which is the result of a smearing effect produced by the \ro{time-variable jet position angle}. \ro{Since the stack time is more than twice the length of the obtained period}, the precession is a feasible interpretation of the observed phenomena. The \ro{parameters are derived assuming ballistic trajectories, implying the conical geometry of the jet's precession cone}. However, if the expansion parameter of the fit model (Eq.~\ref{eq:helix_rotate}) is free, the fitting yields $k = 0.8 \pm 0.1$ (Table~\ref{tab:helix_fits}), that deviates from the conical shape, but provides the best fit to the data. Nevertheless, it does not change obtained \ro{precession period} significantly. On the one hand, deviation from the conical shape can be induced by limited and localized data points with high accuracy, providing measurements of less than one wavelength. \ro{Additionally}, real physical phenomena like jet nutation or non-uniform local hydrodynamic conditions can provide an apparent or real distortion of the jet shape.

  
In the second scenario, we assume the Kelvin-Helmholtz instability plays the leading role in forming the helical pattern. From the fitting results, the expansion index $k = 0.5 \pm 0.2$ (Table~\ref{tab:helix_fits}), does not correspond to the local visible hyperbolic expansion, but agrees with the global parabolic geometry of the jet at 15\,GHz \citep{2017MNRAS.468.4992P}. We used linear analysis \citep{2000ApJ...533..176H, LOBANOV2003629, vega2020multiband} to identify and obtain the basic physical parameters of the jet. 
\ro{The} measurements made in Sec.~\ref{sec:fitting_pr_instab} show a single wiggling ridge line that can be identified as helical surface instability mode with a wavelength $\lambda/R_{\textrm{jet}} = 13 \pm 10$. Using the apparent speed, the viewing angle listed in the previous paragraph and pattern speed $\beta_{w\textrm{app}}$, we estimate the Mach number $ M_{\textrm{j}}$ and the ratio of the jet to the ambient density $\eta$ \citep{vega2020multiband}. However, the pattern speed $\beta_{w\textrm{app}}$ is complicated to obtain using existing observations since a much greater sensitivity to a faint extended structure and resolution is needed. In the case of M87, the pattern speed was found using wavelet-based analysis and $\beta_{w\textrm{app}} \approx 0.3$ \citep{mertens2016kinematics}. In the case of \nrao, the same type of analysis is unfeasible, thus the best estimation for the pattern speed will be sub-relativistic speeds $\sim 0.5c$ \citep{hardee1988spatial, norman1988spatial, 2000ApJ...533..176H}. Using this set of parameters, the unrealistically large $\eta \approx 0.8$ was found. That means, the $\beta_{w\textrm{app}}$ estimation is too large. For example in a relativistic jet simulations M$_{\textrm{jet}} \sim 3$, $\eta \sim 0.01\textrm{--}0.1$  \citep{norman1988spatial, 2005MmSAI..76..110P, perucho2007}. To achieve physical $\eta \sim 0.01\textrm{--}0.1$, $\beta_{w\textrm{app}} \approx 0.1\textrm{--}0.2$ is needed. Finally, using $\beta_{w\textrm{app}} = 0.1$ we obtained the Mach number $M_{\textrm{j}} = 4 \pm 3$, and the density ratio $\eta = 0.03 \pm 0.07$. 

To show how KH pattern corresponds to the observed stacked image, we calculated the apparent pattern speed $\beta_{w \textrm{app}} \approx 0.006$. Using the same $d_{\textrm{proj}}$ from the precession case, we obtained an observed period of instability pattern $P_{\textrm{obs}} \sim 5000$\,years. The analysis shows \ro{that} KH instability cannot provide such a rapid pattern change as precession. Even if $\beta_{w\textrm{app}} = 0.5$ is considered, $P_{\textrm{obs}} \sim 600$\,years. In any case, the instability pattern will not create a smeared picture within 14 years. 

It should be noted that even though the precession scenario is better supported with our observations, it has its own shortcomings. For instance, if precession is the only mechanism driving the jet wobbling, then all model components should follow ballistic trajectories. As it was shown by \cite{Weaver22}, in the innermost 1~mas from the core components move along straight trajectories, see Fig.~\ref{fig:bu_component_map}. However, farther from the core trajectories of some components are clearly non-ballistic, which is also supported by \cite{2019ApJ...874...43L} at 15~GHz. This discrepancy should be investigated further. We will dedicate a future publication to a detailed analysis of temporal variations of the jet structure and its direction.

\subsection{Magnetic field in and around the jet}


Polarimetric images, Fig.~\ref{fig:pol_images} in this paper and those presented by \cite{Jorstad2023}, show that at 15~GHz EVPA is perpendicular to the jet direction in the core region while at higher frequencies EVPA are approximately parallel to the jet. It could be interpreted within the assumption that the toroidal component of magnetic field dominates in the innermost portion of the jet in \nrao and $90^{\circ}$ rotation of the electric vector is solely ascribed to different opacity of the jet matter at different frequencies. Namely at 86~GHz and 227~GHz the core is shown to be optically thin, see Sect.~\ref{sec:bfield}. It is well supported by  findings of \cite{gomez_probing_2016} who also found magnetic field to remain toroidal in the inner jet of BL~Lac. In this case we assume that the toroidal configuration of the magnetic field could persist even close to the central black hole and hence one can extrapolate the magnetic field strength towards the black hole using $B\propto r^{-1}$ relation. 

Within this assumption, magnetic field strength at $5\,R_\mathrm{g}$ is $B = 3000 - 30000$~G depending on the black hole mass in the range $2 - 0.2\times 10^9 M_\odot$ \citep{2003MNRAS.340..632L, MKeck2019}. \ro{This range of black hole masses corresponds to $5\,R_\mathrm{g} = 100 - 10$~AU.} The values of magnetic field are at least two orders of magnitude higher than those reported by \cite{akiyama_first_2021} in M87 at the same distance. If we compare the jet luminosity  $L_\nu = 4 \pi D_\mathrm{L} \nu F_\nu $ at some frequency where synchrotron emission of the jet dominates and Doppler boosting can be measured, say at 15~GHz, \nrao appears 1000 times brighter than M87. This difference could largely be attributed to the difference in magnetic field strength. 

On the other hand, \cite{2022ApJ...939...83K} estimated the magnetic field of $200 - 5000$~G at the horizon of the black hole in M87. Magnetic fields of the order of 1000~G are also required in the arguments provided by \cite{2019ARA&A..57..467B}. Future polarimetric observations of  M87 and \nrao at both 227~GHz and 345~GHz will help to better account for the magnetic field difference in the models of jet launching \citep{2023Galax..11...61J}.


Linear polarization structure of \nrao downstream the jet at 15 and 43~GHz shows limb brightening structure with the electric vectors pointing toward jet edges. The fractional polarization along the jet spine is almost zero which is accompanied by the high variability of the EVPAs along the jet center that is observed over a two-decade time interval  \citep{2023MNRAS.523.3615Z}.
Such behavior can be well explained in a model of jet threaded by the helical magnetic field precessing at a time interval of ten years \citep{2024AJrus.100.12}.
Due to the small viewing angle, averaging of jet images that captured different jet direction will lead to the overlapping of polarization with different orientations. As a result, the observed polarized intensity in the center of the jet will be smeared and will be accompanied by an increased variability of EVPA in this area. 

Substantial variations in the position angle of the jet components observed in the significant number of AGN jets \citep{2021ApJ...923...30L} and large apparent jet opening angles \citep{2017MNRAS.468.4992P} serve as the basis for this scenario.
Moreover, there is an increasing number of sources for which a possible connection between the variation of intensity or jet position angle with precession was shown \citep[e.g.,][]{2018MNRAS.478.3199B, 2019ApJ...886...85A, cui_etal2023, 2023ApJ...951..106B}.

We also note that the values of the Faraday Rotation Measure at 43-227\,GHz increase in absolute values downstream the jet, see Fig.~\ref{fig:frm_43230}. This behavior is in contrast with expected gradual decreasing of the RM magnitude due to decreasing magnetic field strength and particle density of the rotating medium surrounding the jet. However, according to the model of precession, in the region of interest the jet viewing angle is increasing approximately from $2.3^\circ$ to $3.1^\circ$. Therefore, the path length in the rotating medium is increased, thus, increasing rotation measure values. Alternatively, we might be able to resolve small scale inhomogeneities in the Faraday screen which likely occur at the probed distance of only 1\,pc from the central engine. Multifrequency observations of \nrao at other precession phases are required to test these hypotheses.

\subsection{Jet beginning}
According to our multifrequency model of the jet in \nrao we can associate its real starting point with component C0b at 227~\,GHz, see Fig.~\ref{fig:nrao530_c2}. As we show in Sect.~\ref{sec:spectral}, the jet is transparent at 227\,GHz (460\,GHz in the source frame) and we can indeed detect emission down to the black hole. The flux \ro{density} of this component $F_\mathrm{C0b} = 67$~mJy can be compared with the total flux of the M87 ring from ETH observations \cite{EHT_M87_P3}, $F\mathrm{ring}\approx1$~Jy. According to \cite{2023Natur.616..686L}, the ring is indeed the brightest feature in the innermost jet. 
Taking into account the difference in distance, frequency shift, and spectrum, intrinsic luminosity of the C0b feature of \nrao is $10^4$ times higher than the total flux of the M87 ring. Given the difference of the same order in the magnetic field strength near the black holes of M87 and \nrao we might indeed probe the same region around the black hole with our observations at 227\,GHz. Future observations of \nrao at 227 and 345\,GHz will be required to answer this question.

\section{Conclusions}
\label{sec:conclusions}
We have analyzed quasi-simultaneous observations of the blazar \nrao at frequencies 15, 22, 43, 86, and 227~GHz performed around April~2017.

\begin{itemize}
\item The relatively dim component C2 identified at 227\,GHz by \cite{Jorstad2023} spatially coincides with the apparent core at 43\,GHz and optically thin component at 86\,GHz.
\item We found that the jet exhibits a spiral pattern, which is most likely attributed to the precession of the jet nozzle with a period of $6 \pm 4$~years.
\item With core-shift \ro{analysis}, we show that the matter of the jet might be completely transparent at frequencies above 86~GHz. However, further EHT observations at 345~GHz are required to test this result. The apparent core at 345~GHz should have flux density of $S_{345}\approx 200$~mJy and should be located not farther than 4~$\mu$as from the core at 227~GHz.
\item We estimate the strength of the magnetic field close to the central black hole as $3\times(10^3 - 10^4)$~G, much stronger than that derived by direct observations of the M87* black hole shadow \ro{at the same distance}.
\item For the first time, we have revealed the structure of Faraday rotation measure and obtained its amplitude using VLBI polarimetric observations between 43, 86, and 227~GHz. The largest value we detected is $-48000$~rad/m$^2$.
\item With fine resolution at 227\,GHz, core shift analysis, and estimates of the magnetic field strength we conclude that the component C0b \ro{with the flux density of 67~mJy} could represent the disk around the black hole \rt{or a bright feature in the innermost jet, for instance a recollimation shock wave}.
\end{itemize}


\begin{acknowledgements}
The authors thank Andrei Lobanov and two anonymous referees for reviewing the manuscript. 

M2FINDERS project has received funding from the European Research Council (ERC) under the European Union’s Horizon 2020 research and innovation programme (grant agreement No 101018682).

This research has made use of data obtained with the Global Millimeter VLBI Array (GMVA), coordinated by the VLBI group at the Max-Planck-Institut für Radioastronomie (MPIfR). The GMVA consists of telescopes operated by the MPIfR, IRAM, Onsala, Metsähovi, Yebes, the Korean VLBI Network, the Green Bank Observatory and the Very Long Baseline Array (VLBA). The VLBA and the GBT are a facility of the National Science Foundation operated under cooperative agreement by Associated Universities, Inc. The data were correlated at the VLBI correlator of the MPIfR in Bonn, Germany. \rt{This paper uses data obtained with the 100\,m radiotelescope of the MPIfR at Effelsberg.}

This research has made use of data from the MOJAVE database that is maintained by the MOJAVE team \citep{2018ApJS..234...12L}.

This study makes use of VLBA data from the VLBA-BU Blazar Monitoring Program (BEAM-ME and VLBA-BU-BLAZAR \footnote{\url{http://www.bu.edu/blazars/VLBA\_GLAST/1730.html}}), funded by NASA through the Fermi Guest Investigator Program. The VLBA is an instrument of the National Radio Astronomy Observatory. The National Radio Astronomy Observatory is a facility of the National Science Foundation operated by Associated Universities, Inc.

ALMA is a partnership of the European Southern Observatory (ESO; Europe, representing its member states), NSF, and National Institutes of Natural Sciences of Japan, together with National Research Council (Canada), Ministry of Science and Technology (MOST; Taiwan), Academia Sinica Institute of Astronomy and Astrophysics (ASIAA; Taiwan), and Korea Astronomy and Space Science Institute (KASI; Republic of Korea), in cooperation with the Republic of Chile. The Joint ALMA Observatory is operated by ESO, Associated Universities, Inc. (AUI)/NRAO, and the National Astronomical Observatory of Japan (NAOJ). We also gratefully acknowledge the support provided by the extended staff of the ALMA, both from the inception of the ALMA Phasing Project through the observational campaigns since 2017. 

\end{acknowledgements}

\bibliographystyle{aa}
\bibliography{bibliography}

\begin{thebibliography}{93}
\expandafter\ifx\csname natexlab\endcsname\relax\def\natexlab#1{#1}\fi

\bibitem[{{Abdollahi} {et~al.}(2020){Abdollahi}, {Acero}, {Ackermann}, \& {et
  al.}}]{4FGL}
{Abdollahi}, S., {Acero}, F., {Ackermann}, M., \& {et al.} 2020, \apjs, 247, 33

\bibitem[{{Algaba}(2013)}]{2013MNRAS.429.3551A}
{Algaba}, J.~C. 2013, \mnras, 429, 3551

\bibitem[{{Algaba} {et~al.}(2019){Algaba}, {Rani}, {Lee}, {Kino}, {Park}, \&
  {Kim}}]{2019ApJ...886...85A}
{Algaba}, J.~C., {Rani}, B., {Lee}, S.~S., {et~al.} 2019, ApJ, 886, 85

\bibitem[{{Aller} {et~al.}(1985){Aller}, {Aller}, \& {et al.}}]{Aller1985}
{Aller}, H., {Aller}, M., \& {et al.} 1985, \apjs, 59, 533

\bibitem[{{An} {et~al.}(2013){An}, {Baan}, \& {et al.}}]{AN2013}
{An}, T., {Baan}, W., \& {et al.} 2013, \mnras, 434, 3487

\bibitem[{{Blackburn} {et~al.}(2019){Blackburn}, {Chan}, {Crew}, {Fish},
  {Issaoun}, {Johnson}, {Wielgus}, {Akiyama}, {Barrett}, {Bouman}, {Cappallo},
  {Chael}, {Janssen}, {Lonsdale}, \& {Doeleman}}]{Blackburn_2019}
{Blackburn}, L., {Chan}, C.-k., {Crew}, G.~B., {et~al.} 2019, \apj, 882, 23

\bibitem[{{Blandford} {et~al.}(2019){Blandford}, {Meier}, \&
  {Readhead}}]{2019ARA&A..57..467B}
{Blandford}, R., {Meier}, D., \& {Readhead}, A. 2019, \araa, 57, 467

\bibitem[{{Bower} {et~al.}(2018){Bower}, {Broderick}, {Dexter}, {Doeleman},
  {Falcke}, {Fish}, {Johnson}, {Marrone}, {Moran}, {Moscibrodzka}, {Peck},
  {Plambeck}, \& {Rao}}]{2018ApJ...868..101B}
{Bower}, G.~C., {Broderick}, A., {Dexter}, J., {et~al.} 2018, \apj, 868, 101

\bibitem[{{Britzen} {et~al.}(2018){Britzen}, {Fendt}, {Witzel}, {Qian},
  {Pashchenko}, {Kurtanidze}, {Zajacek}, {Martinez}, {Karas}, {Aller}, {Aller},
  {Eckart}, {Nilsson}, {Ar{\'e}valo}, {Cuadra}, {Subroweit}, \&
  {Witzel}}]{2018MNRAS.478.3199B}
{Britzen}, S., {Fendt}, C., {Witzel}, G., {et~al.} 2018, \mnras, 478, 3199

\bibitem[{{Britzen} {et~al.}(2023){Britzen}, {Zaja{\v{c}}ek}, {Gopal-Krishna},
  {Fendt}, {Kun}, {Jaron}, {Sillanp{\"a}{\"a}}, \&
  {Eckart}}]{2023ApJ...951..106B}
{Britzen}, S., {Zaja{\v{c}}ek}, M., {Gopal-Krishna}, {et~al.} 2023, \apj, 951,
  106

\bibitem[{{Burd} {et~al.}(2022){Burd}, {Kadler}, {Mannheim}, {Baczko},
  {Ringholz}, \& {Ros}}]{burd2022}
{Burd}, P.~R., {Kadler}, M., {Mannheim}, K., {et~al.} 2022, \aap, 660, A1

\bibitem[{{Caproni} {et~al.}(2004){Caproni}, {Mosquera Cuesta}, \&
  {Abraham}}]{2004ApJ...616L..99C}
{Caproni}, A., {Mosquera Cuesta}, H.~J., \& {Abraham}, Z. 2004, \apjl, 616, L99

\bibitem[{Casadio {et~al.}(2019)Casadio, Marscher, Jorstad, \&
  et~al.}]{Casadio2019}
Casadio, C., Marscher, A., Jorstad, S., \& et~al. 2019, \aap, 622, A158

\bibitem[{{Chael} {et~al.}(2018){Chael}, {Johnson}, {Bouman}, {Blackburn},
  {Akiyama}, \& {Narayan}}]{Chael_2018}
{Chael}, A.~A., {Johnson}, M.~D., {Bouman}, K.~L., {et~al.} 2018, \apj, 857, 23

\bibitem[{{Chen} {et~al.}(2010){Chen}, {Shen}, \& {Feng}}]{2010MNRAS.408..841C}
{Chen}, Y.~J., {Shen}, Z.~Q., \& {Feng}, S.~W. 2010, \mnras, 408, 841

\bibitem[{Cho {et~al.}(2022)Cho, Zhao, Kawashima, Kino, Akiyama, Johnson,
  Issaoun, Moriyama, Cheng, Algaba, Jung, Sohn, Krichbaum, Wielgus, Hada, Lu,
  Cui, Sawada-Satoh, Shen, Park, Jiang, Ro, Yi, Wajima, Lee, Hodgson, Tazaki,
  Honma, Niinuma, Trippe, An, Zhang, Lee, Oh, Byun, Lee, Kim, Oh, Koyama,
  Asada, Wang, Cui, Hagiwara, Nakamura, Takamura, Hirota, Sugiyama, Kawaguchi,
  Kobayashi, Oyama, Yonekura, Kim, Hwang, Jung, Kim, Kim, Oh, Roh, Yeom, Xia,
  Zhong, Li, Zhao, Wang, Liu, \& Chen}]{Cho2022}
Cho, I., Zhao, G.-Y., Kawashima, T., {et~al.} 2022, \apj, 926, 108

\bibitem[{{Cui} {et~al.}(2023){Cui}, {Hada}, {Kawashima}, {Kino}, {Lin},
  {Mizuno}, \& {Ro}}]{cui_etal2023}
{Cui}, Y., {Hada}, K., {Kawashima}, T., {et~al.} 2023, Nature, 621, 711

\bibitem[{{Cui} {et~al.}(2021){Cui}, {Hada}, {Kino}, {Sohn}, {Park}, {Ro},
  {Sawada-Satoh}, {Jiang}, {Cui}, {Honma}, {Shen}, {Tazaki}, {An}, {Cho},
  {Zhao}, {Cheng}, {Niinuma}, {Wajima}, {Zhang}, {Kawaguchi}, {Algaba},
  {Koyama}, {Hirota}, {Yonekura}, {Sakai}, {Xia}, {Jiang}, {Yu}, {Gou},
  {Hwang}, {Jiang}, {Sun}, {Jung}, {Kim}, {Kim}, {Kobayashi}, {Lee}, {Lee},
  {Zhang}, {Li}, {Xu}, {Li}, {Oh}, {Oh}, {Oh}, {Oyama}, {Roh}, {Shibata},
  {Guo}, {Zhao}, {Zhong}, {Wang}, {Yang}, {Yan}, {Yeom}, {Li}, {Li}, {Yuan},
  {Dong}, {Chen}, {Akiyama}, {Asada}, {Byun}, {Hagiwara}, {Hodgson}, {Jung},
  {Kim}, {Lee}, {Yi}, {Liu}, {Liu}, {Lu}, {Nakamura}, {Trippe}, {Wang}, {Wang},
  \& {Zhang}}]{Cui_2021}
{Cui}, Y.-Z., {Hada}, K., {Kino}, M., {et~al.} 2021, Research in Astronomy and
  Astrophysics, 21, 205

\bibitem[{{Doeleman} {et~al.}(2023){Doeleman}, {Barrett}, {Blackburn},
  {Bouman}, {Broderick}, {Chaves}, {Fish}, {Fitzpatrick}, {Freeman}, {Fuentes},
  {G{\'o}mez}, {Haworth}, {Houston}, {Issaoun}, {Johnson}, {Kettenis},
  {Loinard}, {Nagar}, {Narayanan}, {Oppenheimer}, {Palumbo}, {Patel}, {Pesce},
  {Raymond}, {Roelofs}, {Srinivasan}, {Tiede}, {Weintroub}, \&
  {Wielgus}}]{Doeleman2023}
{Doeleman}, S.~S., {Barrett}, J., {Blackburn}, L., {et~al.} 2023, Galaxies, 11,
  107

\bibitem[{{EHTC} {et~al.}(2022{\natexlab{a}}){EHTC}, {Akiyama}, {Alberdi},
  {Alef}, {Algaba}, {Anantua}, {Asada}, {Azulay}, {Bach}, {Baczko}, {Ball},
  {Balokovi{\'c}}, {Barrett}, {Baub{\"o}ck}, {Benson}, {Bintley}, {Blackburn},
  {Blundell}, {Bouman}, {Bower}, {Boyce}, {Bremer}, {Brinkerink}, {Brissenden},
  {Britzen}, {Broderick}, {Broguiere}, {Bronzwaer}, {Bustamante}, {Byun},
  {Carlstrom}, {Ceccobello}, {Chael}, {Chan}, {Chatterjee}, {Chatterjee},
  {Chen}, {Chen}, {Cheng}, {Cho}, {Christian}, {Conroy}, {Conway}, {Cordes},
  {Crawford}, {Crew}, {Cruz-Osorio}, {Cui}, {Davelaar}, {De Laurentis},
  {Deane}, {Dempsey}, {Desvignes}, {Dexter}, {Dhruv}, {Doeleman}, {Dougal},
  {Dzib}, {Eatough}, {Emami}, {Falcke}, {Farah}, {Fish}, {Fomalont}, {Ford},
  {Fraga-Encinas}, {Freeman}, {Friberg}, {Fromm}, {Fuentes}, {Galison},
  {Gammie}, {Garc{\'\i}a}, {Gentaz}, {Georgiev}, {Goddi}, {Gold},
  {G{\'o}mez-Ruiz}, {G{\'o}mez}, {Gu}, {Gurwell}, {Hada}, {Haggard}, {Haworth},
  {Hecht}, {Hesper}, {Heumann}, {Ho}, {Ho}, {Honma}, {Huang}, {Huang},
  {Hughes}, {Ikeda}, {Impellizzeri}, {Inoue}, {Issaoun}, {James}, {Jannuzi},
  {Janssen}, {Jeter}, {Jiang}, {Jim{\'e}nez-Rosales}, {Johnson}, {Jorstad},
  {Joshi}, {Jung}, {Karami}, {Karuppusamy}, {Kawashima}, {Keating}, {Kettenis},
  {Kim}, {Kim}, {Kim}, {Kim}, {Kino}, {Koay}, {Kocherlakota}, {Kofuji}, {Koch},
  {Koyama}, {Kramer}, {Kramer}, {Krichbaum}, {Kuo}, {La Bella}, {Lauer}, {Lee},
  {Lee}, {Leung}, {Levis}, {Li}, {Lico}, {Lindahl}, {Lindqvist}, {Lisakov},
  {Liu}, {Liu}, {Liuzzo}, {Lo}, {Lobanov}, {Loinard}, {Lonsdale}, {Lu}, {Mao},
  {Marchili}, {Markoff}, {Marrone}, {Marscher}, {Mart{\'\i}-Vidal},
  {Matsushita}, {Matthews}, {Medeiros}, {Menten}, {Michalik}, {Mizuno},
  {Mizuno}, {Moran}, {Moriyama}, {Moscibrodzka}, {M{\"u}ller}, {Mus}, {Musoke},
  {Myserlis}, {Nadolski}, {Nagai}, {Nagar}, {Nakamura}, {Narayan}, {Narayanan},
  {Natarajan}, {Nathanail}, {Fuentes}, {Neilsen}, {Neri}, {Ni}, {Noutsos},
  {Nowak}, {Oh}, {Okino}, {Olivares}, {Ortiz-Le{\'o}n}, {Oyama}, {{\"O}zel},
  {Palumbo}, {Paraschos}, {Park}, {Parsons}, {Patel}, {Pen}, {Pesce},
  {Pi{\'e}tu}, {Plambeck}, {PopStefanija}, {Porth}, {P{\"o}tzl}, {Prather},
  {Preciado-L{\'o}pez}, {Psaltis}, {Pu}, {Ramakrishnan}, {Rao}, {Rawlings},
  {Raymond}, {Rezzolla}, {Ricarte}, {Ripperda}, {Roelofs}, {Rogers}, {Ros},
  {Romero-Ca{\~n}izales}, {Roshanineshat}, {Rottmann}, {Roy}, {Ruiz},
  {Ruszczyk}, {Rygl}, {S{\'a}nchez}, {S{\'a}nchez-Arg{\"u}elles},
  {S{\'a}nchez-Portal}, {Sasada}, {Satapathy}, {Savolainen}, {Schloerb},
  {Schonfeld}, {Schuster}, {Shao}, {Shen}, {Small}, {Sohn}, {SooHoo},
  {Souccar}, {Sun}, {Tazaki}, {Tetarenko}, {Tiede}, {Tilanus}, {Titus},
  {Torne}, {Traianou}, {Trent}, {Trippe}, {Turk}, {van Bemmel}, {van
  Langevelde}, {van Rossum}, {Vos}, {Wagner}, {Ward-Thompson}, {Wardle},
  {Weintroub}, {Wex}, {Wharton}, {Wielgus}, {Wiik}, {Witzel}, {Wondrak},
  {Wong}, {Wu}, {Yamaguchi}, {Yoon}, {Young}, {Young}, {Younsi}, {Yuan},
  {Yuan}, {Zensus}, {Zhang}, {Zhao}, {Zhao}, {Agurto}, {Allardi}, {Amestica},
  {Araneda}, {Arriagada}, {Berghuis}, {Bertarini}, {Berthold}, {Blanchard},
  {Brown}, {C{\'a}rdenas}, {Cantzler}, {Caro}, {Castillo-Dom{\'\i}nguez},
  {Chan}, {Chang}, {Chang}, {Chang}, {Chang}, {Chen}, {Chilson}, {Chuter},
  {Ciechanowicz}, {Colin-Beltran}, {Coulson}, {Crowley}, {Degenaar},
  {Dornbusch}, {Dur{\'a}n}, {Everett}, {Faber}, {Forster}, {Fuchs}, {Gale},
  {Geertsema}, {Gonz{\'a}lez}, {Graham}, {Gueth}, {Halverson}, {Han}, {Han},
  {Hasegawa}, {Hern{\'a}ndez-Rebollar}, {Herrera}, {Herrero-Illana},
  {Heyminck}, {Hirota}, {Hoge}, {Hostler Schimpf}, {Howie}, {Huang}, {Jiang},
  {Jinchi}, {John}, {Kimura}, {Klein}, {Kubo}, {Kuroda}, {Kwon}, {Lacasse},
  {Laing}, {Leitch}, {Li}, {Liu}, {Liu}, {Lin}, {Lu}, {Mac-Auliffe},
  {Martin-Cocher}, {Matulonis}, {Maute}, {Messias}, {Meyer-Zhao},
  {Monta{\~n}a}, {Montenegro-Montes}, {Montgomerie}, {Moreno Nolasco},
  {Muders}, {Nishioka}, {Norton}, {Nystrom}, {Ogawa}, {Olivares}, {Oshiro},
  {P{\'e}rez-Beaupuits}, {Parra}, {Phillips}, {Poirier}, {Pradel}, {Qiu},
  {Raffin}, {Rahlin}, {Ram{\'\i}rez}, {Ressler}, {Reynolds},
  {Rodr{\'\i}guez-Montoya}, {Saez-Madain}, {Santana}, {Shaw}, {Shirkey},
  {Silva}, {Snow}, {Sousa}, {Sridharan}, {Stahm}, {Stark}, {Test},
  {Torstensson}, {Venegas}, {Walther}, {Wei}, {White}, {Wieching}, {Wijnands},
  {Wouterloot}, {Yu}, {Yu (于威)}, \& {Zeballos}}]{SgraP1}
{EHTC}, {Akiyama}, K., {Alberdi}, A., {et~al.} 2022{\natexlab{a}}, \apjl, 930,
  L12

\bibitem[{{EHTC} {et~al.}(2022{\natexlab{b}}){EHTC}, {Akiyama}, {Alberdi},
  {Alef}, {Algaba}, {Anantua}, {Asada}, {Azulay}, {Bach}, {Baczko}, {Ball},
  {Balokovi{\'c}}, {Barrett}, {Baub{\"o}ck}, {Benson}, {Bintley}, {Blackburn},
  {Blundell}, {Bouman}, {Bower}, {Boyce}, {Bremer}, {Brinkerink}, {Brissenden},
  {Britzen}, {Broderick}, {Broguiere}, {Bronzwaer}, {Bustamante}, {Byun},
  {Carlstrom}, {Ceccobello}, {Chael}, {Chan}, {Chatterjee}, {Chatterjee},
  {Chen}, {Chen}, {Cheng}, {Cho}, {Christian}, {Conroy}, {Conway}, {Cordes},
  {Crawford}, {Crew}, {Cruz-Osorio}, {Cui}, {Davelaar}, {De Laurentis},
  {Deane}, {Dempsey}, {Desvignes}, {Dexter}, {Dhruv}, {Doeleman}, {Dougal},
  {Dzib}, {Eatough}, {Emami}, {Falcke}, {Farah}, {Fish}, {Fomalont}, {Ford},
  {Fraga-Encinas}, {Freeman}, {Friberg}, {Fromm}, {Fuentes}, {Galison},
  {Gammie}, {Garc{\'\i}a}, {Gentaz}, {Georgiev}, {Goddi}, {Gold},
  {G{\'o}mez-Ruiz}, {G{\'o}mez}, {Gu}, {Gurwell}, {Hada}, {Haggard}, {Haworth},
  {Hecht}, {Hesper}, {Heumann}, {Ho}, {Ho}, {Honma}, {Huang}, {Huang},
  {Hughes}, {Ikeda}, {Impellizzeri}, {Inoue}, {Issaoun}, {James}, {Jannuzi},
  {Janssen}, {Jeter}, {Jiang}, {Jim{\'e}nez-Rosales}, {Johnson}, {Jorstad},
  {Joshi}, {Jung}, {Karami}, {Karuppusamy}, {Kawashima}, {Keating}, {Kettenis},
  {Kim}, {Kim}, {Kim}, {Kim}, {Kino}, {Koay}, {Kocherlakota}, {Kofuji}, {Koch},
  {Koyama}, {Kramer}, {Kramer}, {Krichbaum}, {Kuo}, {La Bella}, {Lauer}, {Lee},
  {Lee}, {Leung}, {Levis}, {Li}, {Lico}, {Lindahl}, {Lindqvist}, {Lisakov},
  {Liu}, {Liu}, {Liuzzo}, {Lo}, {Lobanov}, {Loinard}, {Lonsdale}, {Lu}, {Mao},
  {Marchili}, {Markoff}, {Marrone}, {Marscher}, {Mart{\'\i}-Vidal},
  {Matsushita}, {Matthews}, {Medeiros}, {Menten}, {Michalik}, {Mizuno},
  {Mizuno}, {Moran}, {Moriyama}, {Moscibrodzka}, {M{\"u}ller}, {Mus}, {Musoke},
  {Myserlis}, {Nadolski}, {Nagai}, {Nagar}, {Nakamura}, {Narayan}, {Narayanan},
  {Natarajan}, {Nathanail}, {Fuentes}, {Neilsen}, {Neri}, {Ni}, {Noutsos},
  {Nowak}, {Oh}, {Okino}, {Olivares}, {Ortiz-Le{\'o}n}, {Oyama}, {{\"O}zel},
  {Palumbo}, {Paraschos}, {Park}, {Parsons}, {Patel}, {Pen}, {Pesce},
  {Pi{\'e}tu}, {Plambeck}, {PopStefanija}, {Porth}, {P{\"o}tzl}, {Prather},
  {Preciado-L{\'o}pez}, {Psaltis}, {Pu}, {Ramakrishnan}, {Rao}, {Rawlings},
  {Raymond}, {Rezzolla}, {Ricarte}, {Ripperda}, {Roelofs}, {Rogers}, {Ros},
  {Romero-Ca{\~n}izales}, {Roshanineshat}, {Rottmann}, {Roy}, {Ruiz},
  {Ruszczyk}, {Rygl}, {S{\'a}nchez}, {S{\'a}nchez-Arg{\"u}elles},
  {S{\'a}nchez-Portal}, {Sasada}, {Satapathy}, {Savolainen}, {Schloerb},
  {Schonfeld}, {Schuster}, {Shao}, {Shen}, {Small}, {Sohn}, {SooHoo},
  {Souccar}, {Sun}, {Tazaki}, {Tetarenko}, {Tiede}, {Tilanus}, {Titus},
  {Torne}, {Traianou}, {Trent}, {Trippe}, {Turk}, {van Bemmel}, {van
  Langevelde}, {van Rossum}, {Vos}, {Wagner}, {Ward-Thompson}, {Wardle},
  {Weintroub}, {Wex}, {Wharton}, {Wielgus}, {Wiik}, {Witzel}, {Wondrak},
  {Wong}, {Wu}, {Yamaguchi}, {Yoon}, {Young}, {Young}, {Younsi}, {Yuan},
  {Yuan}, {Zensus}, {Zhang}, {Zhao}, {Zhao}, {Agurto}, {Araneda}, {Arriagada},
  {Bertarini}, {Berthold}, {Blanchard}, {Brown}, {C{\'a}rdenas}, {Cantzler},
  {Caro}, {Chuter}, {Ciechanowicz}, {Coulson}, {Crowley}, {Degenaar},
  {Dornbusch}, {Dur{\'a}n}, {Forster}, {Geertsema}, {Gonz{\'a}lez}, {Graham},
  {Gueth}, {Han}, {Herrera}, {Herrero-Illana}, {Heyminck}, {Hoge}, {Huang},
  {Jiang}, {John}, {Klein}, {Kubo}, {Kuroda}, {Kwon}, {Laing}, {Liu}, {Liu},
  {Mac-Auliffe}, {Martin-Cocher}, {Matulonis}, {Messias}, {Meyer-Zhao},
  {Montenegro-Montes}, {Montgomerie}, {Muders}, {Nishioka}, {Norton},
  {Olivares}, {P{\'e}rez-Beaupuits}, {Parra}, {Poirier}, {Pradel}, {Raffin},
  {Ram{\'\i}rez}, {Reynolds}, {Saez-Madain}, {Santana}, {Silva}, {Sousa},
  {Stahm}, {Torstensson}, {Venegas}, {Walther}, {Wieching}, {Wijnands}, \&
  {Wouterloot}}]{SgraP2}
{EHTC}, {Akiyama}, K., {Alberdi}, A., {et~al.} 2022{\natexlab{b}}, \apjl, 930,
  L13

\bibitem[{{EHTC} {et~al.}(2019){EHTC}, {Akiyama}, {Alberdi}, {Alef}, {Asada},
  {Azulay}, {Baczko}, {Ball}, {Balokovi{\'c}}, {Barrett}, {Bintley},
  {Blackburn}, {Boland}, {Bouman}, {Bower}, {Bremer}, {Brinkerink},
  {Brissenden}, {Britzen}, {Broderick}, {Broguiere}, {Bronzwaer}, {Byun},
  {Carlstrom}, {Chael}, {Chan}, {Chatterjee}, {Chatterjee}, {Chen}, {Chen},
  {Cho}, {Christian}, {Conway}, {Cordes}, {Crew}, {Cui}, {Davelaar}, {De
  Laurentis}, {Deane}, {Dempsey}, {Desvignes}, {Dexter}, {Doeleman}, {Eatough},
  {Falcke}, {Fish}, {Fomalont}, {Fraga-Encinas}, {Friberg}, {Fromm},
  {G{\'o}mez}, {Galison}, {Gammie}, {Garc{\'\i}a}, {Gentaz}, {Georgiev},
  {Goddi}, {Gold}, {Gu}, {Gurwell}, {Hada}, {Hecht}, {Hesper}, {Ho}, {Ho},
  {Honma}, {Huang}, {Huang}, {Hughes}, {Ikeda}, {Inoue}, {Issaoun}, {James},
  {Jannuzi}, {Janssen}, {Jeter}, {Jiang}, {Johnson}, {Jorstad}, {Jung},
  {Karami}, {Karuppusamy}, {Kawashima}, {Keating}, {Kettenis}, {Kim}, {Kim},
  {Kim}, {Kino}, {Koay}, {Koch}, {Koyama}, {Kramer}, {Kramer}, {Krichbaum},
  {Kuo}, {Lauer}, {Lee}, {Li}, {Li}, {Lindqvist}, {Liu}, {Liuzzo}, {Lo},
  {Lobanov}, {Loinard}, {Lonsdale}, {Lu}, {MacDonald}, {Mao}, {Markoff},
  {Marrone}, {Marscher}, {Mart{\'\i}-Vidal}, {Matsushita}, {Matthews},
  {Medeiros}, {Menten}, {Mizuno}, {Mizuno}, {Moran}, {Moriyama},
  {Moscibrodzka}, {M{\"u}ller}, {Nagai}, {Nagar}, {Nakamura}, {Narayan},
  {Narayanan}, {Natarajan}, {Neri}, {Ni}, {Noutsos}, {Okino}, {Olivares},
  {Ortiz-Le{\'o}n}, {Oyama}, {{\"O}zel}, {Palumbo}, {Patel}, {Pen}, {Pesce},
  {Pi{\'e}tu}, {Plambeck}, {PopStefanija}, {Porth}, {Prather},
  {Preciado-L{\'o}pez}, {Psaltis}, {Pu}, {Ramakrishnan}, {Rao}, {Rawlings},
  {Raymond}, {Rezzolla}, {Ripperda}, {Roelofs}, {Rogers}, {Ros}, {Rose},
  {Roshanineshat}, {Rottmann}, {Roy}, {Ruszczyk}, {Ryan}, {Rygl},
  {S{\'a}nchez}, {S{\'a}nchez-Arguelles}, {Sasada}, {Savolainen}, {Schloerb},
  {Schuster}, {Shao}, {Shen}, {Small}, {Sohn}, {SooHoo}, {Tazaki}, {Tiede},
  {Tilanus}, {Titus}, {Toma}, {Torne}, {Trent}, {Trippe}, {Tsuda}, {van
  Bemmel}, {van Langevelde}, {van Rossum}, {Wagner}, {Wardle}, {Weintroub},
  {Wex}, {Wharton}, {Wielgus}, {Wong}, {Wu}, {Young}, {Young}, {Younsi},
  {Yuan}, {Yuan}, {Zensus}, {Zhao}, {Zhao}, {Zhu}, {Cappallo}, {Farah},
  {Folkers}, {Meyer-Zhao}, {Michalik}, {Nadolski}, {Nishioka}, {Pradel},
  {Primiani}, {Souccar}, {Vertatschitsch}, \& {Yamaguchi}}]{EHT_M87_P3}
{EHTC}, {Akiyama}, K., {Alberdi}, A., {et~al.} 2019, \apjl, 875, L3

\bibitem[{{EHTC} {et~al.}(2021{\natexlab{a}}){EHTC}, {Akiyama}, {Algaba},
  {Alberdi}, {Alef}, {Anantua}, {Asada}, {Azulay}, \& {Baczko}}]{PaperVII}
{EHTC}, {Akiyama}, K., {Algaba}, J.~C., {et~al.} 2021{\natexlab{a}}, \apjl,
  910, L12

\bibitem[{{EHTC} {et~al.}(2021{\natexlab{b}}){EHTC}, {Akiyama}, {Algaba},
  {Alberdi}, {Alef}, {Anantua}, {Asada}, {Azulay}, {Baczko}, {Ball}, \&
  et~al.}]{akiyama_first_2021}
{EHTC}, {Akiyama}, K., {Algaba}, J.~C., {et~al.} 2021{\natexlab{b}}, \apjl,
  910, L13

\bibitem[{{Fendt} \& {Yardimci}(2022)}]{2022ApJ...933...71F}
{Fendt}, C. \& {Yardimci}, M. 2022, \apj, 933, 71

\bibitem[{{Ferrari} {et~al.}(1978){Ferrari}, {Trussoni}, \&
  {Zaninetti}}]{1978A&A....64...43F}
{Ferrari}, A., {Trussoni}, E., \& {Zaninetti}, L. 1978, \aap, 64, 43

\bibitem[{{Fomalont}(1999)}]{1999ASPC..180..301F}
{Fomalont}, E.~B. 1999, in Astronomical Society of the Pacific Conference
  Series, Vol. 180, Synthesis Imaging in Radio Astronomy II, ed. G.~B.
  {Taylor}, C.~L. {Carilli}, \& R.~A. {Perley}, 301

\bibitem[{{Goddi} {et~al.}(2021){Goddi}, {Mart{\'\i}-Vidal}, {Messias},
  {Bower}, {Broderick}, {Dexter}, {Marrone}, {Moscibrodzka}, {Nagai}, {Algaba},
  {Asada}, {Crew}, {G{\'o}mez}, {Impellizzeri}, {Janssen}, {Kadler},
  {Krichbaum}, {Lico}, {Matthews}, {Nathanail}, {Ricarte}, {Ros}, {Younsi},
  {Akiyama}, {Alberdi}, {Alef}, {Anantua}, {Azulay}, {Baczko}, {Ball},
  {Balokovi{\'c}}, {Barrett}, {Benson}, {Bintley}, {Blackburn}, {Blundell},
  {Boland}, {Bouman}, {Boyce}, {Bremer}, {Brinkerink}, {Brissenden}, {Britzen},
  {Broguiere}, {Bronzwaer}, {Byun}, {Carlstrom}, {Chael}, {Chan}, {Chatterjee},
  {Chatterjee}, {Chen}, {Chen}, {Chesler}, {Cho}, {Christian}, {Conway},
  {Cordes}, {Crawford}, {Cruz-Osorio}, {Cui}, {Davelaar}, {De Laurentis},
  {Deane}, {Dempsey}, {Desvignes}, {Doeleman}, {Eatough}, {Falcke}, {Farah},
  {Fish}, {Fomalont}, {Ford}, {Fraga-Encinas}, {Freeman}, {Friberg}, {Fromm},
  {Fuentes}, {Galison}, {Gammie}, {Garc{\'\i}a}, {Gentaz}, {Georgiev}, {Gold},
  {G{\'o}mez-Ruiz}, {Gu}, {Gurwell}, {Hada}, {Haggard}, {Hecht}, {Hesper},
  {Ho}, {Ho}, {Honma}, {Huang}, {Huang}, {Hughes}, {Inoue}, {Issaoun}, {James},
  {Jannuzi}, {Jeter}, {Jiang}, {Jimenez-Rosales}, {Johnson}, {Jorstad}, {Jung},
  {Karami}, {Karuppusamy}, {Kawashima}, {Keating}, {Kettenis}, {Kim}, {Kim},
  {Kim}, {Kim}, {Kino}, {Koay}, {Kofuji}, {Koch}, {Koyama}, {Kramer}, {Kramer},
  {Kuo}, {Lauer}, {Lee}, {Levis}, {Li}, {Li}, {Lindqvist}, {Lindahl}, {Liu},
  {Liu}, {Liuzzo}, {Lo}, {Lobanov}, {Loinard}, {Lonsdale}, {Lu}, {MacDonald},
  {Mao}, {Marchili}, {Markoff}, {Marscher}, {Matsushita}, {Medeiros}, {Menten},
  {Mizuno}, {Mizuno}, {Moran}, {Moriyama}, {M{\"u}ller}, {Musoke},
  {Mej{\'\i}as}, {Nagar}, {Nakamura}, {Narayan}, {Narayanan}, {Natarajan},
  {Neilsen}, {Neri}, {Ni}, {Noutsos}, {Nowak}, {Okino}, {Olivares},
  {Ortiz-Le{\'o}n}, {Oyama}, {{\"O}zel}, {Palumbo}, {Park}, {Patel}, {Pen},
  {Pesce}, {Pi{\'e}tu}, {Plambeck}, {PopStefanija}, {Porth}, {P{\"o}tzl},
  {Prather}, {Preciado-L{\'o}pez}, {Psaltis}, {Pu}, {Ramakrishnan}, {Rao},
  {Rawlings}, {Raymond}, {Rezzolla}, {Ripperda}, {Roelofs}, {Rogers}, {Rose},
  {Roshanineshat}, {Rottmann}, {Roy}, {Ruszczyk}, {Rygl}, {S{\'a}nchez},
  {S{\'a}nchez-Arguelles}, {Sasada}, {Savolainen}, {Schloerb}, {Schuster},
  {Shao}, {Shen}, {Small}, {Sohn}, {SooHoo}, {Sun}, {Tazaki}, {Tetarenko},
  {Tiede}, {Tilanus}, {Titus}, {Toma}, {Torne}, {Trent}, {Traianou}, {Trippe},
  {van Bemmel}, {van Langevelde}, {van Rossum}, {Wagner}, {Ward-Thompson},
  {Wardle}, {Weintroub}, {Wex}, {Wharton}, {Wielgus}, {Wong}, {Wu}, {Yoon},
  {Young}, {Young}, {Yuan}, {Yuan}, {Zensus}, {Zhao}, {Zhao}, {Bruni},
  {Gopakumar}, {Hern{\'a}ndez-G{\'o}mez}, {Herrero-Illana}, {Ingram},
  {Komossa}, {Kovalev}, {Muders}, {Perucho}, {R{\"o}sch}, \&
  {Valtonen}}]{2021ApJ...910L..14G}
{Goddi}, C., {Mart{\'\i}-Vidal}, I., {Messias}, H., {et~al.} 2021, \apjl, 910,
  L14

\bibitem[{{Goddi} {et~al.}(2019){Goddi}, {Mart{\'\i}-Vidal}, {Messias}, {Crew},
  {Herrero-Illana}, {Impellizzeri}, {Rottmann}, {Wagner}, {Fomalont},
  {Matthews}, {Petry}, {Phillips}, {Tilanus}, {Villard}, {Blackburn},
  {Janssen}, \& {Wielgus}}]{Goddi2019}
{Goddi}, C., {Mart{\'\i}-Vidal}, I., {Messias}, H., {et~al.} 2019, \pasp, 131,
  075003

\bibitem[{{G\'omez}(2002)}]{Gomez2002}
{G\'omez}, J.~l. 2002, VLBA Scientific MEMO (Socorro: NRAO), 30, 1

\bibitem[{{G{\'o}mez} {et~al.}(2022){G{\'o}mez}, {Traianou}, {Krichbaum},
  {Lobanov}, {Fuentes}, {Lico}, {Zhao}, {Bruni}, {Kovalev},
  {L{\"a}hteenm{\"a}ki}, {Voitsik}, {Lisakov}, {Angelakis}, {Bach}, {Casadio},
  {Cho}, {Dey}, {Gopakumar}, {Gurvits}, {Jorstad}, {Kovalev}, {Lister},
  {Marscher}, {Myserlis}, {Pushkarev}, {Ros}, {Savolainen}, {Tornikoski},
  {Valtonen}, \& {Zensus}}]{2022ApJ...924..122G}
{G{\'o}mez}, J.~L., {Traianou}, E., {Krichbaum}, T.~P., {et~al.} 2022, \apj,
  924, 122

\bibitem[{Gómez {et~al.}(2016)Gómez, Lobanov, Bruni, Kovalev, Marscher,
  Jorstad, Mizuno, Bach, Sokolovsky, Anderson, Galindo, Kardashev, \&
  Lisakov}]{gomez_probing_2016}
Gómez, J.~L., Lobanov, A.~P., Bruni, G., {et~al.} 2016, \apj, 817, 96

\bibitem[{{Hada} {et~al.}(2017){Hada}, {Park}, {Kino}, {Niinuma}, {Sohn}, {Ro},
  {Jung}, {Algaba}, {Zhao}, {Lee}, {Akiyama}, {Trippe}, {Wajima},
  {Sawada-Satoh}, {Tazaki}, {Cho}, {Hodgson}, {Lee}, {Hagiwara}, {Honma},
  {Koyama}, {Oh}, {Lee}, {Yoo}, {Kawaguchi}, {Roh}, {Oh}, {Yeom}, {Jung}, {Oh},
  {Kim}, {Hwang}, {Byun}, {Cho}, {Kim}, {Kobayashi}, \& {Shibata}}]{Hada_2017}
{Hada}, K., {Park}, J.~H., {Kino}, M., {et~al.} 2017, \pasj, 69, 71

\bibitem[{{Hardee}(2000)}]{2000ApJ...533..176H}
{Hardee}, P.~E. 2000, \apj, 533, 176

\bibitem[{Hardee \& Norman(1988)}]{hardee1988spatial}
Hardee, P.~E. \& Norman, M.~L. 1988, \apj, 334, 70

\bibitem[{{Hovatta} {et~al.}(2014){Hovatta}, {Lister}, \& {et
  al.}}]{Hovatta2014}
{Hovatta}, T., {Lister}, M., \& {et al.} 2014, \aj, 147, 143

\bibitem[{{Hovatta} {et~al.}(2012){Hovatta}, {Lister}, {Aller}, {Aller},
  {Homan}, {Kovalev}, {Pushkarev}, \& {Savolainen}}]{2012AJ....144..105H}
{Hovatta}, T., {Lister}, M.~L., {Aller}, M.~F., {et~al.} 2012, \aj, 144, 105

\bibitem[{{Issaoun} {et~al.}(2019){Issaoun}, {Johnson}, {Blackburn},
  {Brinkerink}, {Mo{\'s}cibrodzka}, {Chael}, {Goddi}, {Mart{\'{\i}}-Vidal},
  {Wagner}, {Doeleman}, {Falcke}, {Krichbaum}, {Akiyama}, {Bach}, {Bouman},
  {Bower}, {Broderick}, {Cho}, {Crew}, {Dexter}, {Fish}, {Gold}, {G{\'o}mez},
  {Hada}, {Hern{\'a}ndez-G{\'o}mez}, {Jan{\ss}en}, {Kino}, {Kramer}, {Loinard},
  {Lu}, {Markoff}, {Marrone}, {Matthews}, {Moran}, {M{\"u}ller}, {Roelofs},
  {Ros}, {Rottmann}, {Sanchez}, {Tilanus}, {de Vicente}, {Wielgus}, {Zensus},
  \& {Zhao}}]{Issaoun_2019}
{Issaoun}, S., {Johnson}, M.~D., {Blackburn}, L., {et~al.} 2019, \apj, 871, 30

\bibitem[{{Issaoun} {et~al.}(2022){Issaoun}, {Wielgus}, {Jorstad}, {Krichbaum},
  {Blackburn}, {Janssen}, {Chan}, {Pesce}, {G{\'o}mez}, {Akiyama},
  {Mo{\'s}cibrodzka}, {Mart{\'\i}-Vidal}, {Chael}, {Lico}, {Liu},
  {Ramakrishnan}, {Lisakov}, {Fuentes}, {Zhao}, {Moriyama}, {Broderick},
  {Tiede}, {MacDonald}, {Mizuno}, {Traianou}, {Loinard}, {Davelaar}, {Gurwell},
  {Lu}, {Alberdi}, {Alef}, {Algaba}, {Anantua}, {Asada}, {Azulay}, {Bach},
  {Baczko}, {Ball}, {Balokovi{\'c}}, {Barrett}, {Baub{\"o}ck}, {Benson},
  {Bintley}, {Blundell}, {Boland}, {Bouman}, {Bower}, {Boyce}, {Bremer},
  {Brinkerink}, {Brissenden}, {Britzen}, {Broguiere}, {Bronzwaer},
  {Bustamante}, {Byun}, {Carlstrom}, {Ceccobello}, {Chatterjee}, {Chatterjee},
  {Chen}, {Chen}, {Cho}, {Christian}, {Conroy}, {Conway}, {Cordes}, {Crawford},
  {Crew}, {Cruz-Osorio}, {Cui}, {De Laurentis}, {Deane}, {Dempsey},
  {Desvignes}, {Dexter}, {Doeleman}, {Dhruv}, {Dzib Quijano}, {Eatough},
  {Emami}, {Falcke}, {Farah}, {Fish}, {Fomalont}, {Ford}, {Fraga-Encinas},
  {Freeman}, {Friberg}, {Fromm}, {Galison}, {Gammie}, {Garc{\'\i}a}, {Gentaz},
  {Georgiev}, {Goddi}, {Gold}, {G{\'o}mez-Ruiz}, {Gu}, {Hada}, {Haggard},
  {Hecht}, {Hesper}, {Ho}, {Ho}, {Honma}, {Huang}, {Huang}, {Hughes}, {Ikeda},
  {Impellizzeri}, {Inoue}, {James}, {Jannuzi}, {Jeter}, {Jiang},
  {Jimenez-Rosales}, {Johnson}, {Joshi}, {Jung}, {Karami}, {Karuppusamy},
  {Kawashima}, {Keating}, {Kettenis}, {Kim}, {Kim}, {Kim}, {Kim}, {Kino},
  {Koay}, {Kocherlakota}, {Kofuji}, {Koch}, {Koyama}, {Kramer}, {Kramer},
  {Kuo}, {La Bella}, {Lauer}, {Lee}, {Lee}, {Leung}, {Levis}, {Li}, {Lico},
  {Lindahl}, {Lindqvist}, {Liu}, {Liuzzo}, {Lo}, {Lobanov}, {Lonsdale}, {Mao},
  {Marchili}, {Markoff}, {Marrone}, {Marscher}, {Matsushita}, {Matthews},
  {Medeiros}, {Menten}, {Michalik}, {Mizuno}, {Mizuno}, {Moran}, {M{\"u}ller},
  {Mus}, {Musoke}, {Myserlis}, {Nadolski}, {Nagai}, {Nagar}, {Nakamura},
  {Narayan}, {Narayanan}, {Natarajan}, {Nathanail}, {Neilsen}, {Neri}, {Ni},
  {Noutsos}, {Nowak}, {Oh}, {Okino}, {Olivares}, {Ortiz-Le{\'o}n}, {Oyama},
  {{\"O}zel}, {Palumbo}, {Paraschos}, {Park}, {Parsons}, {Patel}, {Pen},
  {Pi{\'e}tu}, {Plambeck}, {PopStefanija}, {Porth}, {P{\"o}tzl}, {Prather},
  {Preciado-L{\'o}pez}, {Psaltis}, {Pu}, {Rao}, {Rawlings}, {Raymond},
  {Rezzolla}, {Ricarte}, {Ripperda}, {Roelofs}, {Rogers}, {Ros},
  {Romero-Canizales}, {Roshanineshat}, {Rottmann}, {Roy}, {Ruiz}, {Ruszczyk},
  {Rygl}, {S{\'a}nchez}, {S{\'a}nchez-Arguelles}, {Sanchez-Portal}, {Sasada},
  {Satapathy}, {Savolainen}, {Schloerb}, {Schuster}, {Shao}, {Shen}, {Small},
  {Sohn}, {SooHoo}, {Souccar}, {Sun}, {Tazaki}, {Tetarenko}, {Tiede},
  {Tilanus}, {Titus}, {Torne}, {Trent}, {Trippe}, {van Bemmel}, {van
  Langevelde}, {van Rossum}, {Vos}, {Wagner}, {Ward-Thompson}, {Wardle},
  {Weintroub}, {Wex}, {Wharton}, {Wiik}, {Witzel}, {Wondrak}, {Wong}, {Wu},
  {Yamaguchi}, {Yoon}, {Young}, {Young}, {Younsi}, {Yuan}, {Yuan}, {Zensus},
  {Zhang}, \& {Zhao}}]{Issaoun2022}
{Issaoun}, S., {Wielgus}, M., {Jorstad}, S., {et~al.} 2022, \apj, 934, 145

\bibitem[{Istomin \& Pariev(1996)}]{istomin1996stability}
Istomin, Y.~N. \& Pariev, V. 1996, \mnras, 281, 1

\bibitem[{{Janssen} {et~al.}(2019){Janssen}, {Goddi}, {van Bemmel}, {Kettenis},
  {Small}, {Liuzzo}, {Rygl}, {Mart{\'\i}-Vidal}, {Blackburn}, {Wielgus}, \&
  {Falcke}}]{Janssen2019}
{Janssen}, M., {Goddi}, C., {van Bemmel}, I.~M., {et~al.} 2019, \aap, 626, A75

\bibitem[{{Johnson} {et~al.}(2023){Johnson}, {Akiyama}, {Blackburn}, {Bouman},
  {Broderick}, {Cardoso}, {Fender}, {Fromm}, {Galison}, {G{\'o}mez}, {Haggard},
  {Lister}, {Lobanov}, {Markoff}, {Narayan}, {Natarajan}, {Nichols}, {Pesce},
  {Younsi}, {Chael}, {Chatterjee}, {Chaves}, {Doboszewski}, {Dodson},
  {Doeleman}, {Elder}, {Fitzpatrick}, {Haworth}, {Houston}, {Issaoun},
  {Kovalev}, {Levis}, {Lico}, {Marcoci}, {Martens}, {Nagar}, {Oppenheimer},
  {Palumbo}, {Ricarte}, {Rioja}, {Roelofs}, {Thresher}, {Tiede}, {Weintroub},
  \& {Wielgus}}]{2023Galax..11...61J}
{Johnson}, M.~D., {Akiyama}, K., {Blackburn}, L., {et~al.} 2023, Galaxies, 11,
  61

\bibitem[{{Jorstad} {et~al.}(2023){Jorstad}, {Wielgus}, {Lico}, {Issaoun},
  {Broderick}, {Pesce}, {Liu}, {Zhao}, {Krichbaum}, {Blackburn}, {Chan},
  {Janssen}, {Ramakrishnan}, {Akiyama}, {Alberdi}, {Algaba}, {Bouman}, {Cho},
  {Fuentes}, {G{\'o}mez}, {Gurwell}, {Johnson}, {Kim}, {Lu},
  {Mart{\'\i}-Vidal}, {Moscibrodzka}, {P{\"o}tzl}, {Traianou}, {van Bemmel},
  {Alef}, {Anantua}, {Asada}, {Azulay}, {Bach}, {Baczko}, {Ball},
  {Balokovi{\'c}}, {Barrett}, {Baub{\"o}ck}, {Benson}, {Bintley}, {Blundell},
  {Bower}, {Boyce}, {Bremer}, {Brinkerink}, {Brissenden}, {Britzen},
  {Broguiere}, {Bronzwaer}, {Bustamante}, {Byun}, {Carlstrom}, {Ceccobello},
  {Chael}, {Chatterjee}, {Chatterjee}, {Chen}, {Chen}, {Cheng}, {Christian},
  {Conroy}, {Conway}, {Cordes}, {Crawford}, {Crew}, {Cruz-Osorio}, {Cui},
  {Davelaar}, {De Laurentis}, {Deane}, {Dempsey}, {Desvignes}, {Dexter},
  {Dhruv}, {Doeleman}, {Dougal}, {Dzib}, {Eatough}, {Emami}, {Falcke}, {Farah},
  {Fish}, {Fomalont}, {Ford}, {Fraga-Encinas}, {Freeman}, {Friberg}, {Fromm},
  {Galison}, {Gammie}, {Garc{\'\i}a}, {Gentaz}, {Georgiev}, {Goddi}, {Gold},
  {G{\'o}mez-Ruiz}, {Gu}, {Hada}, {Haggard}, {Haworth}, {Hecht}, {Hesper},
  {Heumann}, {Ho}, {Ho}, {Honma}, {Huang}, {Huang}, {Hughes}, {Ikeda},
  {Impellizzeri}, {Inoue}, {James}, {Jannuzi}, {Jeter}, {Jiang},
  {Jim{\'e}nez-Rosales}, {Joshi}, {Jung}, {Karami}, {Karuppusamy}, {Kawashima},
  {Keating}, {Kettenis}, {Kim}, {Kim}, {Kim}, {Kino}, {Koay}, {Kocherlakota},
  {Kofuji}, {Koyama}, {Kramer}, {Kramer}, {Kuo}, {La Bella}, {Lauer}, {Lee},
  {Lee}, {Leung}, {Levis}, {Li}, {Lindahl}, {Lindqvist}, {Lisakov}, {Liu},
  {Liuzzo}, {Lo}, {Lobanov}, {Loinard}, {Lonsdale}, {MacDonald}, {Mao},
  {Marchili}, {Markoff}, {Marrone}, {Marscher}, {Matsushita}, {Matthews},
  {Medeiros}, {Menten}, {Michalik}, {Mizuno}, {Mizuno}, {Moran}, {Moriyama},
  {M{\"u}ller}, {Mus}, {Musoke}, {Myserlis}, {Nadolski}, {Nagai}, {Nagar},
  {Nakamura}, {Narayan}, {Narayanan}, {Natarajan}, {Nathanail}, {Fuentes},
  {Neilsen}, {Neri}, {Ni}, {Noutsos}, {Nowak}, {Oh}, {Okino}, {Olivares},
  {Ortiz-Le{\'o}n}, {Oyama}, {{\"O}zel}, {Palumbo}, {Paraschos}, {Park},
  {Parsons}, {Patel}, {Pen}, {Pi{\'e}tu}, {Plambeck}, {PopStefanija}, {Porth},
  {Prather}, {Preciado-L{\'o}pez}, {Psaltis}, {Pu}, {Rao}, {Rawlings},
  {Raymond}, {Rezzolla}, {Ricarte}, {Ripperda}, {Roelofs}, {Rogers}, {Ros},
  {Romero-Ca{\~n}izales}, {Roshanineshat}, {Rottmann}, {Roy}, {Ruiz},
  {Ruszczyk}, {Rygl}, {S{\'a}nchez}, {S{\'a}nchez-Arg{\"u}elles},
  {S{\'a}nchez-Portal}, {Sasada}, {Satapathy}, {Savolainen}, {Schloerb},
  {Schonfeld}, {Schuster}, {Shao}, {Shen}, {Small}, {Sohn}, {SooHoo},
  {Souccar}, {Sun}, {Tazaki}, {Tetarenko}, {Tiede}, {Tilanus}, {Titus},
  {Torne}, {Trent}, {Trippe}, {Turk}, {van Langevelde}, {van Rossum}, {Vos},
  {Wagner}, {Ward-Thompson}, {Wardle}, {Weintroub}, {Wex}, {Wharton}, {Wiik},
  {Witzel}, {Wondrak}, {Wong}, {Wu}, {Yamaguchi}, {Yoon}, {Young}, {Young},
  {Younsi}, {Yuan}, {Yuan}, {Zensus}, {Zhang}, \& {Zhao}}]{Jorstad2023}
{Jorstad}, S., {Wielgus}, M., {Lico}, R., {et~al.} 2023, \apj, 943, 170

\bibitem[{{Jorstad} {et~al.}(2017){Jorstad}, {Marscher}, {Morozova}, \& {et
  al.}}]{Jorstad2017}
{Jorstad}, S.~G., {Marscher}, A.~P., {Morozova}, D.~A., \& {et al.} 2017, \apj,
  846, 98

\bibitem[{{Jorstad} {et~al.}(2007){Jorstad}, {Marscher}, {Stevens}, {Smith},
  {Forster}, {Gear}, {Cawthorne}, {Lister}, {Stirling}, {G{\'o}mez}, {Greaves},
  \& {Robson}}]{2007AJ....134..799J}
{Jorstad}, S.~G., {Marscher}, A.~P., {Stevens}, J.~A., {et~al.} 2007, \aj, 134,
  799

\bibitem[{{Junkkarinen}(1984)}]{Junkkarinen1984}
{Junkkarinen}, V. 1984, \pasp, 96, 539

\bibitem[{Keck(2019)}]{MKeck2019}
Keck, M. 2019, PhD thesis, Boston University

\bibitem[{{Kellermann} {et~al.}(1998){Kellermann}, {Vermeulen}, {Zensus}, \&
  {Cohen}}]{1998AJ....115.1295K}
{Kellermann}, K.~I., {Vermeulen}, R.~C., {Zensus}, J.~A., \& {Cohen}, M.~H.
  1998, \aj, 115, 1295

\bibitem[{{Kharb} {et~al.}(2010){Kharb}, {Lister}, \& {Cooper}}]{Kharb2010}
{Kharb}, P., {Lister}, M.~L., \& {Cooper}, N.~J. 2010, \apj, 710, 764

\bibitem[{{Kino} {et~al.}(2022){Kino}, {Takahashi}, {Kawashima}, {Park},
  {Hada}, {Ro}, \& {Cui}}]{2022ApJ...939...83K}
{Kino}, M., {Takahashi}, M., {Kawashima}, T., {et~al.} 2022, \apj, 939, 83

\bibitem[{{Konigl}(1981)}]{1981ApJ...243..700K}
{Konigl}, A. 1981, \apj, 243, 700

\bibitem[{{Kravchenko} {et~al.}(2017){Kravchenko}, {Kovalev}, \&
  {Sokolovsky}}]{2017MNRAS.467...83K}
{Kravchenko}, E.~V., {Kovalev}, Y.~Y., \& {Sokolovsky}, K.~V. 2017, \mnras,
  467, 83

\bibitem[{{Kutkin} {et~al.}(2014){Kutkin}, {Sokolovsky}, {Lisakov}, {Kovalev},
  {Savolainen}, {Voytsik}, {Lobanov}, {Aller}, {Aller}, {Lahteenmaki},
  {Tornikoski}, {Volvach}, \& {Volvach}}]{2014MNRAS.437.3396K}
{Kutkin}, A.~M., {Sokolovsky}, K.~V., {Lisakov}, M.~M., {et~al.} 2014, \mnras,
  437, 3396

\bibitem[{Lepp\"anen {et~al.}(1995)Lepp\"anen, Zensus, \&
  Diamond}]{Leppanen1995}
Lepp\"anen, K., Zensus, J., \& Diamond, P. 1995, \aj, 110, 2479

\bibitem[{{Liang} \& {Liu}(2003)}]{2003MNRAS.340..632L}
{Liang}, E.~W. \& {Liu}, H.~T. 2003, \mnras, 340, 632

\bibitem[{Lisakov {et~al.}(2017)Lisakov, Kovalev, Savolainen, Hovatta, \&
  Kutkin}]{lisakov_connection_2017}
Lisakov, M.~M., Kovalev, Y.~Y., Savolainen, T., Hovatta, T., \& Kutkin, A.~M.
  2017, \mnras, 468, 4478

\bibitem[{Lisakov {et~al.}(2021)Lisakov, Kravchenko, Pushkarev, Kovalev,
  Savolainen, \& Lister}]{lisakov_oversized_2021}
Lisakov, M.~M., Kravchenko, E.~V., Pushkarev, A.~B., {et~al.} 2021, \apj, 910,
  35

\bibitem[{{Lister} {et~al.}(2018){Lister}, {Aller}, {Aller}, {Hodge}, {Homan},
  {Kovalev}, {Pushkarev}, \& {Savolainen}}]{2018ApJS..234...12L}
{Lister}, M.~L., {Aller}, M.~F., {Aller}, H.~D., {et~al.} 2018, \apjs, 234, 12

\bibitem[{{Lister} {et~al.}(2019){Lister}, {Homan}, {Hovatta}, {Kellermann},
  {Kiehlmann}, {Kovalev}, {Max-Moerbeck}, {Pushkarev}, {Readhead}, {Ros}, \&
  {Savolainen}}]{2019ApJ...874...43L}
{Lister}, M.~L., {Homan}, D.~C., {Hovatta}, T., {et~al.} 2019, \apj, 874, 43

\bibitem[{{Lister} {et~al.}(2021){Lister}, {Homan}, {Kellermann}, {Kovalev},
  {Pushkarev}, {Ros}, \& {Savolainen}}]{2021ApJ...923...30L}
{Lister}, M.~L., {Homan}, D.~C., {Kellermann}, K.~I., {et~al.} 2021, \apj, 923,
  30

\bibitem[{Lobanov {et~al.}(2003)Lobanov, Hardee, \& Eilek}]{LOBANOV2003629}
Lobanov, A., Hardee, P., \& Eilek, J. 2003, New Astronomy Reviews, 47,
  629–632, the physics of relativistic jets in the CHANDRA and XMM era

\bibitem[{{Lobanov}(1998)}]{lobanov_ultracompact_1998}
{Lobanov}, A.~P. 1998, \aap, 330, 79

\bibitem[{Lobanov(2005)}]{lobanov_resolution_2005}
Lobanov, A.~P. 2005, arXiv:astro-ph/0503225, arXiv: astro-ph/0503225

\bibitem[{Lu(2010)}]{Rusen2010}
Lu, R.-S. 2010, PhD thesis, der Mathematisch-Naturwissenschaftlichen Fakult\"at
  der Universit\"at zu K\"oln

\bibitem[{{Lu} {et~al.}(2023){Lu}, {Asada}, {Krichbaum}, {Park}, {Tazaki},
  {Pu}, {Nakamura}, {Lobanov}, {Hada}, {Akiyama}, {Kim}, {Marti-Vidal},
  {G{\'o}mez}, {Kawashima}, {Yuan}, {Ros}, {Alef}, {Britzen}, {Bremer},
  {Broderick}, {Doi}, {Giovannini}, {Giroletti}, {Ho}, {Honma}, {Hughes},
  {Inoue}, {Jiang}, {Kino}, {Koyama}, {Lindqvist}, {Liu}, {Marscher},
  {Matsushita}, {Nagai}, {Rottmann}, {Savolainen}, {Schuster}, {Shen}, {de
  Vicente}, {Walker}, {Yang}, {Zensus}, {Algaba}, {Allardi}, {Bach},
  {Berthold}, {Bintley}, {Byun}, {Casadio}, {Chang}, {Chang}, {Chang}, {Chen},
  {Chen}, {Chilson}, {Chuter}, {Conway}, {Crew}, {Dempsey}, {Dornbusch},
  {Faber}, {Friberg}, {Garc{\'\i}a}, {Garrido}, {Han}, {Han}, {Hasegawa},
  {Herrero-Illana}, {Huang}, {Huang}, {Impellizzeri}, {Jiang}, {Jinchi},
  {Jung}, {Kallunki}, {Kirves}, {Kimura}, {Koay}, {Koch}, {Kramer}, {Kraus},
  {Kubo}, {Kuo}, {Li}, {Lin}, {Liu}, {Liu}, {Lo}, {Lu}, {MacDonald},
  {Martin-Cocher}, {Messias}, {Meyer-Zhao}, {Minter}, {Nair}, {Nishioka},
  {Norton}, {Nystrom}, {Ogawa}, {Oshiro}, {Patel}, {Pen}, {Pidopryhora},
  {Pradel}, {Raffin}, {Rao}, {Ruiz}, {Sanchez}, {Shaw}, {Snow}, {Sridharan},
  {Srinivasan}, {Tercero}, {Torne}, {Traianou}, {Wagner}, {Walther}, {Wei},
  {Yang}, \& {Yu}}]{2023Natur.616..686L}
{Lu}, R.-S., {Asada}, K., {Krichbaum}, T.~P., {et~al.} 2023, \nat, 616, 686

\bibitem[{{Lu} {et~al.}(2011){Lu}, {Krichbaum}, \& {Zensus}}]{Lu2011}
{Lu}, R.~S., {Krichbaum}, T.~P., \& {Zensus}, J.~A. 2011, \mnras, 418, 2260

\bibitem[{{Mart{\'\i}-Vidal} {et~al.}(2011){Mart{\'\i}-Vidal}, {Marcaide},
  {Alberdi}, {P{\'e}rez-Torres}, {Ros}, \& {Guirado}}]{2011A&A...533A.111M}
{Mart{\'\i}-Vidal}, I., {Marcaide}, J.~M., {Alberdi}, A., {et~al.} 2011, \aap,
  533, A111

\bibitem[{Mertens {et~al.}(2016)Mertens, Lobanov, Walker, \&
  Hardee}]{mertens2016kinematics}
Mertens, F., Lobanov, A., Walker, R., \& Hardee, P. 2016, Astronomy \&
  Astrophysics, 595, A54

\bibitem[{Mizuno {et~al.}(2012)Mizuno, Lyubarsky, Nishikawa, \&
  Hardee}]{mizuno2012three}
Mizuno, Y., Lyubarsky, Y., Nishikawa, K.-I., \& Hardee, P.~E. 2012, \apj, 757,
  16

\bibitem[{Narayan {et~al.}(2009)Narayan, Li, \& Tchekhovskoy}]{Narayan_2009}
Narayan, R., Li, J., \& Tchekhovskoy, A. 2009, \apj, 697, 1681

\bibitem[{{Niinuma} {et~al.}(2015){Niinuma}, {Lee}, {Kino}, \&
  {Sohn}}]{Niinuma_2015}
{Niinuma}, K., {Lee}, S.-S., {Kino}, M., \& {Sohn}, B.~W. 2015, Publication of
  Korean Astronomical Society, 30, 637

\bibitem[{Norman \& Hardee(1988)}]{norman1988spatial}
Norman, M.~L. \& Hardee, P.~E. 1988, \apj, 334, 80

\bibitem[{{Park} {et~al.}(2019){Park}, {Hada}, {Kino}, {Nakamura}, {Hodgson},
  {Ro}, {Cui}, {Asada}, {Algaba}, {Sawada-Satoh}, {Lee}, {Cho}, {Shen},
  {Jiang}, {Trippe}, {Niinuma}, {Sohn}, {Jung}, {Zhao}, {Wajima}, {Tazaki},
  {Honma}, {An}, {Akiyama}, {Byun}, {Kim}, {Zhang}, {Cheng}, {Kobayashi},
  {Shibata}, {Lee}, {Roh}, {Oh}, {Yeom}, {Jung}, {Oh}, {Kim}, {Hwang}, \&
  {Hagiwara}}]{Park_2019}
{Park}, J., {Hada}, K., {Kino}, M., {et~al.} 2019, \apj, 887, 147

\bibitem[{{Perucho} {et~al.}(2004){Perucho}, {Hanasz}, {Mart{\'\i}}, \&
  {Sol}}]{2004A&A...427..415P}
{Perucho}, M., {Hanasz}, M., {Mart{\'\i}}, J.~M., \& {Sol}, H. 2004, \aap, 427,
  415

\bibitem[{{Perucho} {et~al.}(2005){Perucho}, {Lobanov}, \&
  {Mart{\'\i}}}]{2005MmSAI..76..110P}
{Perucho}, M., {Lobanov}, A.~P., \& {Mart{\'\i}}, J.~M. 2005, \memsai, 76, 110

\bibitem[{Perucho \& Martí(2007)}]{perucho2007}
Perucho, M. \& Martí, J.~M. 2007, \mnras, 382, 526

\bibitem[{{Planck Collaboration} {et~al.}(2016){Planck Collaboration}, {Ade},
  {Aghanim}, {Arnaud}, {Ashdown}, {Aumont}, {Baccigalupi}, {Banday},
  {Barreiro}, {Bartlett}, {Bartolo}, {Battaner}, {Battye}, {Benabed},
  {Beno{\^\i}t}, {Benoit-L{\'e}vy}, {Bernard}, {Bersanelli}, {Bielewicz},
  {Bock}, {Bonaldi}, {Bonavera}, {Bond}, {Borrill}, {Bouchet}, {Boulanger},
  {Bucher}, {Burigana}, {Butler}, {Calabrese}, {Cardoso}, {Catalano},
  {Challinor}, {Chamballu}, {Chary}, {Chiang}, {Chluba}, {Christensen},
  {Church}, {Clements}, {Colombi}, {Colombo}, {Combet}, {Coulais}, {Crill},
  {Curto}, {Cuttaia}, {Danese}, {Davies}, {Davis}, {de Bernardis}, {de Rosa},
  {de Zotti}, {Delabrouille}, {D{\'e}sert}, {Di Valentino}, {Dickinson},
  {Diego}, {Dolag}, {Dole}, {Donzelli}, {Dor{\'e}}, {Douspis}, {Ducout},
  {Dunkley}, {Dupac}, {Efstathiou}, {Elsner}, {En{\ss}lin}, {Eriksen},
  {Farhang}, {Fergusson}, {Finelli}, {Forni}, {Frailis}, {Fraisse},
  {Franceschi}, {Frejsel}, {Galeotta}, {Galli}, {Ganga}, {Gauthier}, {Gerbino},
  {Ghosh}, {Giard}, {Giraud-H{\'e}raud}, {Giusarma}, {Gjerl{\o}w},
  {Gonz{\'a}lez-Nuevo}, {G{\'o}rski}, {Gratton}, {Gregorio}, {Gruppuso},
  {Gudmundsson}, {Hamann}, {Hansen}, {Hanson}, {Harrison}, {Helou},
  {Henrot-Versill{\'e}}, {Hern{\'a}ndez-Monteagudo}, {Herranz}, {Hildebrand t},
  {Hivon}, {Hobson}, {Holmes}, {Hornstrup}, {Hovest}, {Huang}, {Huffenberger},
  {Hurier}, {Jaffe}, {Jaffe}, {Jones}, {Juvela}, {Keih{\"a}nen}, {Keskitalo},
  {Kisner}, {Kneissl}, {Knoche}, {Knox}, {Kunz}, {Kurki-Suonio}, {Lagache},
  {L{\"a}hteenm{\"a}ki}, {Lamarre}, {Lasenby}, {Lattanzi}, {Lawrence}, {Leahy},
  {Leonardi}, {Lesgourgues}, {Levrier}, {Lewis}, {Liguori}, {Lilje},
  {Linden-V{\o}rnle}, {L{\'o}pez-Caniego}, {Lubin}, {Mac{\'\i}as-P{\'e}rez},
  {Maggio}, {Maino}, {Mandolesi}, {Mangilli}, {Marchini}, {Maris}, {Martin},
  {Martinelli}, {Mart{\'\i}nez-Gonz{\'a}lez}, {Masi}, {Matarrese}, {McGehee},
  {Meinhold}, {Melchiorri}, {Melin}, {Mendes}, {Mennella}, {Migliaccio},
  {Millea}, {Mitra}, {Miville-Desch{\^e}nes}, {Moneti}, {Montier}, {Morgante},
  {Mortlock}, {Moss}, {Munshi}, {Murphy}, {Naselsky}, {Nati}, {Natoli},
  {Netterfield}, {N{\o}rgaard-Nielsen}, {Noviello}, {Novikov}, {Novikov},
  {Oxborrow}, {Paci}, {Pagano}, {Pajot}, {Paladini}, {Paoletti}, {Partridge},
  {Pasian}, {Patanchon}, {Pearson}, {Perdereau}, {Perotto}, {Perrotta},
  {Pettorino}, {Piacentini}, {Piat}, {Pierpaoli}, {Pietrobon}, {Plaszczynski},
  {Pointecouteau}, {Polenta}, {Popa}, {Pratt}, {Pr{\'e}zeau}, {Prunet},
  {Puget}, {Rachen}, {Reach}, {Rebolo}, {Reinecke}, {Remazeilles}, {Renault},
  {Renzi}, {Ristorcelli}, {Rocha}, {Rosset}, {Rossetti}, {Roudier},
  {Rouill{\'e} d'Orfeuil}, {Rowan-Robinson}, {Rubi{\~n}o-Mart{\'\i}n},
  {Rusholme}, {Said}, {Salvatelli}, {Salvati}, {Sandri}, {Santos},
  {Savelainen}, {Savini}, {Scott}, {Seiffert}, {Serra}, {Shellard}, {Spencer},
  {Spinelli}, {Stolyarov}, {Stompor}, {Sudiwala}, {Sunyaev}, {Sutton},
  {Suur-Uski}, {Sygnet}, {Tauber}, {Terenzi}, {Toffolatti}, {Tomasi},
  {Tristram}, {Trombetti}, {Tucci}, {Tuovinen}, {T{\"u}rler}, {Umana},
  {Valenziano}, {Valiviita}, {Van Tent}, {Vielva}, {Villa}, {Wade}, {Wandelt},
  {Wehus}, {White}, {White}, {Wilkinson}, {Yvon}, {Zacchei}, \&
  {Zonca}}]{Planck2016}
{Planck Collaboration}, {Ade}, P.~A.~R., {Aghanim}, N., {et~al.} 2016, \aap,
  594, A13

\bibitem[{{Pushkarev} {et~al.}(2023){Pushkarev}, {Aller}, {Aller}, {Homan},
  {Kovalev}, {Lister}, {Pashchenko}, {Savolainen}, \&
  {Zobnina}}]{2023MNRAS.520.6053P}
{Pushkarev}, A.~B., {Aller}, H.~D., {Aller}, M.~F., {et~al.} 2023, \mnras, 520,
  6053

\bibitem[{Pushkarev {et~al.}(2012)Pushkarev, Hovatta, Kovalev, Lister, Lobanov,
  Savolainen, \& Zensus}]{pushkarev_mojave_2012}
Pushkarev, A.~B., Hovatta, T., Kovalev, Y.~Y., {et~al.} 2012, Astronomy \&
  Astrophysics, 545, A113

\bibitem[{{Pushkarev} {et~al.}(2009){Pushkarev}, {Kovalev}, {Lister}, \&
  {Savolainen}}]{2009A&A...507L..33P}
{Pushkarev}, A.~B., {Kovalev}, Y.~Y., {Lister}, M.~L., \& {Savolainen}, T.
  2009, \aap, 507, L33

\bibitem[{{Pushkarev} {et~al.}(2017){Pushkarev}, {Kovalev}, {Lister}, \&
  {Savolainen}}]{2017MNRAS.468.4992P}
{Pushkarev}, A.~B., {Kovalev}, Y.~Y., {Lister}, M.~L., \& {Savolainen}, T.
  2017, \mnras, 468, 4992

\bibitem[{{Raymond} {et~al.}(2024){Raymond}, {Doeleman}, {Asada}, {Blackburn},
  {Bower}, {Bremer}, {Broguiere}, {Chen}, {Crew}, {Dornbusch}, {Fish},
  {Garc{\'\i}a}, {Gentaz}, {Goddi}, {Han}, {Hecht}, {Huang}, {Janssen},
  {Keating}, {Koay}, {Krichbaum}, {Lo}, {Matsushita}, {Matthews}, {Moran},
  {Norton}, {Patel}, {Pesce}, {Ramakrishnan}, {Rottmann}, {Roy}, {S{\'a}nchez},
  {Tilanus}, {Titus}, {Torne}, {Wagner}, {Weintroub}, {Wielgus}, \&
  {Young}}]{Raymond2024}
{Raymond}, A.~W., {Doeleman}, S.~S., {Asada}, K., {et~al.} 2024, \aj, 168, 130

\bibitem[{{Shepherd}(1997)}]{Difmap}
{Shepherd}, M.~C. 1997, Astronomical Data Analysis Software and Systems VI,
  A.S.P. Conference Series, eds. G. Hunt \& H.E. Payne, 125, 77

\bibitem[{{Sillanpaa} {et~al.}(1988){Sillanpaa}, {Haarala}, {Valtonen},
  {Sundelius}, \& {Byrd}}]{1988ApJ...325..628S}
{Sillanpaa}, A., {Haarala}, S., {Valtonen}, M.~J., {Sundelius}, B., \& {Byrd},
  G.~G. 1988, \apj, 325, 628

\bibitem[{Thirring(1918)}]{thirring1918effect}
Thirring, H. 1918, Phys. Z, 19, 33

\bibitem[{{Todorov} {et~al.}(2024){Todorov}, {Kravchenko}, {Pashchenko}, \&
  {Pushkarev}}]{2024AJrus.100.12}
{Todorov}, R.~V., {Kravchenko}, E.~V., {Pashchenko}, I.~N., \& {Pushkarev},
  A.~B. 2024, Astronomy Reports, 100, 1132

\bibitem[{{Valtonen} {et~al.}(2009){Valtonen}, {Nilsson}, {Villforth}, {Lehto},
  {Takalo}, {Lindfors}, {Sillanp{\"a}{\"a}}, {Hentunen}, {Mikkola}, {Zola},
  {Drozdz}, {Koziel}, {Ogloza}, {Kurpinska-Winiarska}, {Siwak}, {Winiarski},
  {Heidt}, {Kidger}, {Pursimo}, {Wu}, {Zhou}, {Sadakane}, {Marchev},
  {Nissinen}, {Niarchos}, {Liakos}, {Gazeas}, {Dogru}, {Poyner}, {Dietrich},
  {Assef}, {Atlee}, {Bird}, {DePoy}, {Eastman}, {Peeples}, {Prieto}, {Watson},
  {Yee}, {Mattingly}, \& {Ohlert}}]{2009ApJ...698..781V}
{Valtonen}, M.~J., {Nilsson}, K., {Villforth}, C., {et~al.} 2009, \apj, 698,
  781

\bibitem[{Vega-Garc{\'\i}a {et~al.}(2020)Vega-Garc{\'\i}a, Lobanov, Perucho,
  Bruni, Ros, Anderson, Agudo, Davis, G{\'o}mez, Kovalev,
  {et~al.}}]{vega2020multiband}
Vega-Garc{\'\i}a, L., Lobanov, A., Perucho, M., {et~al.} 2020, Astronomy \&
  Astrophysics, 641, A40

\bibitem[{{von Fellenberg} {et~al.}(2023){von Fellenberg}, {Janssen},
  {Davelaar}, {Zaja{\v{c}}ek}, {Britzen}, {Falcke}, {K{\"o}rding}, \&
  {Ros}}]{2023A&A...672L...5V}
{von Fellenberg}, S.~D., {Janssen}, M., {Davelaar}, J., {et~al.} 2023, \aap,
  672, L5

\bibitem[{{Weaver} {et~al.}(2022){Weaver}, {Jorstad}, {Marscher}, {Morozova},
  {Troitsky}, {Agudo}, {G{\'o}mez}, {L{\"a}hteenm{\"a}ki}, {Tammi}, \&
  {Tornikoski}}]{Weaver22}
{Weaver}, Z.~R., {Jorstad}, S.~G., {Marscher}, A.~P., {et~al.} 2022, \apjs,
  260, 12

\bibitem[{{Wright}(2006)}]{CosmoCalc2006}
{Wright}, E.~L. 2006, \pasp, 118, 1711

\bibitem[{Zhao {et~al.}(2022)Zhao, G\'omez, Fuentes, \& et~al.}]{Zhao2022}
Zhao, G.-Y., G\'omez, J., Fuentes, A., \& et~al. 2022, \apj, 932, 72

\bibitem[{{Zobnina} {et~al.}(2023){Zobnina}, {Aller}, {Aller}, {Homan},
  {Kovalev}, {Lister}, {Pashchenko}, {Pushkarev}, \&
  {Savolainen}}]{2023MNRAS.523.3615Z}
{Zobnina}, D.~I., {Aller}, H.~D., {Aller}, M.~F., {et~al.} 2023, \mnras, 523,
  3615

\end{thebibliography}

\appendix
\section{Pairwise model alignment}
\label{app:image_alignment}

\begin{figure*}
    \centering
    \includegraphics[width=0.3\linewidth]{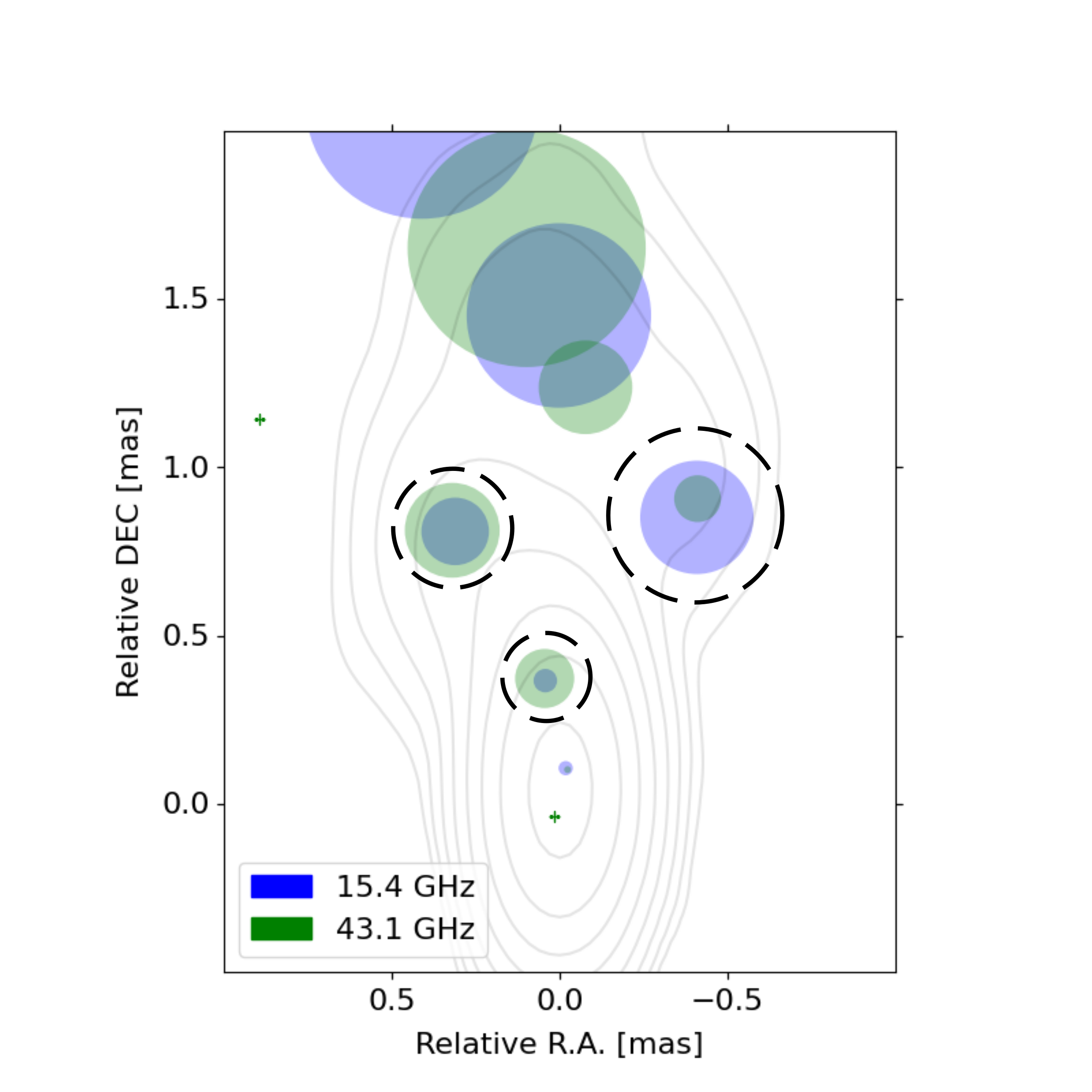}
    \includegraphics[width=0.3\linewidth]{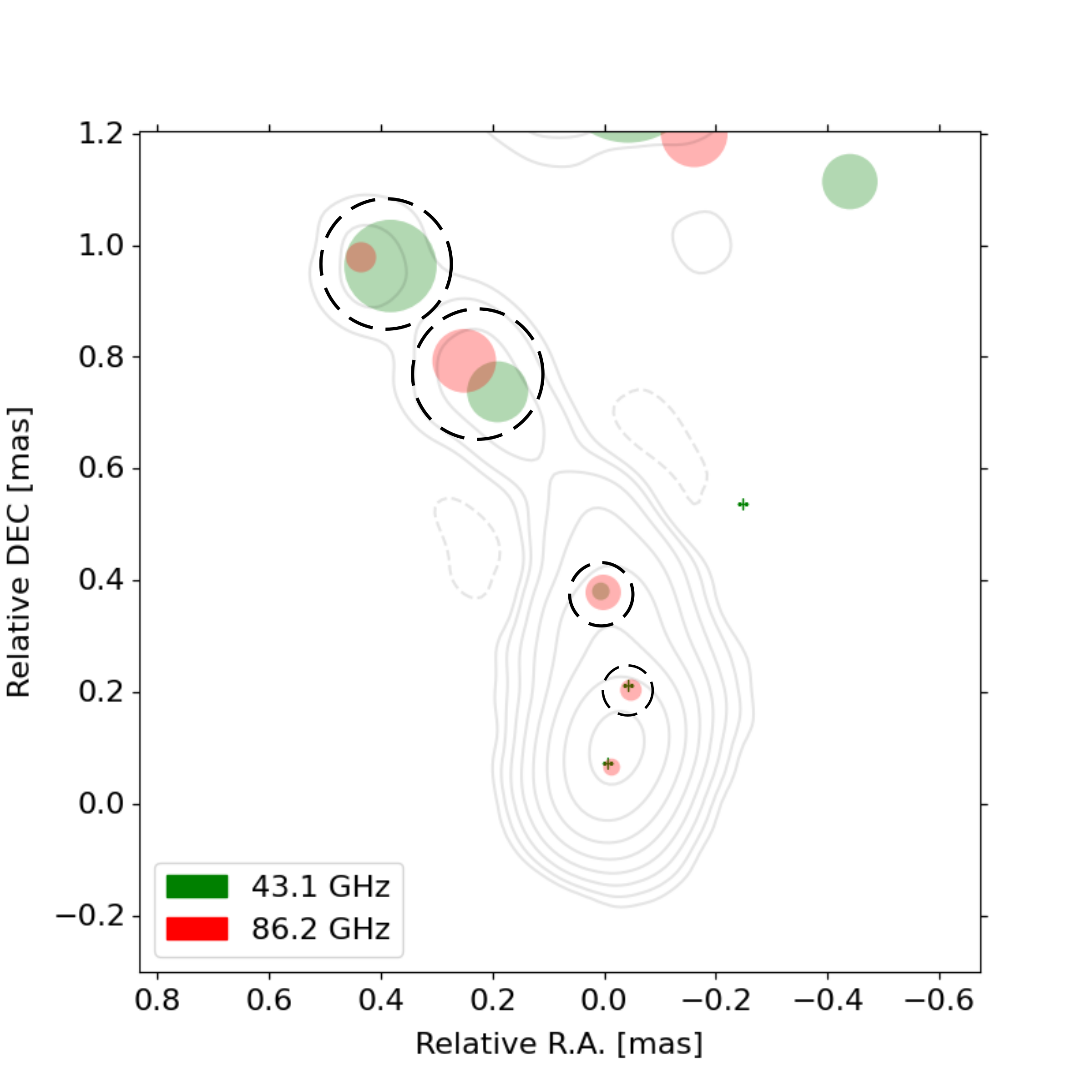}
    \includegraphics[width=0.3\linewidth]{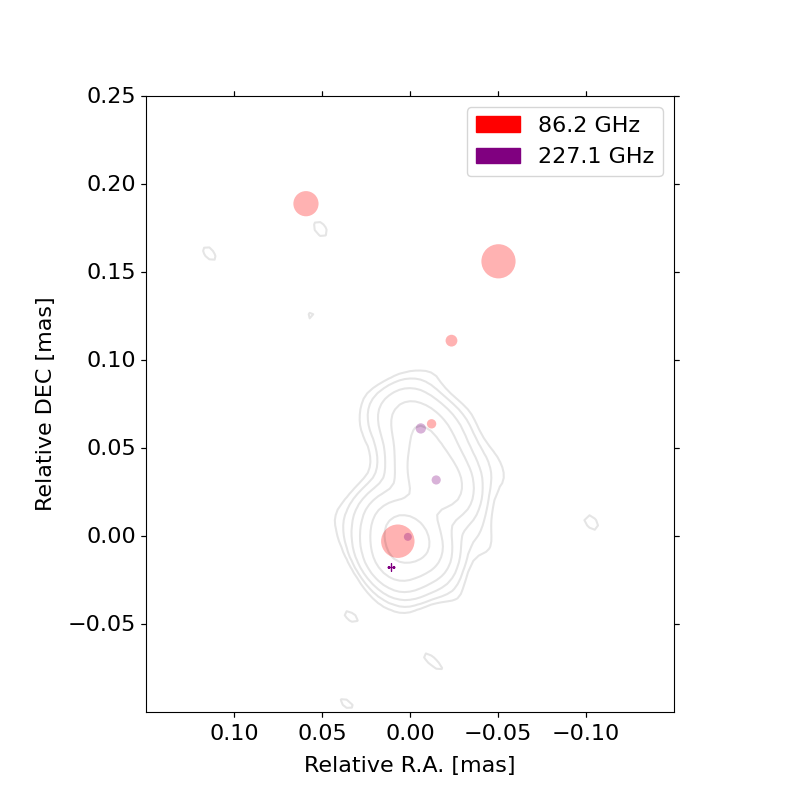}
    \caption{Pairwise image alignment based on same-uv-range modelfit models. \rt{Dash-line circles indicate the pairs of cross-identified components that were used to derive the image shift.} Left to right: 15-43\,GHz. 43-86\,GHz, and 86-227\,GHz. \ro{In each image, contours represent total intensity at the higher frequency.}}
    \label{fig:alignment}
\end{figure*}

Pairwise image alignment based on modelfit models is presented in Fig.~\ref{fig:alignment}. For each frequency there were several model components that had counterparts in the model at another frequency\rt{, indicated by dash-line circles}. All these pairs of components contributed to the derived shift proportionally to their flux \rt{-to-size ratio}. 15-43\,GHz and 43-86\,GHz shifts are based on several pairs of cross-identified components and are robust. \ro{Moreover, at 15-43~GHz and 43-86~GHz, uv-data were cropped to the same range of projected baselines before modelfitting. This approach allowed us to obtain very similar models at both frequencies in each pair.} 86-227\,GHz shift is measured based on the only one component pair. However, with this alignment shift the core at 86\,GHz coincides with the core at 227\,GHz, which is the case if both are optically thin as discussed in the main text. 

\begin{figure*}
    \centering
    \includegraphics[width=0.3\linewidth]{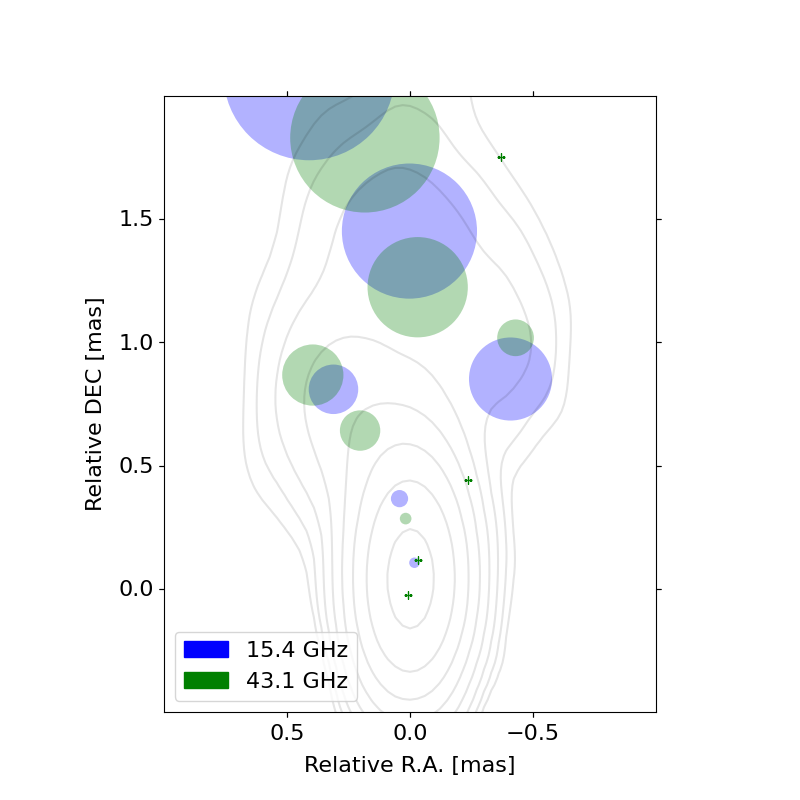}
    \includegraphics[width=0.3\linewidth]{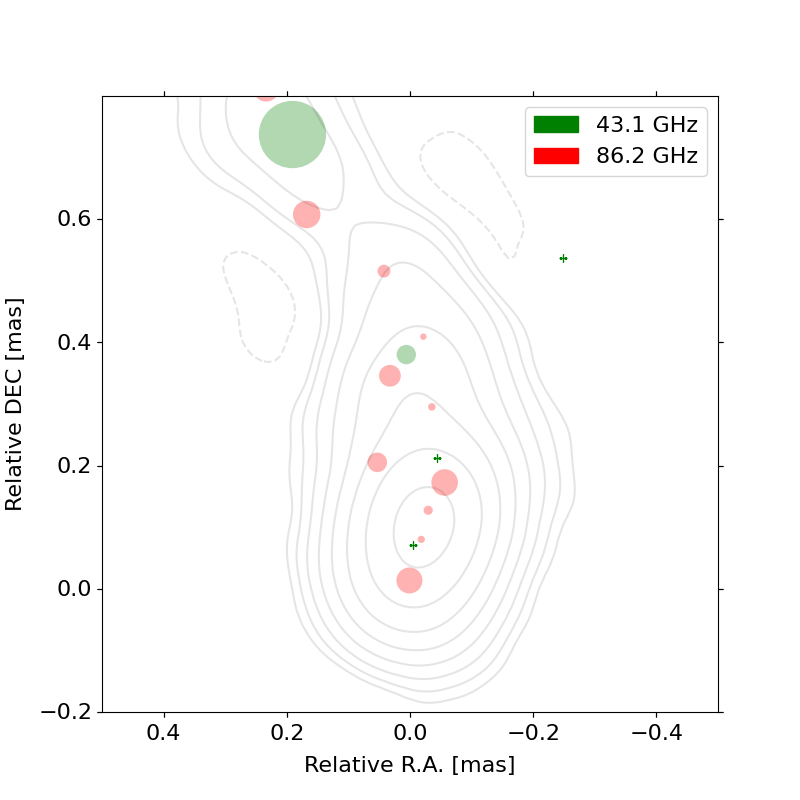}
    \includegraphics[width=0.3\linewidth]{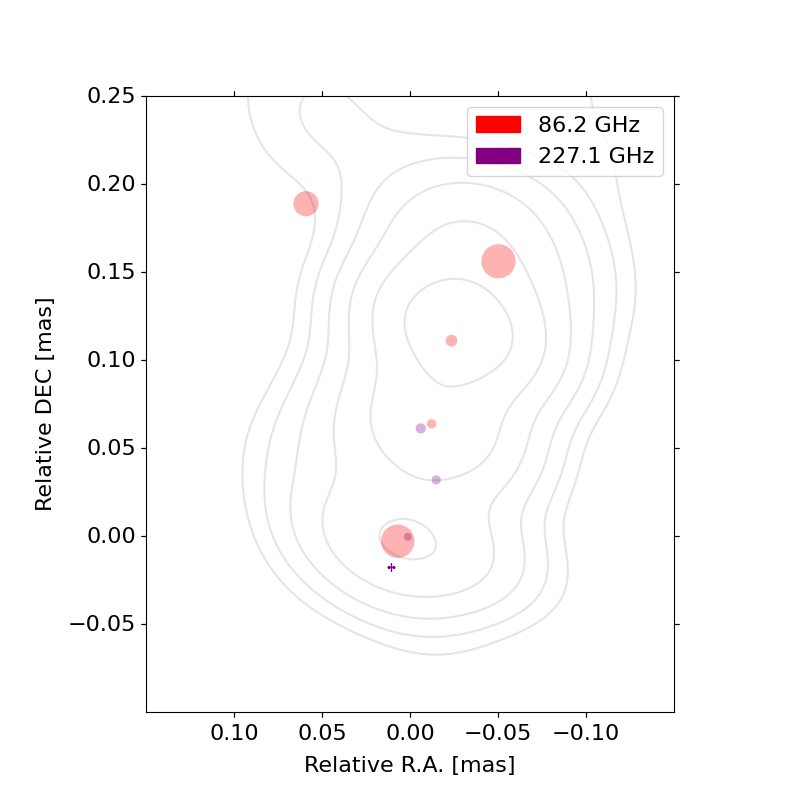}
    \caption{\ro{Best modelfit models after aligning the images. Left to right: 15-43\,GHz. 43-86\,GHz, and 86-227\,GHz. The last pair represents the same model, as that shown in Fig.~\ref{fig:alignment}, but with intensity contours from the lower frequency.}}
    \label{fig:alignment_original}
\end{figure*}

\ro{After deriving the shift between images, we have used the best models at each frequency, without cropping uv-range, to derive the core shift. These models are shown in Fig.~\ref{fig:alignment_original}. The two-step approach allowed us to obtain the best possible accuracy of the core-shift measurement with this data set.}

\begin{figure*}
    \centering
    \includegraphics[width=0.3\linewidth]{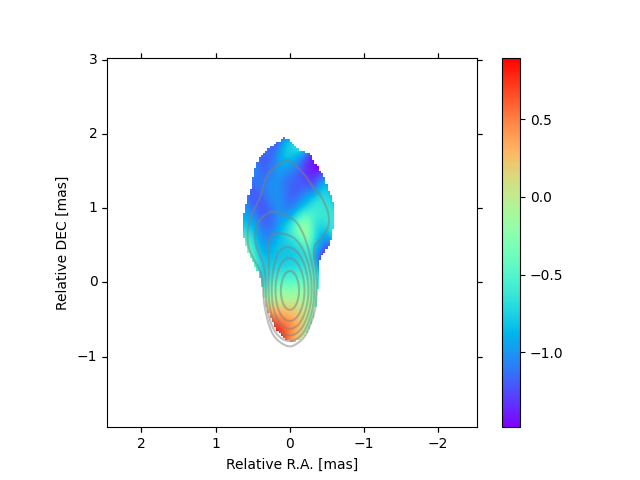}
    \includegraphics[width=0.3\linewidth]{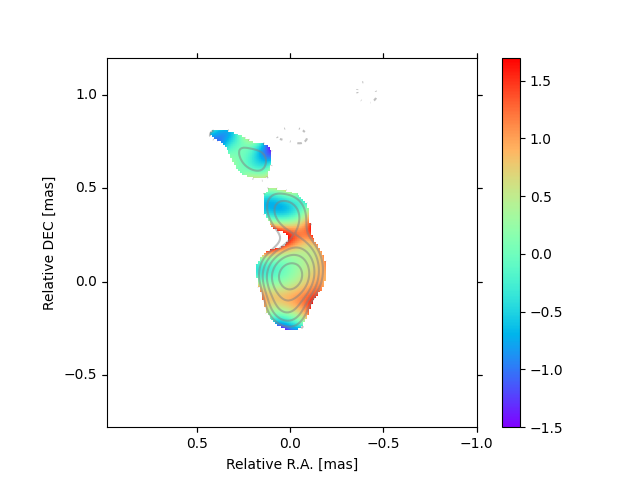}
    \includegraphics[width=0.3\linewidth]{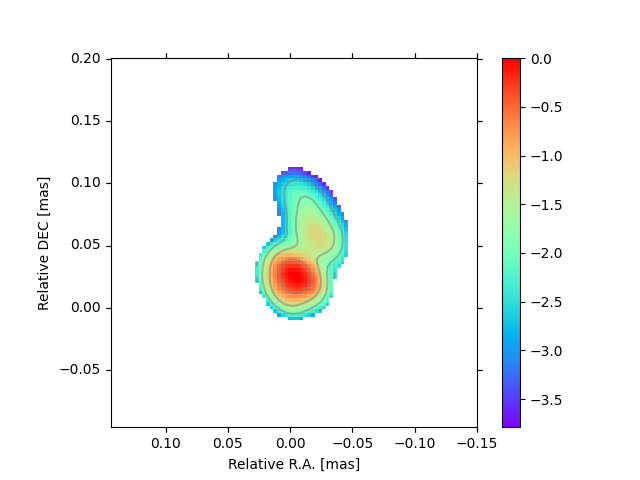}
    \caption{\ro{Pairwise spectral index maps. Left to right: 15-43\,GHz. 43-86\,GHz, and 86-227\,GHz. Contours indicate total intensity at the higher frequency while colors represent spectral index $\alpha$, defined as $S_\nu\propto\nu^\alpha$.}}
    \label{fig:spix}
\end{figure*}

\ro{To test the correctness of derived shifts, we have produced pairwise spectral index maps for frequency pairs 15--43, 43--86, 86--227~GHz. Spectral index maps are presented in Fig.~\ref{fig:spix}. It is clearly visible that the 15--43~GHz spectral index distribution looks as expected, with optically thick core and optically thin jet. At higher frequencies, however, spectral index maps, especially that at 43--86~GHz, are more complex, reflecting intrinsically complex distribution of plasma opacity and parameters. At 86--227~GHz, spectral index varies across the map but always stays $\alpha\leq0$, hinting at the optically thin emission. However, it should be taken with caution since the total flux at 86~GHz was manually scaled, as stated in Sect.~\ref{sec:data}.}

\section{Data interpolation using modelfit models}
\label{app:data_interpolation}

\begin{figure}[H]
    \centering
    \includegraphics[width=0.95\columnwidth]{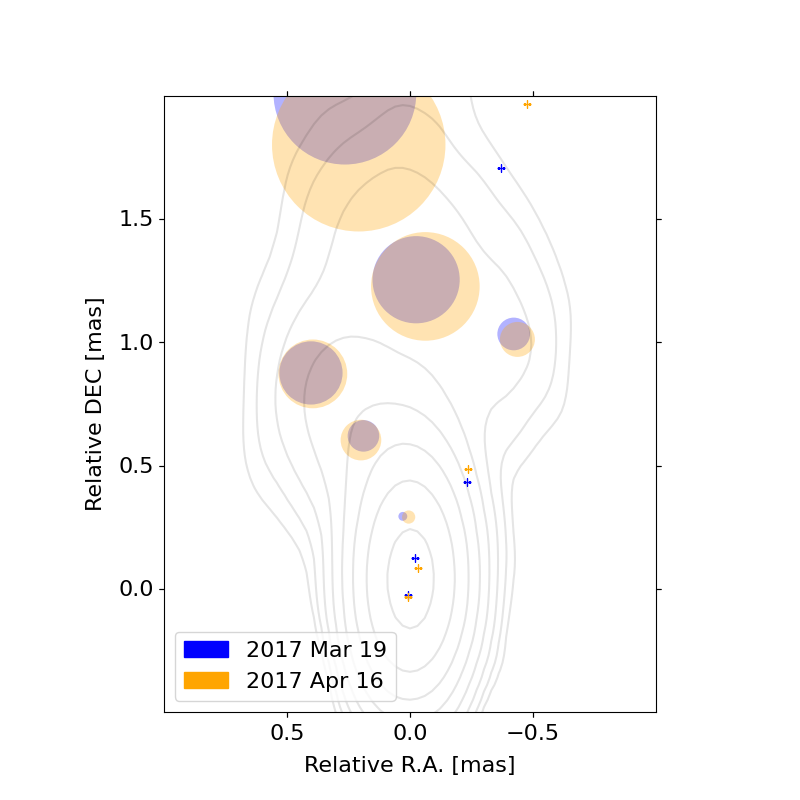}
    \caption{\nrao 43~GHz models for observations taken on 19~Mar and 16~Apr, 2017. The structure consists of the same number of components and these components are easy to cross-identify.}
    \label{fig:43ghz_model_comparison}
\end{figure}

When going to higher and higher frequencies, it is more important to work with simultaneous multifrequency data since structural changes in the source might be significant even on a timescale of a week. For \nrao, the typical apparent velocity of its moving features is $\approx 1 \mu\textrm{as}$ per day. 15~GHz and 43~GHz data were taken during MOJAVE and BU campaigns two months and two weeks apart from the EHT 2017 observations respectively. 
Structural changes are detectable but modelfit model taken before and after the EHT campaign look similar, see Fig.~\ref{fig:43ghz_model_comparison}. Therefore, we have taken the components identified in both observations and, for each pair of components, linearly interpolated their parameters to a given date, Apr~4, 2017 in this case. Since we were using only circular Gaussian components and $\delta$-functions, we have only interpolated flux density, position, and size of components.


\begin{figure}[H]
    \centering
    \includegraphics[width=0.95\columnwidth]{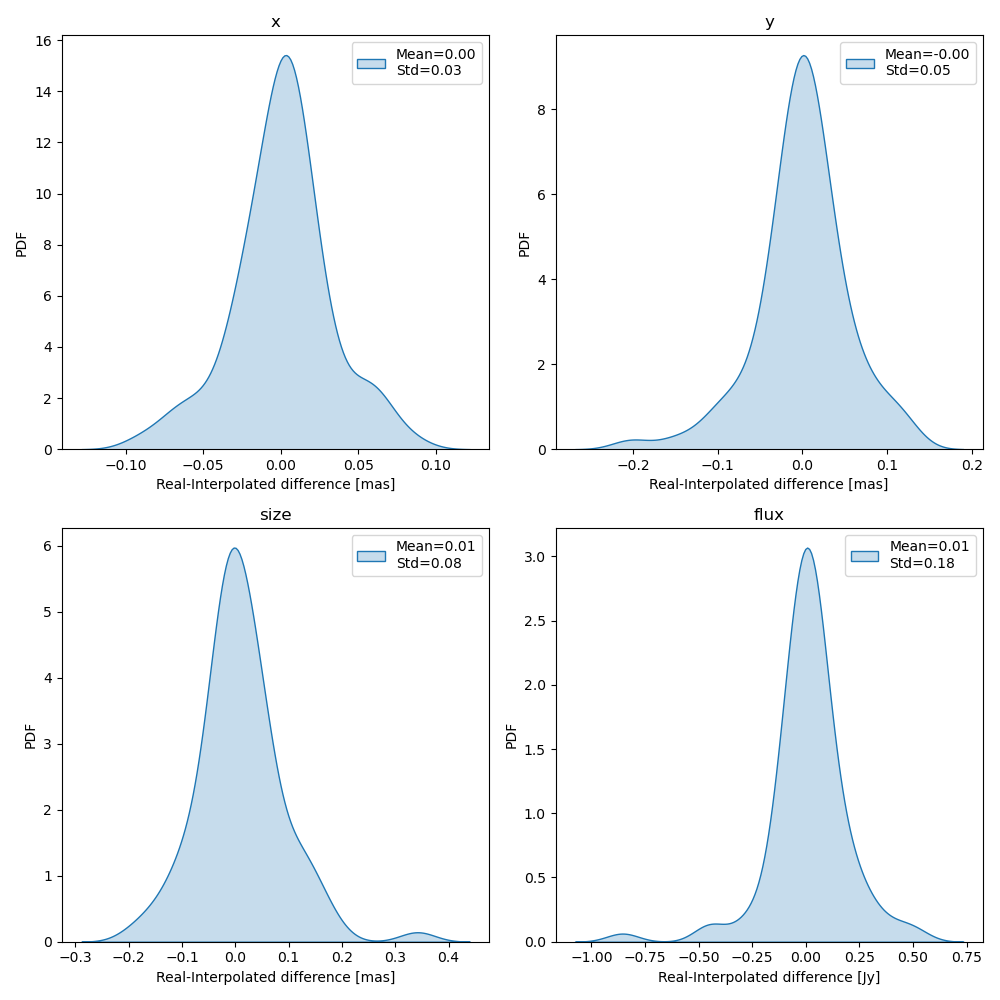}
    \caption{Probability density function for the difference between the real parameters of the model components and the interpolated values. Top row: component position in R.A.(x) and DEC(y) directions. Bottom row: size and flux density differences. X, Y, and size difference is measured in mas, while flux density difference is measured in Jy.}
    \label{fig:interpol_difference}
\end{figure}

  \begin{figure*}[h!]
    \centering
    \includegraphics[width=0.42\textwidth,angle=270]{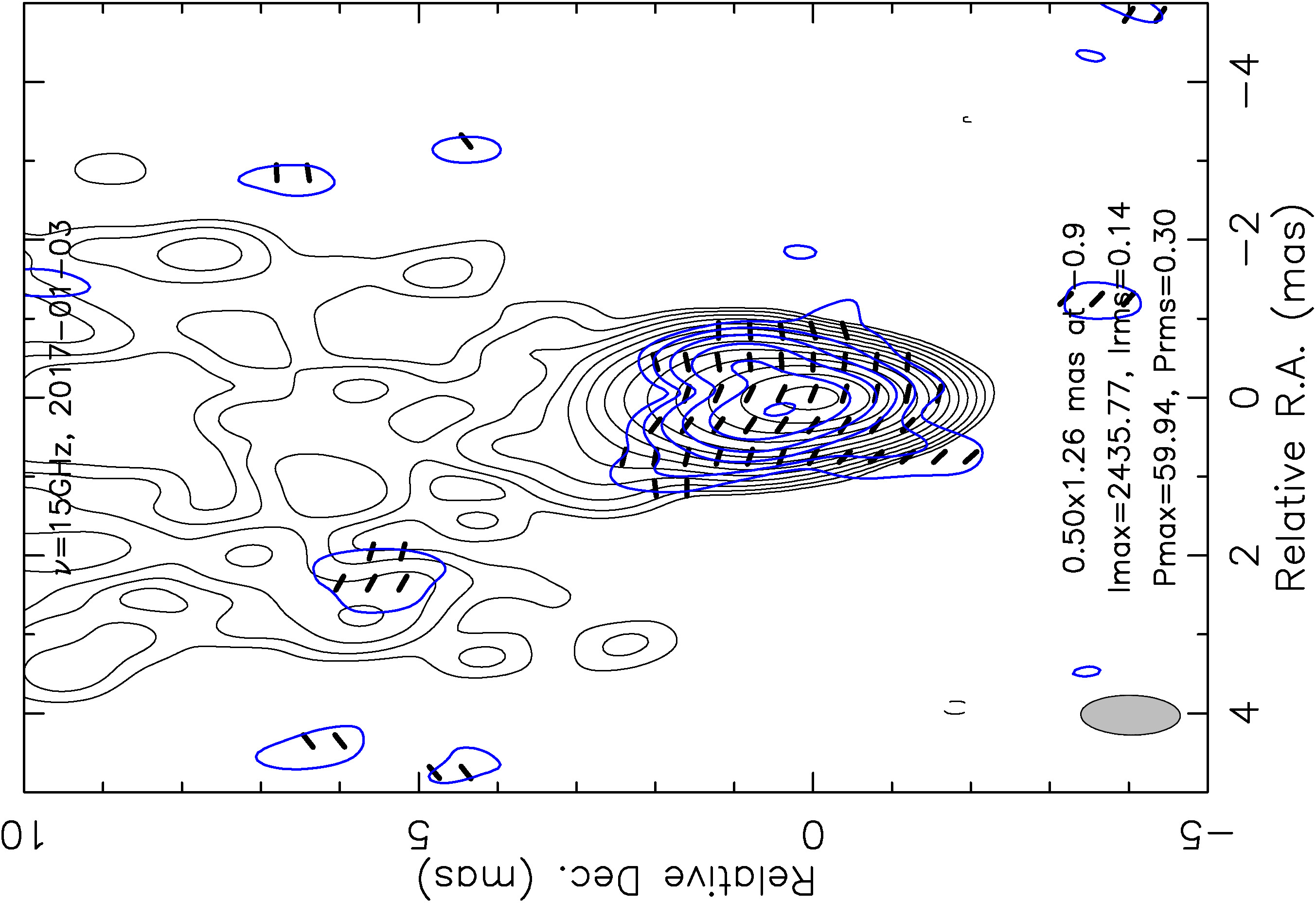}
    \includegraphics[width=0.42\textwidth,angle=270]{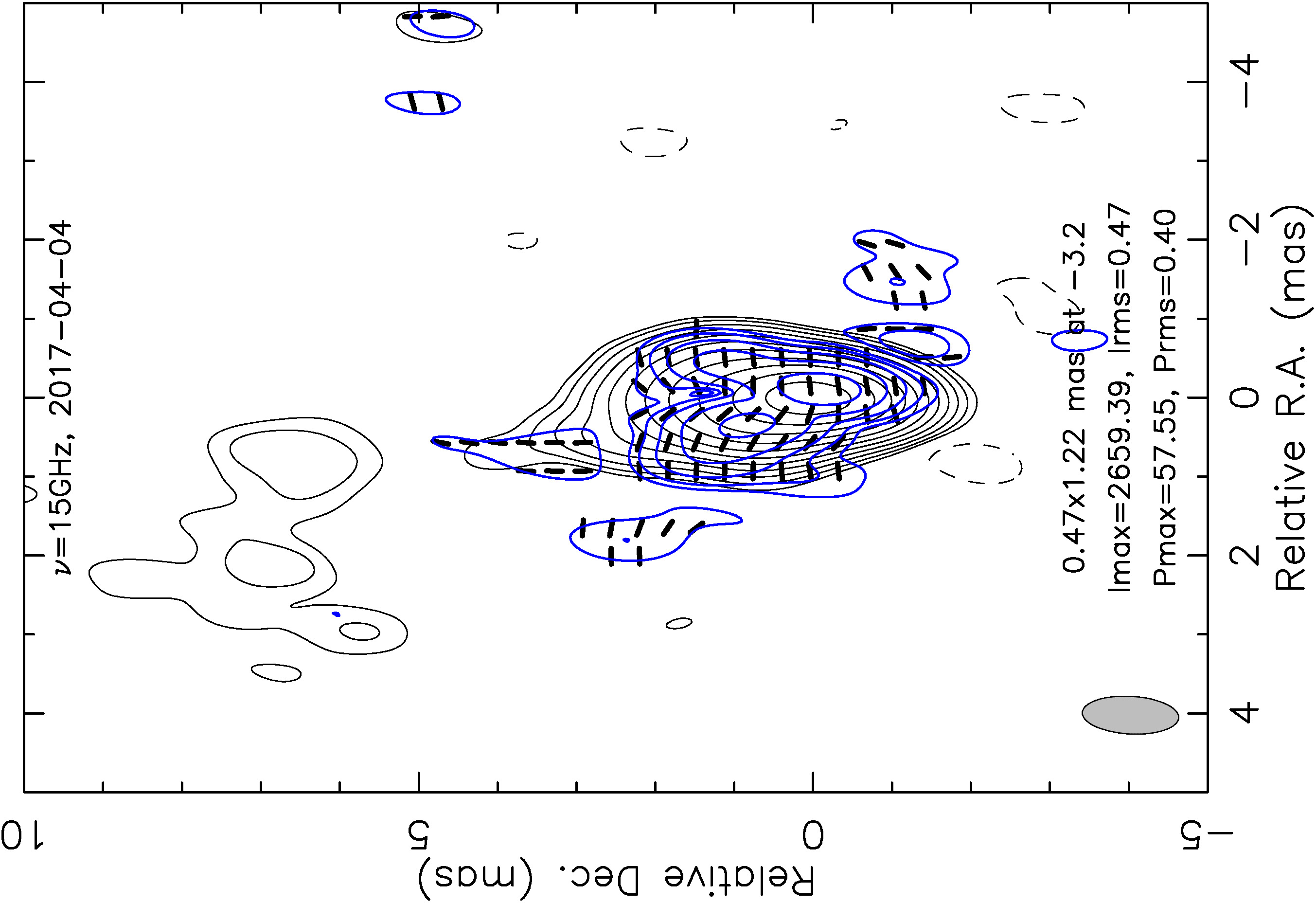}
    \includegraphics[width=0.42\textwidth,angle=270]{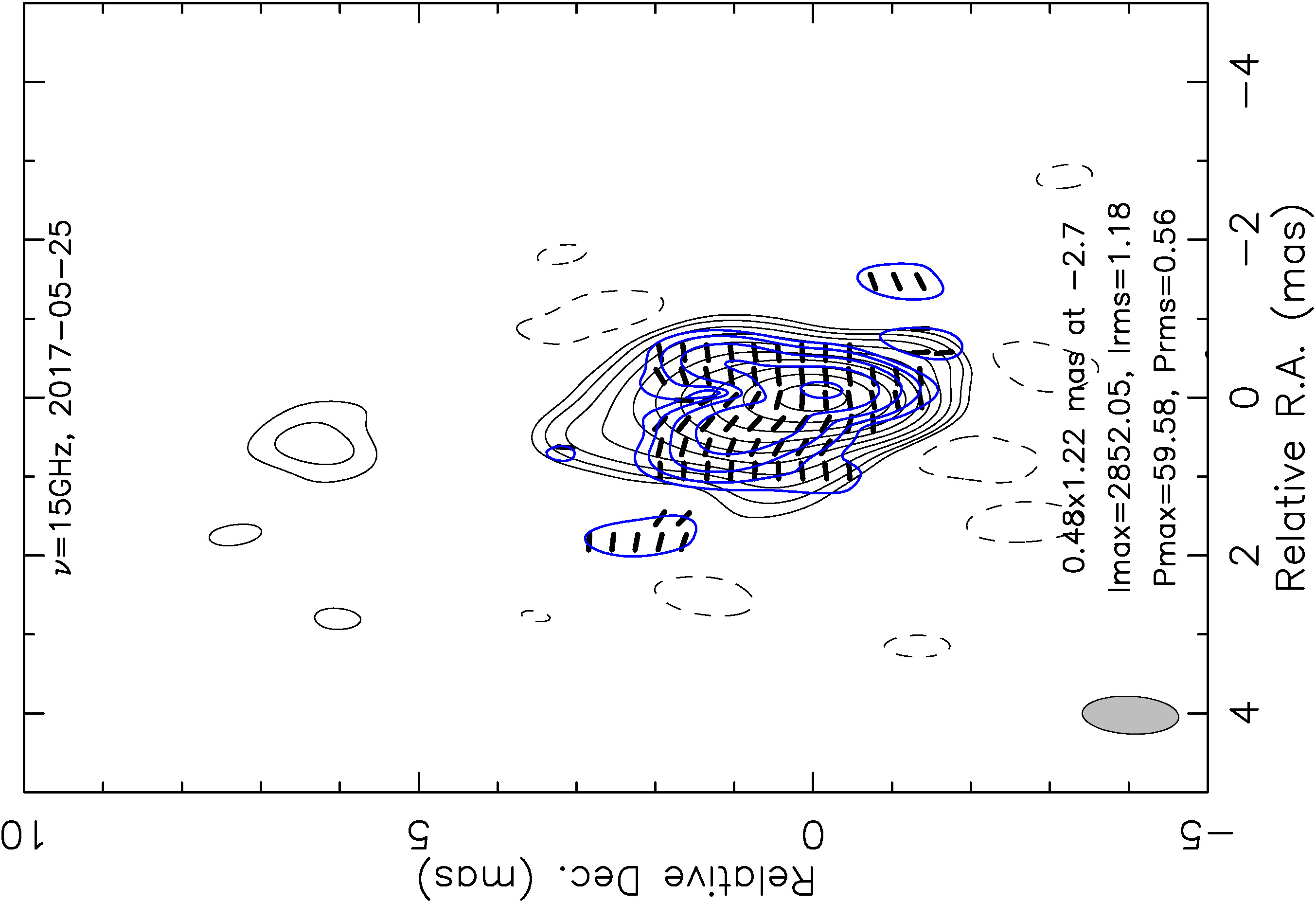}\\
    \includegraphics[width=0.39\textwidth,angle=270]{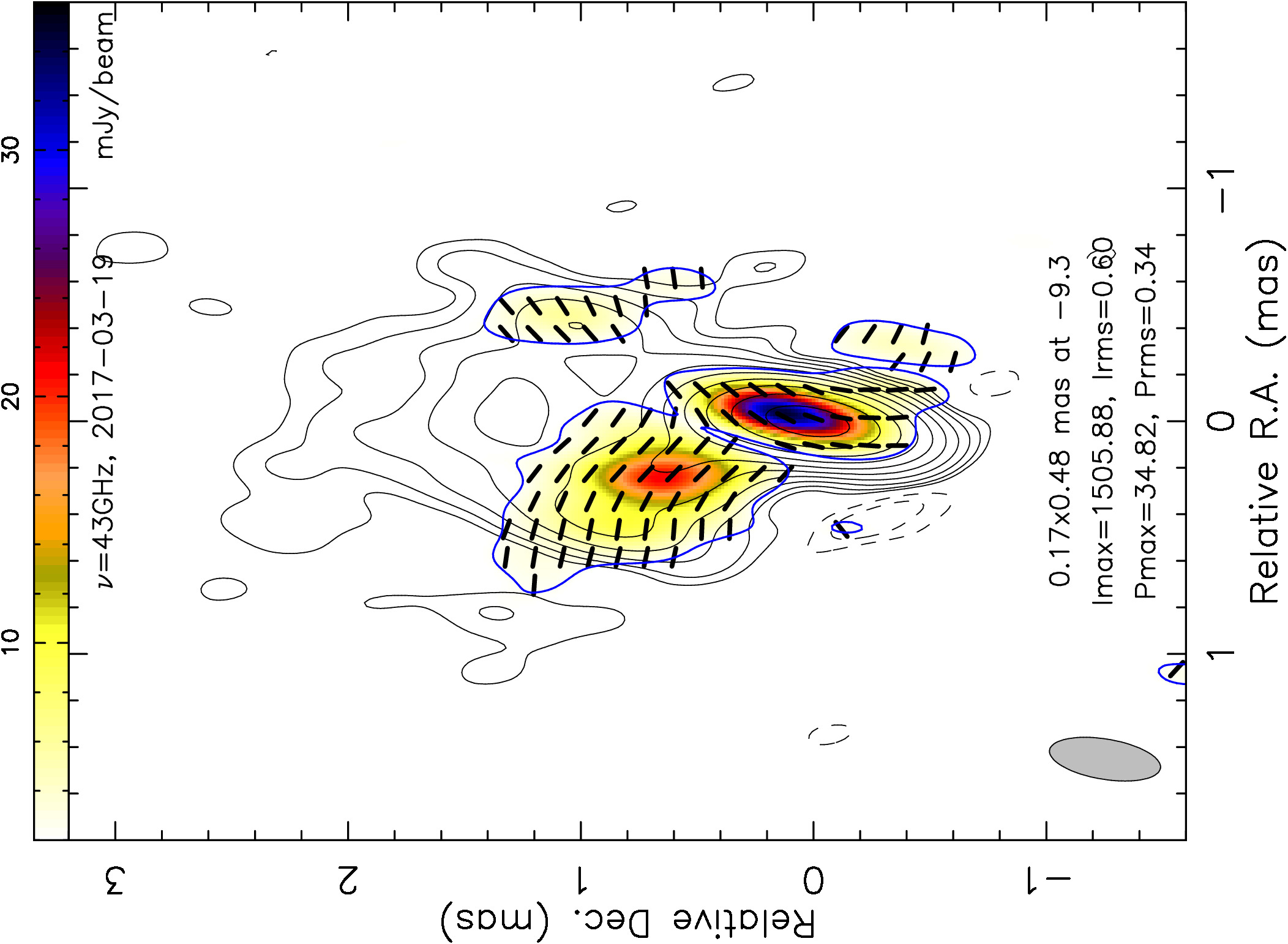}
    \includegraphics[width=0.39\textwidth,angle=270]{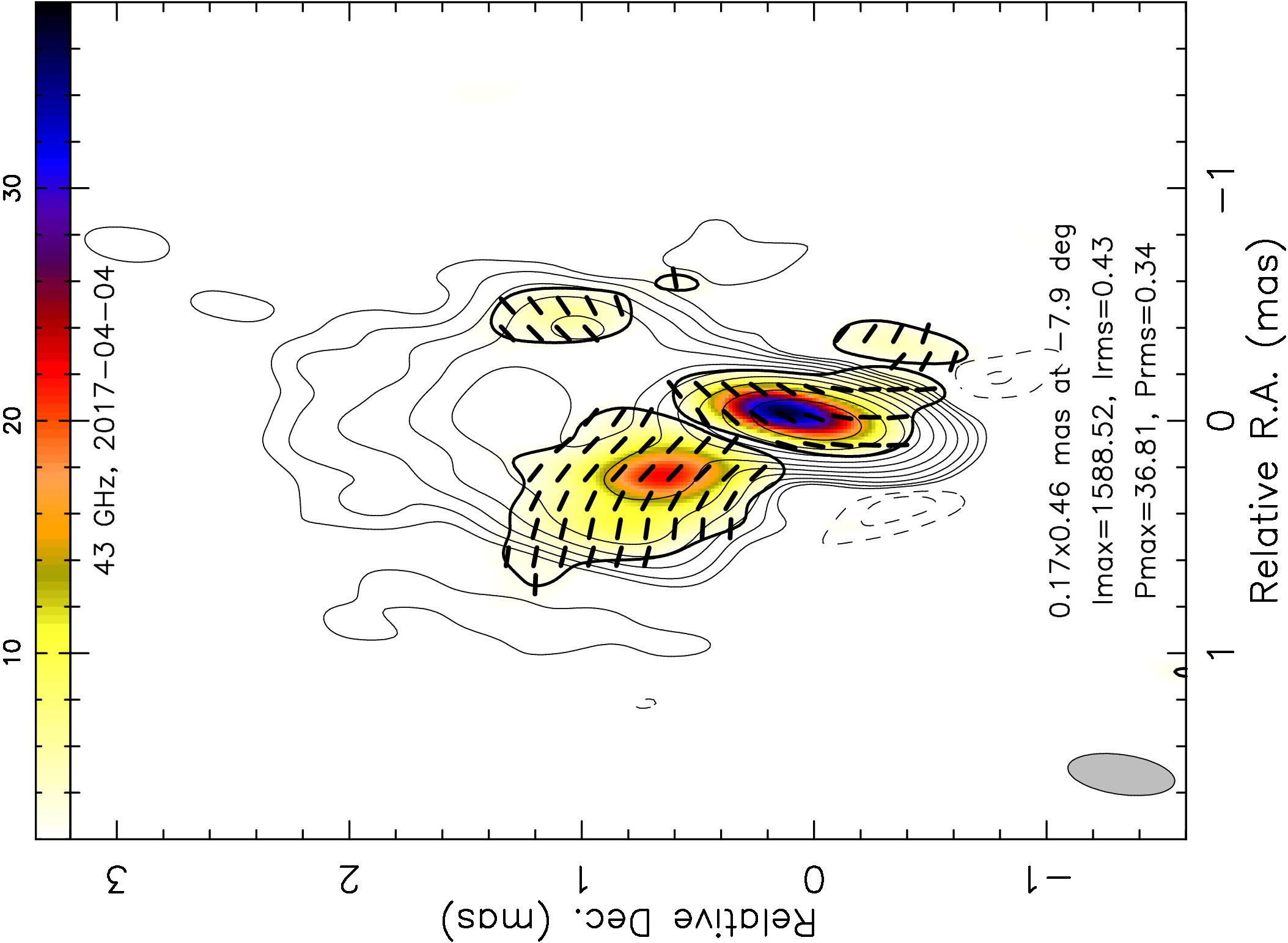}
    \includegraphics[width=0.39\textwidth,angle=270]{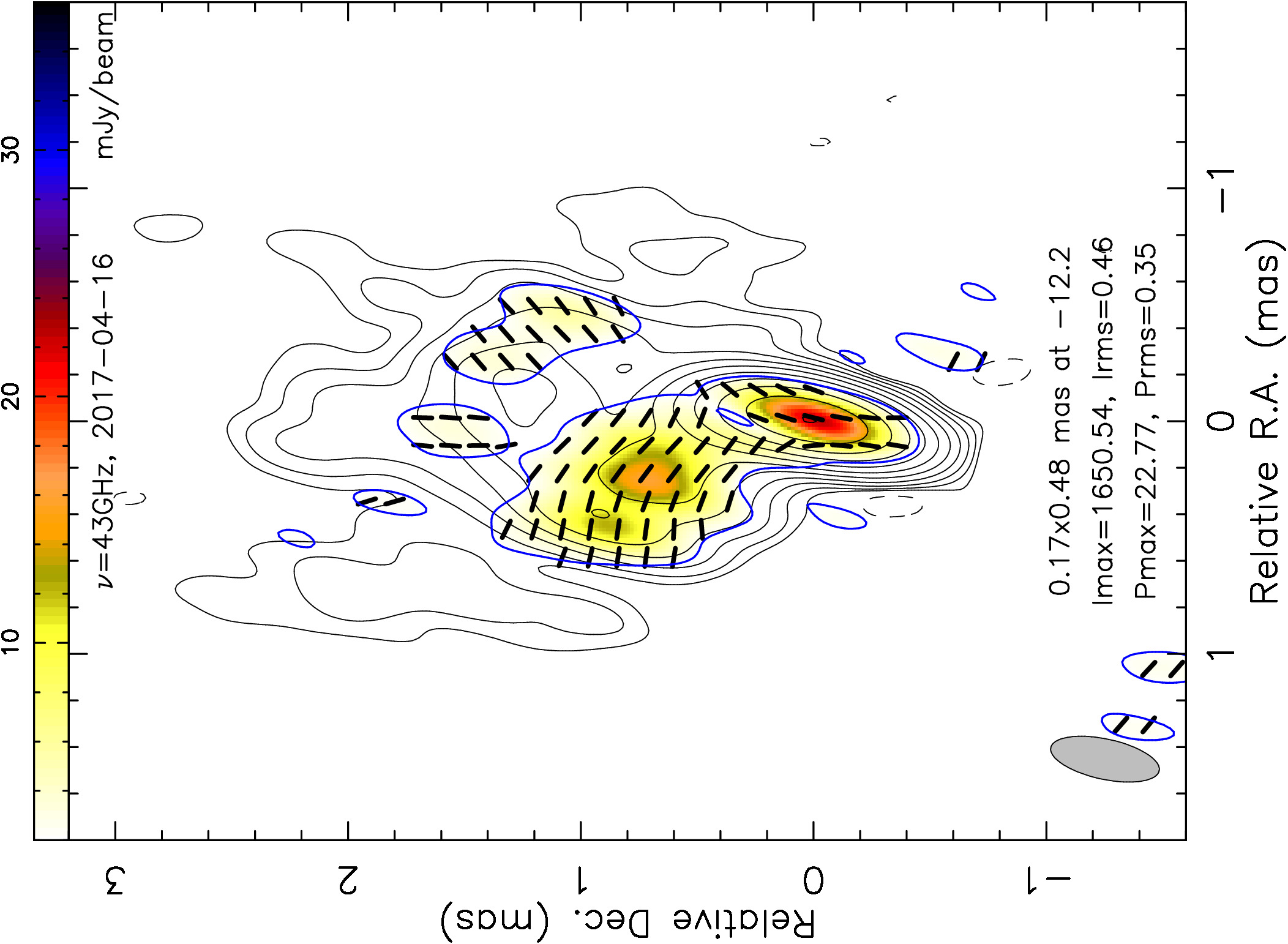}
    \caption{(top row) 15\,GHz polarimetric images obtained on 2017-01-03 (left) and 2017-05-25 (right), and the interpolated one for the epoch of the EHT observations (middle). (bottom row) 43\,GHz polarimetric images obtained on 2017-03-19 (left) and 2017-04-16 (right), and the interpolated one for the epoch of the EHT observations (middle). Natural weighting. Total and polarized intensity contours and EVPAs are plotted. Synthesized beams (shaded ellipse) and image parameters (in mJy/beam) are given.}
    \label{fig:app_pol_moj}
 \end{figure*}

We have conducted a test to estimate the level of uncertainty introduced by the interpolation procedure. Namely, we have compared the difference between the observed parameters of model components for a given observation with those obtained by interpolation between adjacent epochs. Specifically, we have taken all published models for \nrao at 43~GHz \citep{Weaver22} for many epochs. All data were organized into triplets of consecutive models. The first and the last models were used to make an interpolated model for the middle epoch which was compared to the real observed one.

For the comparison we have cross-identified model components between the real and the interpolated models and used only close pairs of components, lying within 20\% of their size from each other. The core components were used to align models and hence were excluded from this analysis. Resulting distributions in Fig.~\ref{fig:interpol_difference} show that the mean interpolated values for component positions are close to zero and standard deviation is of the order of 35~$\mu$as. This value is comparable to the uncertainty in component position for most of model components in each observation.
We also have not found any strong degradation of the interpolation accuracy with the time difference between the epochs used for interpolation.

As noted in Sect.~\ref{sec:data}, reduced $\chi^2$ of these modelfit models is close to that of the CLEAN models. Therefore we self-calibrated the data for epoch 2017~Mar~19 to the interpolated model. 
This approach allowed us to use image-specific analysis, like producing spectral index maps, using interpolated simultaneous data and keeping data errors taken from real a observation. 
The same procedure was applied to both 15~GHz and 43~GHz data. 

According to the kinematic studies \citep{2021ApJ...923...30L, Weaver22}, no new jet component was ejected in 2017. Assuming the structural changes are small at 15 and 43\,GHz in the time span of a few months, the interpolation is justified and resultant rotation measure maps are reliable. In Fig.~\ref{fig:app_pol_moj} we plot the original 15 and 43\,GHz images at available epochs (See Table~\ref{tab:observ}) together with the interpolated images.

\section{Model components}
\label{app:all_models}
In Table~\ref{tab:models} we list all model components at all frequencies used in the paper: 15, 22, 43, 86, and 227\,GHz. 
These models provide a good fit to the data with $\chi^2$ comparable to that of \textsc{CLEAN}. At 227\,GHz the naming of components is adopted from \cite{Jorstad2023}. \ro{The component-specific errors were determined in the image plane following eq.~14-5 from \cite{1999ASPC..180..301F}.}

\begin{table*}[t]
    \caption{Components of the models at all frequencies}
    \label{tab:models}
    \centering
    \begin{tabular}{c|c|c|c|c|c}
ID & Flux density &  Radius & Theta & Size & Frequency \\
 &  (Jy) & (mas) & (degrees) & (mas) & (GHz) \\
\hline
U0 & 1.854 $\pm$ 0.001 & 0.063 & -165.5 & 0.105 & 15.4  \\
U1 & 0.832 $\pm$ 0.003 & 0.205 & 12.7 & 0.105 $\pm$ 0.001 & 15.4  \\
U2 & 0.364 $\pm$ 0.006 & 0.715 $\pm$ 0.003 & 25.9 & 0.201 $\pm$ 0.006 & 15.4  \\
U3 & 0.130 $\pm$ 0.003 & 0.796 $\pm$ 0.003 & -30.6 & 0.337 $\pm$ 0.007 & 15.4  \\
U4 & 0.483 $\pm$ 0.011 & 1.285 $\pm$ 0.002 & 0.2 & 0.549 $\pm$ 0.004 & 15.4  \\
U5 & 0.054 $\pm$ 0.010 & 1.964 $\pm$ 0.017 & 12.1 & 0.695 $\pm$ 0.035 & 15.4  \\
U6 & 0.071 $\pm$ 0.051 & 6.823 $\pm$ 0.177 & 12.5 & 3.013 $\pm$ 0.442 & 15.4  \\
U7 & 0.101 $\pm$ 0.060 & 11.781 $\pm$ 0.204 & 2.3 & 3.741 $\pm$ 0.527 & 15.4  \\
U8 & 0.146 $\pm$ 0.087 & 24.486 $\pm$ 0.535 & -0.8 & 5.483 $\pm$ 1.474 & 15.4  \\
\hline
K0 & 2.070 & 0.071 & -164.3 & 0.146 & 22.2  \\
K1 & 0.805 $\pm$ 0.019 & 0.724 $\pm$ 0.002 & 4.3 & 0.632 $\pm$ 0.004 & 22.2  \\
K2 & 0.133 $\pm$ 0.020 & 1.713 $\pm$ 0.011 & -0.8 & 0.752 $\pm$ 0.022 & 22.2  \\
\hline
Q0 & 1.144 & 0.024 & 164.9 & 0.037 & 43.1  \\
Q1 & 0.467 $\pm$ 0.002 & 0.121 & -15.2 & 0.037 & 43.1  \\
Q2 & 0.233 & 0.285 & 3.6 & 0.047 & 43.1  \\
Q3 & 0.025 $\pm$ 0.003 & 0.501 $\pm$ 0.007 & -28.1 & 0.037 $\pm$ 0.014 & 43.1  \\
Q4 & 0.080 $\pm$ 0.005 & 0.673 $\pm$ 0.002 & 17.5 & 0.164 $\pm$ 0.004 & 43.1  \\
Q5 & 0.104 $\pm$ 0.007 & 0.953 $\pm$ 0.002 & 24.5 & 0.248 $\pm$ 0.004 & 43.1  \\
Q6 & 0.044 $\pm$ 0.004 & 1.105 $\pm$ 0.004 & -22.8 & 0.149 $\pm$ 0.007 & 43.1  \\
Q7 & 0.134 $\pm$ 0.013 & 1.224 $\pm$ 0.003 & -1.5 & 0.407 $\pm$ 0.005 & 43.1  \\
Q8 & 0.007 $\pm$ 0.001 & 1.789 $\pm$ 0.009 & -11.9 & 0.037 $\pm$ 0.018 & 43.1  \\
Q9 & 0.046 $\pm$ 0.016 & 1.839 $\pm$ 0.011 & 5.7 & 0.606 $\pm$ 0.023 & 43.1  \\
\hline
W0 & 0.553 $\pm$ 0.005 & 0.014 & 3.9 & 0.063 & 86.2  \\
W1 & 0.580 $\pm$ 0.002 & 0.082 & -12.8 & 0.019 & 86.2  \\
W2 & 0.778 $\pm$ 0.001 & 0.131 & -13.1 & 0.022 & 86.2  \\
W3 & 0.206 $\pm$ 0.005 & 0.182 & -18.1 & 0.065 & 86.2  \\
W4 & 0.061 $\pm$ 0.003 & 0.212 & 14.5 & 0.048 $\pm$ 0.001 & 86.2  \\
W5 & 0.160 $\pm$ 0.002 & 0.297 & -6.8 & 0.019 & 86.2  \\
W6 & 0.141 $\pm$ 0.004 & 0.347 & 5.4 & 0.053 & 86.2  \\
W7 & 0.063 $\pm$ 0.001 & 0.410 & -3.0 & 0.019 & 86.2  \\
W8 & 0.040 $\pm$ 0.003 & 0.517 & 4.7 & 0.031 $\pm$ 0.002 & 86.2  \\
W9 & 0.051 $\pm$ 0.005 & 0.630 $\pm$ 0.001 & 15.4 & 0.067 $\pm$ 0.002 & 86.2  \\
W10 & 0.074 $\pm$ 0.005 & 0.845 & 16.1 & 0.062 $\pm$ 0.001 & 86.2  \\
W11 & 0.025 $\pm$ 0.003 & 1.158 $\pm$ 0.001 & -6.0 & 0.054 $\pm$ 0.003 & 86.2  \\
W12 & 0.017 $\pm$ 0.001 & 1.161 $\pm$ 0.001 & 0.6 & 0.019 $\pm$ 0.003 & 86.2  \\
W13 & 0.014 $\pm$ 0.002 & 1.235 $\pm$ 0.002 & 2.3 & 0.019 $\pm$ 0.003 & 86.2  \\
\hline
C0b & 0.067 $\pm$ 0.007 & 0.021 & 148.7 & 0.007 & 227.1  \\
C0a & 0.303 $\pm$ 0.016 & 0.001 & 111.3 & 0.015 & 227.1  \\
C1 & 0.188 $\pm$ 0.022 & 0.035 & -25.0 & 0.017 & 227.1  \\
C2 & 0.104 $\pm$ 0.021 & 0.061 & -5.7 & 0.020 & 227.1  \\
BG* & 0.793 $\pm$ 1.031 & 0.000 $\pm$ 5.835 & 0.0 & 1.000 $\pm$ 16.503 & 227.1  \\
\hline
\end{tabular}
\tablefoot{Columns are: ID -- component ID in the corresponding model, Radius and Theta -- polar coordinates of the component, Size -- the maximum of the fitted Gaussian FWHM and the resolution limit at the position of the component, Frequency -- the frequency of the model. (*) component BG was introduced to account for the large-scale flux at 227\,GHz, which is resolved out at most of the baselines except for the shortest ones. For the 227\,GHz component names we used those introduced by \cite{Jorstad2023} for the ease of comparison. \ro{The uncertainty is given only in cases when it exceeds the precision of displayed values.} }
\end{table*}

\section{Jet shape based on component sizes}
\label{app:jet_shape}
\begin{figure}
    \centering
    \includegraphics[width=0.49\textwidth]{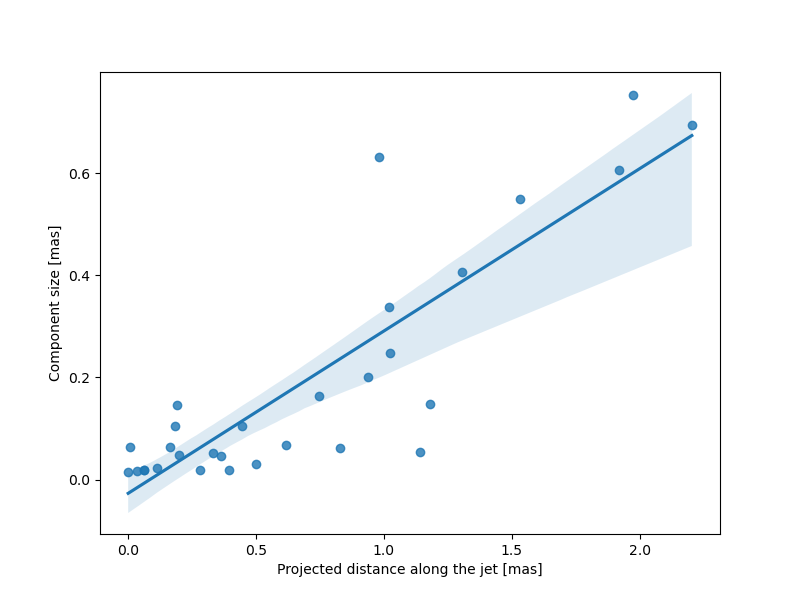}
    \caption{Size of components \ro{at all frequencies} depending on their angular separation from the apparent jet beginning at 227~GHz. The line and the shaded area show the best linear fit to the data and the estimate of uncertainty. This dependency reflects the shape of the one-epoch jet.}
    \label{fig:size_radius}
\end{figure}

To ensure that the assumptions that we made to use the core-shift method hold, we have checked the shape of the one-epoch jet (the one shown in Fig.~\ref{fig:all_models}). We analyzed how the size of individual components depends on their separation from the jet beginning. In Fig.~\ref{fig:size_radius}, we show the measurements and a linear fit. Before fitting, some outliers were removed, namely, unresolved components and those located beyond 10~mas. Linear fit describes the data in the innermost 2~mas quite well, which means that a conical shape of the jet is a valid assumption.
The fit yields the jet apparent opening angle $\phi_\mathrm{app} =17^{\circ +4}_{-8}$. \rt{The value $\phi_\mathrm{app} =10^{\circ}$ ($\phi_\mathrm{int} =0.5^{\circ}$) measured by \cite{2009A&A...507L..33P} falls within our uncertainties and we used it in our calculations.}

\section{Geometry-induced Doppler factor variations}
\label{app:doppler_variations}
\begin{figure}
    \centering
    \includegraphics[width=0.49\textwidth]{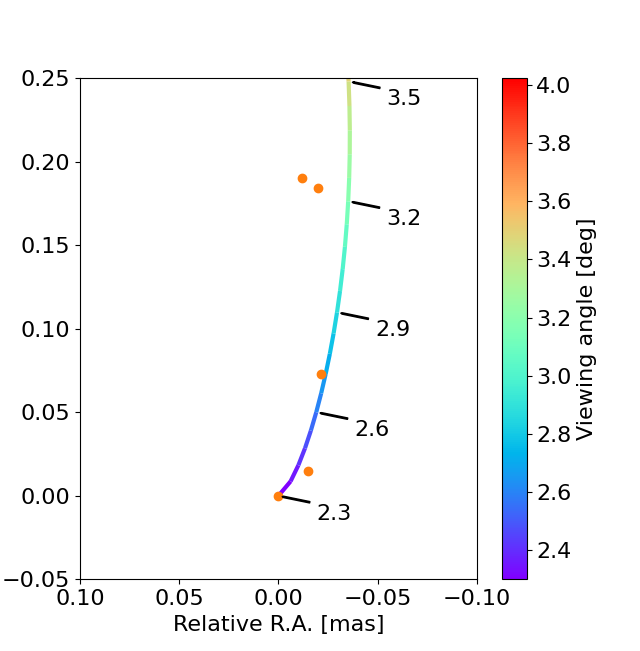}
    \caption{Zoom into the central region of the \nrao. Orange dots show position of the apparent cores at different frequencies. The line shows the precession model and is colored according to the local viewing angle. Several values are marked for convenience.}
    \label{fig:zoom_theta}
\end{figure}

With the model of a precessing jet that we built in Sec.\ref{sec:fitting_pr_instab}, we can estimate the impact of the varying viewing angle on the Doppler-boosting factor in the region $r_\mathrm{deproj} < 5$\,pc, where apparent cores at different frequencies are located. Fig.~\ref{fig:zoom_theta} shows how the viewing angle $\theta$ changes along the jet, as predicted by the fitted precession model. It is clear, that for the region of interest, the jet is turning farther away from the observer. While it can increase the path length through the thermal plasma surrounding the jet and hence increase the absolute value of the Faraday rotation measure, the impact on the Doppler factor is negligible for the jet bulk Lorenz factors $\Gamma < 15$. We note, that $\Gamma = 15$ might be required to explain the rise of the brightness temperature along the jet in this region, but further downstream $\Gamma = 9$ is estimated.

\section{Component trajectories}
\label{app:bu_component_map}
\begin{figure}
    \centering
    \includegraphics[width=0.95\columnwidth]{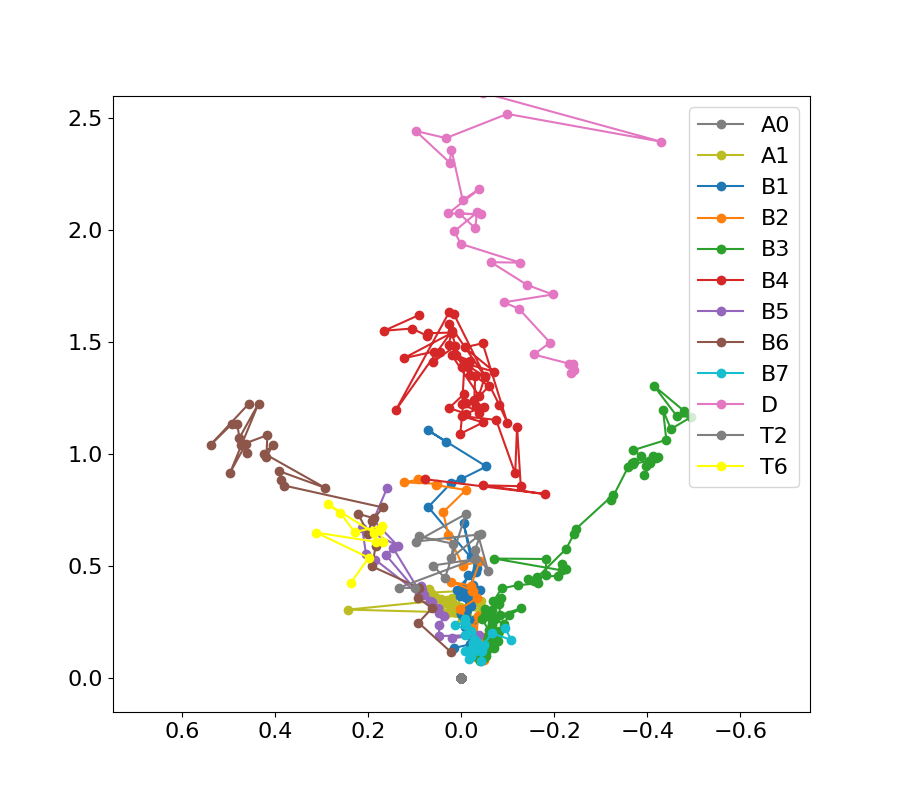}
    \caption{Trajectories of all model components at 43\,GHz covering a period from 2007 to 2018. Plotted based on the data from \cite{Weaver22} .}
    \label{fig:bu_component_map}
\end{figure}

Fig.~\ref{fig:bu_component_map} shows trajectories of all model components in \nrao at 43\,GHz. The measurements were taken between 2007 and 2018 by \cite{Weaver22}. Trajectories of components in the innermost 1~mas from the core show straight trajectories which is consistent with the precession model for the jet position angle variations.

\vfill\eject
\end{document}